\documentclass[10pt,journal,compsoc]{IEEEtran}
\usepackage{dblfloatfix}
\usepackage{setspace}
\usepackage{mathrsfs}
\usepackage[boxed,ruled,vlined,linesnumbered]{algorithm2e}
\usepackage{multirow}
\usepackage{times}
\usepackage{hhline}
\usepackage{url}
\usepackage{framed}
\usepackage{hyperref}
\usepackage{longtable}
\usepackage{mdframed}

\newcommand*\wrapletters[1]{\wr@pletters#1\@nil}
\def\wr@pletters#1#2\@nil{#1\allowbreak\if&#2&\else\wr@pletters#2\@nil\fi}
\usepackage{tikz}
\usetikzlibrary{automata,matrix,shapes,arrows,positioning,chains,calc}
\usetikzlibrary{snakes}
\usetikzlibrary{arrows,scopes}
\usetikzlibrary{positioning,chains,fit,shapes,calc}
\usepackage{caption}
\usepackage{subcaption}
\usepackage{amsmath}
\SetKwHangingKw{Local}{Local}
\SetKwHangingKw{Global}{Global}
\SetKwHangingKw{Constant}{Constant}
\SetKwHangingKw{Input}{Input}
\SetKwHangingKw{Alias}{Alias}
\SetKwHangingKw{Output}{Output}
\SetKwHangingKw{SideEffect}{Side effects}
\SetKwHangingKw{InternalEvent}{Internal event}
\SetKwHangingKw{External}{External}
\SetKwHangingKw{ExternalEvent}{External event}
\SetKwHangingKw{Precondition}{Precondition :}
\SetKwHangingKw{Postcondition}{Postcondition :}
\SetKwHangingKw{Upon}{Upon}
\SetKwHangingKw{DoForever}{Do Forever}
\SetKwHangingKw{PVab}{Persistent variables:}

\usepackage{theorem}
\usepackage{wrapfig}
\usepackage{amssymb}
\usepackage{xspace}

\newtheorem{theorem}{Theorem}

\newcommand{\qed}{\hfill $\blacksquare$}
\newcommand{\qedClaim}{\hfill \ensuremath{\Box}}

\newcommand{\B}{\vspace*{-\smallskipamount}}
\newcommand{\BB}{\vspace*{-\medskipamount}}
\newcommand{\BBB}{\vspace*{-\bigskipamount}}

\newcommand{\remove}[1]{}

\usepackage{csquotes}
\usepackage{enumitem}
\usepackage{tablefootnote}
\usepackage{fnpct}

\ifCLASSOPTIONcompsoc
  \usepackage[nocompress]{cite}
\else
  \usepackage{cite}
\fi
\ifCLASSINFOpdf
\else
\fi

\IEEEoverridecommandlockouts

\begin{document}
\title{{\textsc{{\Huge Obscure}}}: Information-Theoretically Secure, Oblivious, and Verifiable Aggregation Queries on Secret-Shared Outsourced Data\thanks{\textbf{A preliminary version of this work was accepted in VLDB 2019.}\protect\\
\textbf{Corresponding author:} Shantanu Sharma \texttt{shantanu.sharma@uci.edu}\protect\\
P. Gupta, S. Mehrotra, S. Sharma, and S. Almanee are with University of California, Irvine, USA.
Yin Li is with Dongguan University of Technology, P.R. China.
N. Panwar is with Augusta University and University of California, Irvine, USA.\protect\\
Manuscript received 05 Oct. 2019; accepted 22 Mar. 2020. DOI: 10.1109/TKDE.2020.2983932. \copyright 2020 IEEE. Personal use of this material is permitted. Permission from IEEE must be obtained for all other uses, including reprinting/republishing this material for advertising or promotional purposes, collecting new collected works for resale or redistribution to servers or lists, or reuse of any copyrighted component of this work in other works. \protect\\
The final published version of this paper may differ from this accepted version.}}


\author{Peeyush Gupta, Yin Li, Sharad Mehrotra, Nisha Panwar, Shantanu Sharma, and Sumaya Almanee}

\markboth{IEEE Transactions on Knowledge and Data Engineering (TKDE); Accepted March 2020.}%
{Gupta \MakeLowercase{\textit{et al.}}: \textsc{Obscure}: Information-Theoretically Secure, Oblivious, and Verifiable Aggregation Queries on Secret-Shared Outsourced Data}
\IEEEtitleabstractindextext{
\begin{abstract}
Despite exciting progress on cryptography, secure and efficient query processing over outsourced data remains an open challenge. We develop a communication-efficient and information-theoretically secure system, entitled \textsc{Obscure} for aggregation queries with conjunctive or disjunctive predicates, using secret-sharing. \textsc{Obscure} is strongly secure (\textit{i}.\textit{e}., secure regardless of the computational-capabilities of an adversary) and prevents the network, as well as, the (adversarial) servers to learn the user's queries, results, or the database. In addition, \textsc{Obscure} provides additional security features, such as hiding access-patterns (\textit{i}.\textit{e}., hiding the identity of the tuple satisfying a query) and hiding query-patterns (\textit{i}.\textit{e}., hiding which two queries are identical). Also, \textsc{Obscure} does not require any communication between any two servers that store the secret-shared data before/during/after the query execution. Moreover, our techniques deal with the secret-shared data that is outsourced by a single or multiple database owners, as well as, allows a user, which may not be the database owner, to execute the query over secret-shared data. We further develop (non-mandatory) privacy-preserving result verification algorithms that detect malicious behaviors, and experimentally validate the efficiency of \textsc{Obscure} on large datasets, the size of which prior approaches of secret-sharing or multi-party computation systems have not scaled to.
\end{abstract}

\begin{IEEEkeywords}
Computation and data privacy, data and computation outsourcing, multi-party computation, Shamir's secret-sharing, result verification.
\end{IEEEkeywords}}
\maketitle

\IEEEdisplaynontitleabstractindextext
\IEEEpeerreviewmaketitle
\ifCLASSOPTIONcompsoc
\IEEEraisesectionheading{\section{Introduction}\label{sec:introduction}}
\else

\section{Introduction}
\label{sec:introduction}
\fi
Database-as-a-service (DaS)~\cite{DBLP:conf/sigmod/HacigumusILM02} allows authenticated users to execute their queries on an untrusted public cloud. Over the last two decades, several cryptographic techniques (\textit{e}.\textit{g}.,~\cite{DBLP:journals/jcss/GoldwasserM84,DBLP:conf/sp/SongWP00,homo,DBLP:conf/sigmod/AgrawalKSX04,DBLP:journals/cacm/Shamir79}) have been proposed secure and privacy-preserving computations in the DaS model. These techniques can be broadly classified based on cryptographic security into two categories:

\noindent\textbf{Computationally secure techniques} that assume the adversary lacks adequate computational capabilities to break the underlying cryptographic mechanism in polynomial time (\textit{i}.\textit{e}., a practically short amount of time). Non-deterministic encryption~\cite{DBLP:journals/jcss/GoldwasserM84}, homomorphic encryption~\cite{homo}, order-preserving encryption (OPE)~\cite{DBLP:conf/sigmod/AgrawalKSX04}, and searchable-encryption~\cite{DBLP:conf/sp/SongWP00} are examples of such techniques. 

\noindent\textbf{Information-theoretically secure techniques} that are unconditionally secure and independent of adversary's computational capabilities. Shamir's secret-sharing (SSS)~\cite{DBLP:journals/cacm/Shamir79} is a well-known information-theoretically secure protocol. In SSS, multiple (secure) shares of a dataset are kept at mutually suspicious servers, such that a single server cannot learn anything about the data. Secret-sharing-based techniques are secure under the assumption that a majority of the servers (equal to the threshold of the secret-sharing mechanism) do not collude. Secret-sharing mechanisms also have applications in other areas such as Byzantine agreement, secure multiparty computations (MPC), and threshold cryptography, as discussed in~\cite{DBLP:conf/codcry/Beimel11}.

The computationally or information-theoretically secure database techniques can also be broadly classified into two categories, based on the supported queries: (\textit{i}) \textit{Techniques that support selection/join}: Different cryptographic techniques are built for selection queries, \textit{e}.\textit{g}., searchable encryption, deterministic/non-deterministic encryption, and OPE; and (\textit{ii}) \textit{Techniques that support aggregation}: Cryptographic techniques that exploit homomorphic mechanisms such as homomorphic encryption, SSS, or MPC techniques.

While both computationally and information-theoretically secure techniques have been studied extensively in the cryptographic domain, secure
data management has focused disproportionately on computationally secure techniques (\textit{e}.\textit{g}., OPE, homomorphic encryption, searchable-encryption, and bucketization~\cite{DBLP:conf/sigmod/HacigumusILM02}) resulting in systems such as CryptDB~\cite{DBLP:journals/cacm/PopaRZB12}, Monomi~\cite{popa-monomi}, MariaDB~\cite{mariadb}, CorrectDB~\cite{DBLP:journals/pvldb/BajajS13}).
Some exceptions to the above include~\cite{DBLP:conf/icde/EmekciAAG06,DBLP:journals/isci/EmekciMAA14,DBLP:journals/isci/XiangLCGY16,DBLP:journals/corr/abs-1801-10323} that have focused on secret-sharing.

Recently, both academia~\cite{dan,DBLP:journals/isci/EmekciMAA14,DBLP:journals/isci/XiangLCGY16,DBLP:journals/corr/abs-1801-10323} and industries~\cite{DBLP:journals/iacr/ArcherBLKNPSW18,pulsar,Sharemind} have begun to explore information-theoretically secure techniques using MPC that efficiently supports OLAP tasks involving aggregation queries, while achieving higher security than computationally secure techniques.\footnote{{\scriptsize Some of the computationally secure mechanisms are vulnerable to computationally powerful adversaries. For instance, Google, with sufficient computational capabilities, broke SHA-1~\cite{url1}.}} For instance, commercial systems, such as Jana~\cite{DBLP:journals/iacr/ArcherBLKNPSW18} by Galois, Pulsar~\cite{pulsar} by Stealth Software, Sharemind~\cite{Sharemind} by Cybernetica, and products by companies such as Unbound Tech., Partisia, Secret Double Octopus, and SecretSkyDB Ltd. have explored MPC-based databases systems that offer strong security guarantees. Benefits of MPC-based methods in terms of both higher-level security and relatively efficient support for aggregation queries have been extensively discussed in both scientific articles~\cite{DBLP:conf/dev/RajanQABLV18,DBLP:conf/uss/FranklePSGW18,DBLP:books/cu/CDN2015,DBLP:conf/icassp/Orlandi11} and popular media~\cite{media1,media2,media3,media4}.


Much of the above work on MPC-based secure data management requires several servers to collaborate to answer queries. These collaborations
require several rounds of communication among non-colluding
servers. Instead, we explore secure data management based on SSS that does not require servers to collaborate to generate answers and can, hence, be implemented more efficiently. There is prior work on exploring secret-sharing for SQL processing~\cite{DBLP:conf/icde/EmekciAAG06,DBLP:journals/isci/EmekciMAA14,DBLP:journals/isci/XiangLCGY16,DBLP:journals/corr/abs-1801-10323}, but the developed techniques suffer from several drawbacks, \textit{e}.\textit{g}., weak security guarantees such as leakage of access patterns, significant overhead of maintaining polynomials for generating shares at the database (DB) owner, no support for third-party query execution on the secret-shared outsourced database, etc. We discuss the limitations of existing secret-sharing-based data management techniques in details in \S\ref{subsec:Comparison with Existing Work}.

\noindent\textbf{Contribution.} Our contributions in this paper are threefold:
\begin{enumerate}[noitemsep,nolistsep,leftmargin=0.01in]
\item SSS-based algorithms (entitled \textsc{Obscure})
that support a large class of \emph{access-pattern-hiding aggregation queries with selection}. \textsc{Obscure} supports count, sum, average, maximum, minimum, top-k, and reverse top-k, queries, without revealing anything about data/query/results to an adversary. 

\item An oblivious result verification algorithm for aggregation queries such that an adversary does not learn anything from the verification. \textsc{Obscure}'s verification step is not mandatory. A querier may run verification occasionally to confirm the correctness of results.

\item A comprehensive experimental evaluation of \textsc{Obscure} on a variety of queries that clearly highlight its scalability to moderate-size datasets and its efficiency compared to both state-of-the-art MPC-based solutions, as well as, to the simple strategy of downloading encrypted data at the client, decrypting it, and running queries at the (trusted) client.
\end{enumerate}

\noindent\textbf{Applications.} Our proposed algorithms can deal with datasets outsourced by a single or multiple DB owners. Here, we provide examples of each scenario.

\noindent
\textit{\textbf{DB outsourcing by a single DB owner: Hospital database.}} A hospital may outsource its patient database to an (untrusted cloud) server. Given the sensitivity of the patient records, such data needs to be secured cryptographically. The hospital may still wish to execute analytical queries on the sever over such data (\textit{e}.\textit{g}., number of influenza patients seen in the last month) for its own internal logistical planning.

\noindent\textit{\textbf{DB outsourcing by multiple DB owners: Smart metering (or IoT sensors).}} Smart meters' data outsourcing is an example of multiple DB owners and a single querier. In smart meter settings, smart meter devices keep the energy consumptions of a home at given time intervals and send the data to the servers~\cite{DBLP:conf/sac/SilvaMVB17}. This data contains behavioral information of the user; hence, a cryptographic technique should be used to make it secure before outsourcing. Users may execute queries on this secure database for monitoring and comparing their usage to that of others in the neighborhood. Executing such aggregate queries involve count, sum, and maximum operations in an oblivious manner at the server for preventing access to users' behavioral information. Our proposed algorithms prevent an adversarial server to learn the user's behaviors, when storing the database or executing a query. Privacy-preserving data integration~\cite{DBLP:journals/tifs/LimPXC20,barhamgi2019privacy,DBLP:conf/edbt/BarhamgiBACC13,DBLP:conf/caise/BarhamgiBOCCMT13,DBLP:journals/sigmod/BenslimaneBCMDN13}, where different datasets -- owned by different DB owners -- are intergraded into a single dataset, is also an example DB outsourcing by multiple owners. However,~\cite{DBLP:journals/tifs/LimPXC20,barhamgi2019privacy,DBLP:conf/edbt/BarhamgiBACC13,DBLP:conf/caise/BarhamgiBOCCMT13,DBLP:journals/sigmod/BenslimaneBCMDN13} deal with only encrypted data integration; thus, we do not discuss such techniques in detail.

\smallskip
\noindent\textbf{Outline of the paper}: \S\ref{sec:Building Blocks of the Algorithms} provides an overview of secret-sharing techniques and related work. \S\ref{sec:Preliminary} and \S\ref{sec:Data Outsourcing} provide the model, an adversary model, security properties, and data outsourcing model. \S\ref{sec:Count Query} provides conjunctive/disjunctive count queries and their verification algorithm. \S\ref{sec:Sum Queries} provides conjunctive/disjunctive sum queries and their verification algorithm.
\S\ref{sec:maximum} provides an algorithm for fetching tuples having maximum values in some attributes with their verification. \S\ref{sec:Experiments} provides an experimental evaluation.

\noindent\textbf{Appendix.} In appendix, we provide the following: an example of count query verification using secret-shared data, an approach for finding maximum over SSS databases outsourced by multiple DB owners, approaches for the minimum and top-k, an outline for security proofs, and a communication-efficient strategy for knowing tuples that satisfied a query predicate.

\section{Background}
\label{sec:Building Blocks of the Algorithms}
Here, we provide an overview of secret-sharing with an example and compare our proposed approach with existing works.

\subsection{Building Blocks}
\label{subsec:Building Blocks}
\textsc{Obscure} is based on SSS, string-matching operations over SSS, and order-preserving secret-sharing (OP-SS). This section provides an overview of these existing techniques.

\smallskip
\noindent\textbf{Shamir's secret-sharing (SSS).} In SSS~\cite{DBLP:journals/cacm/Shamir79}, the DB owner divides a secret value, say $S$, into $c$ different fragments, called \emph{shares}, and sends each share to a set of $c$ non-communicating participants/servers. These servers cannot know the secret $S$ until they collect $c^{\prime} < c$ shares. In particular, the DB owner randomly selects a polynomial of degree $c^{\prime}$ with $c^{\prime}$ random coefficients, \textit{i}.\textit{e}., $f(x)= a_0+a_1x+a_2x^2+\cdots+a_{c^{\prime}}x^{c^{\prime}}$, where $f(x)\in \mathbb{F}_p[x]$, $p$ is a prime number, $\mathbb{F}_p$ is a finite field of order $p$, $a_0=S$, and $a_i\in \mathbb{N} (1\leq i\leq c^{\prime})$. The DB owner distributes the secret $S$ into $c$ shares by placing $x=1,2,\ldots, c$ into $f(x)$. The secret can be reconstructed based on any $c^{\prime}+1$ shares using Lagrange interpolation~\cite{corless2013graduate}. Note that $c^{\prime} \leq c$, where $c$ is often taken to be larger than $c^{\prime}$ to tolerate malicious adversaries that may modify the value of their shares. For this paper, however, since we are not addressing the availability of data, we will consider $c$ and $c^{\prime}$ to be identical.

SSS allows an \textit{addition} of shares, \textit{i}.\textit{e}., if $s(a)_i$ and $s(b)_i$ are shares of two values $a$ and $b$, respectively, at the server $i$, then the server $i$ can compute an addition of $a$ and $b$ itself, \textit{i}.\textit{e}., $a+b= s(a)+s(b)$, without knowing real values of $a$ and $b$.

\smallskip
\noindent\textbf{String-matching operation on secret-shares. }
Accumulating-Automata (AA)~\cite{DBLP:conf/ccs/DolevGL15} is a new string-matching technique on secret-shares that do not require servers to collaborate to do the operation, unlike MPC-techniques~\cite{sepia,DBLP:conf/tcc/DamgardFKNT06,murat08,DBLP:conf/ccs/BonawitzIKMMPRS17,Sharemind,DBLP:journals/iacr/ArcherBLKNPSW18}. Here, we explain AA to show how string-matching can be performed on secret-shares.

Let $D$ be the cleartext data. Let $S(D)_i$ ($1\leq i\leq c$) be the $i^{\mathit{th}}$ secret-share of $D$ stored at the $i^{\mathit{th}}$ server, and $c$ be the number of \emph{non-communicating} servers. AA allows a user to search a pattern, $\mathit{pt}$, by creating $c$ secret-shares of $\mathit{pt}$ (denoted by $S(\mathit{pt})_i$, $1\leq i\leq c$), so that the $i^{\mathit{th}}$ server can search the secret-shared pattern $S(\mathit{pt})_i$ over $S(D)_i$. The result of the string-matching operation is either $1$ of secret-share form, if $S(\mathit{pt})_i$ matches with a secret-shared string in $S(D)_i$ or $0$ of secret-share form; otherwise. Note that when searching a pattern on the servers, AA uses \textit{multiplication} of shares, as well as, the additive property of SSS, which will be clear by the following example. Thus, if the user wants to search a pattern of length $l$ in only \emph{one communication round}, while the DB owner and the user are using a polynomial of degree one, then due to multiplication of shares, the final degree of the polynomial will be $2l$, and solving such a polynomial will require at least $2l+1$ shares.

\noindent\textbf{\emph{\underline{Example.}}} Assume that the domain of symbols has only three symbols, namely A, B, and C. Thus, A can be represented as $\langle 1,0,0\rangle$. Similarly, B and C can be represented as $\langle 0,1,0\rangle$ and $\langle 0,0,1\rangle$, respectively.

\noindent\emph{DB owner side}. Suppose that the DB owner wants to outsource B to the (cloud) servers. Hence, the DB owner may represent B as its unary representation: $\langle 0,1,0\rangle$. If the DB owner outsources the vector $\langle 0,1,0\rangle$ to the servers, it will reveal the symbol. Thus, the DB owner uses any three polynomials of an identical degree, as shown in Table~\ref{Secret-shares by DB owner}, to create three shares.

\bgroup
\def\arraystretch{.86}
\begin{table}[h]
\centering
  \scriptsize
  \begin{tabular}{|p{1.4cm}|l|p{1.12cm}|p{1.40cm}|p{1.21cm}|}\hline

  Vector values & Polynomials & First shares & Second shares & Third shares\\\hline
  0 & $0+5x$ & 5 & 10 & 15 \\\hline
  1 & $1+9x$ &10 & 19 & 28 \\\hline
  0 & $0+2x$ & 2 &  4 & 6  \\\hline
\end{tabular}
\caption{Secret-shares of a vector $\langle 0,1,0\rangle$, created by the DB owner.}
\label{Secret-shares by DB owner}
\end{table}
\egroup

\noindent\emph{User-side}. Suppose that the user wants to search for a symbol B. The user will first represent B as a unary vector, $\langle0,1,0\rangle$, and then, create secret-shares of B, as shown in Table~\ref{Secret-shares by user}. Note that there is no need to ask the DB owner to send any polynomials to create shares or ask the DB owner to execute the search query.

\bgroup
\def\arraystretch{.86}
\begin{table}[h]
  \centering
  \scriptsize
\begin{tabular}{|p{1.4cm}|l|p{1.12cm}|p{1.40cm}|p{1.21cm}|}\hline
  Vector values & Polynomials & First shares & Second shares & Third shares \\\hline
  0 & $0+x$ & 1 & 2 & 3 \\\hline
  1 & $1+2x$ &3 & 5 & 7 \\\hline
  0 & $0+4x$ &4 &  8 & 12  \\\hline
\end{tabular}
\caption{Secret-shares of a vector $\langle 0,1,0\rangle$, created by the user/querier.}
\label{Secret-shares by user}
\end{table}
\egroup

\noindent\emph{Server-side}. Each server performs position-wise multiplication of the vectors that they have, adds all the multiplication resultants, and sends them to the user, as shown in Table~\ref{cloud multiply}. An important point to note here is that the server cannot deduce the keyword, as well as, the data by observing data/query/results.

\bgroup
\def\arraystretch{.86}
\begin{table}[h]
  \centering
  \scriptsize
  \begin{tabular}{|p{1.3cm}|p{1.6cm}|p{1.5cm}|}\hline
  \multicolumn{3}{|c|}{Computation on}\\\hline

  Server 1             & Server 2           & Server 3          \\\hline
  $5\times  1=  5$    & $10\times 2 = 20$ &  $15\times 3 =45$ \\\hline
  $10\times 3= 30$    & $19\times 5 = 95$ &  $28\times 7 =196$  \\\hline
  $2\times 4 =  8$    & $4\times  8 = 32$ &  $6\times12=72$   \\\hline
  43 &147 & 313 \\\hline
\end{tabular}
\caption{Multiplication of shares and addition of final shares by the servers.}
\label{cloud multiply}
\end{table}
\egroup

\noindent\emph{User-side}. After receiving the outputs ($\langle y_1=43, y_2=147, y_3=313 \rangle$) from the three servers, the user executes Lagrange interpolation~\cite{corless2013graduate} to construct the secret answer, as follows:
\centerline{\scriptsize $
\frac{(x-x_2)(x-x_3)}{(x_1-x_2)(x_1-x_3)}\times y_1 +
\frac{(x-x_1)(x-x_3)}{(x_2-x_1)(x_2-x_3)}\times y_2 +
\frac{(x-x_1)(x-x_2)}{(x_3-x_1)(x_3-x_2)}\times y_3
$}
\centerline{\scriptsize
$=
\frac{(x-2)(x-3)}{(1-2)(1-3)}\times 43 +
\frac{(x-1)(x-3)}{(2-1)(2-3)}\times 147 +
\frac{(x-1)(x-2)}{(3-1)(3-2)}\times 313 =1
$}


The final answer is 1 that confirms that the secret-shares at the servers have B.

\noindent\emph{Note.} In this paper, we use AA that utilizes \emph{unary representation} as a building block. A recent paper Prio~\cite{dan} also uses a unary representation; however, we use significantly fewer number of bits compared to Prio's unary representation. One can use Prio's unary representation too or use a different private string-matching technique over secret-shares that supports string-matching over the shares.


\smallskip
\noindent\textbf{Order-preserving secret-sharing (OP-SS).} The concept of OP-SS was introduced in~\cite{DBLP:conf/icde/EmekciAAG06}. OP-SS maintains the order of the values in secret-shares too, \textit{e}.\textit{g}., if $v_1$ and $v_2$ are two values in cleartext such that $v_1<v_2$, then $S(v_1)< S(v_2)$ at any server. It is clear that finding records with maximum or minimum values using OP-SS are trivial. However, ordering revealed by OP-SS can leak more information about records. Consider, for instance, an employee relation, given in Table~\ref{fig:database1} on page 5. For explanation purpose, we represent Table~\ref{fig:database1} in cleartext. In Table~\ref{fig:database1}, the salary field can be stored using OP-SS. If we know (background knowledge) that employees in the \emph{security} department earn more money than others, we can infer from the representation that the second tuple corresponds to someone from the \emph{security} department. Thus, OP-SS, by itself, offers little security. However, as we will see later in \S\ref{sec:maximum}, by splitting the fields such as \emph{salary} that can be stored using OP-SS, while storing other fields using SSS, we, thus, can benefit from the ordering supported by OP-SS without compromising on security.

\smallskip
\noindent\textbf{2's complement-based sigbit computation.}~\cite{DBLP:conf/algocloud/DolevL15} provided 2's complement-based sigbit computation. We will use signbit to find if two numbers are equal or not, as follows: $A\geq B\textnormal{ if } signbit(A-B)=0$, and $A< B\textnormal{ if } signbit(A-B)=1$. Let $A = [a_n,a_{n-1},\ldots,a_1]$ be a $n$ bit number and $B = [b_n,b_{n-1},\ldots,b_1]$ be a $n$ bit number. 2's complement subtraction converts $B-A$ into $B+\bar{A}+1$, where $\bar{A}+1$ is 2's complement representation of $-A$. We start from the least significant bit (LSB) and go through the rest of the bits. The method inverts $a_i$ (by doing $1-a_i$, where $1\leq i\leq n$), calculates $\bar{a}_0+b_0+1$ and its carry bit. After finishing this on all the $n$ bits, the most significant bit (MSB) keeps the signbit.

\subsection{Comparison with Existing Work}
\label{subsec:Comparison with Existing Work}
\noindent\textbf{Comparison with SSS databases.} In 2006, Emek{\c{c}}i et al.~\cite{DBLP:conf/icde/EmekciAAG06} introduced the first work on SSS data for executing sum, maximum, and minimum queries.
However,~\cite{DBLP:conf/icde/EmekciAAG06} uses a trusted-third-party to perform queries and is not secure, since it uses OP-SS to answer maximum/minimum queries. Another paper by Emek{\c{c}}i et al.~\cite{DBLP:journals/isci/EmekciMAA14}
on OP-SS based aggregation queries requires the database (DB) owner to retain each polynomial, which was used to create database shares, resulting in the DB owner to store $n\times m$ polynomials, where $n$ and $m$ are the numbers of tuples and attributes in a relation.~\cite{DBLP:journals/isci/EmekciMAA14} is also not secure, since it reveals access-patterns (\textit{i}.\textit{e}., the identity of tuples that satisfy a query) and using OP-SS.\footnote{{\scriptsize While~\cite{DBLP:conf/icde/EmekciAAG06,DBLP:journals/isci/EmekciMAA14,DBLP:journals/corr/abs-1801-10323,DBLP:journals/iacr/ArcherBLKNPSW18} have explored mechanisms to support selection and join operations over the secret-shared data, these techniques are not secure (\textit{e}.\textit{g}., leak information from access-patterns), are inefficient (often requiring quadratic computations), and require transmitting entire dataset to users. SS can primarily be used to support OLAP style aggregation queries, which is our focus in this paper.}} Like~\cite{DBLP:journals/isci/EmekciMAA14},~\cite{DBLP:journals/isci/XiangLCGY16} proposed a similar approach and also suffers from similar disadvantages.~\cite{DBLP:conf/pet/ThompsonHHSY09} proposed SSS-based sum and average queries; however, they also require the DB owner to retain tuple-ids of qualifying tuples.~\cite{DBLP:journals/corr/abs-1801-10323} used a novel string-matching operation over the shares at the server,
but it cannot perform general aggregations with selection over complex predicates. In short, all the SSS-based solutions for aggregation queries either overburden the DB owner (by storing enough data related to polynomials and fully participating in a query execution), are insecure due to OP-SS, reveal access-patterns, or support a very limited form of aggregation queries without any selection criteria.

In contrast, \textsc{Obscure} eliminates all such limitations. It provides a fully secure and efficient solution for implementing aggregation queries with selections. Our experimental results will show that \textsc{Obscure} scales to datasets with 6M tuples on TPC-H queries, the size of which prior secret-sharing and/or MPC-based techniques have never scaled to. The key to the efficient performance of \textsc{Obscure} still is exploiting OP-SS -- while OP-SS, in itself, is not secure (it is prone to background knowledge attacks, for instance). The way \textsc{Obscure} uses OP-SS, as will be clear in \S\ref{sec:Data Outsourcing}, it prevents such attacks by appropriately partitioning data, while still being able to exploit OP-SS for efficiency. In addition, to support aggregation with selections, \textsc{Obscure} exploits the string-matching techniques over shares developed in~\cite{DBLP:conf/ccs/DolevGL15}.

Furthermore, as we will see in experimental section (\S\ref{sec:Experiments}), \textsc{Obscure} scales to datasets with 6M tuples on TPC-H queries.

\smallskip
\noindent\textbf{Comparison with MPC-techniques.} \textsc{Obscure} also overcomes several limitations of existing MPC-based solutions. Recent work, Prio~\cite{dan} supports a mechanism for confirming the maximum number, if the maximum number is known; however, Prio~\cite{dan} does not provide any mechanism to compute the maximum/minimum. Also, Prio does not provide methods to execute conjunctive and disjunctive count/sum queries. 
Another recent work~\cite{DBLP:conf/ccs/BonawitzIKMMPRS17} deals with adding shares in an array under malicious servers and malicious users, using the properties of SSS and public-key settings. However,~\cite{DBLP:conf/ccs/BonawitzIKMMPRS17} is unable to execute a single-dimensional, conjunctive, or disjunctive sum query. Note that (as per our assumption) though,~\cite{DBLP:conf/ccs/BonawitzIKMMPRS17} can tolerate malicious users, while \textsc{Obscure} is designed to only handle malicious servers, and it assumes users to be trustworthy.

Other works, \textit{e}.\textit{g}., Sepia~\cite{sepia} and~\cite{DBLP:conf/tcc/DamgardFKNT06}, perform \emph{addition and less than} operations, and use many communication rounds. In contrast, \textsc{Obscure} uses minimal communication rounds between the user and each server, (when having enough shares). Specifically, count, sum, average, and their verification algorithms require at most two rounds between each server and the user. However, maximum/minimum finding algorithms require at most four communication rounds. In addition, our scheme achieves the minimum communication cost for aggregate queries, especially for count, sum, and average queries, by aggregating data locally at each server.

\smallskip
\noindent\textbf{Comparison with MPC/SSS-based verification approaches.}~\cite{murat08} and~\cite{DBLP:conf/pet/ThompsonHHSY09} developed verification approaches for secret-shared data.~\cite{murat08} considered verification process for MPC using a trusted-third-party verifier. While overburdening the DB owner by keeping metadata for each tuple,~\cite{DBLP:conf/pet/ThompsonHHSY09} provided metadata-based operation verification (\textit{i}.\textit{e}., whether all the desired tuples are scanned or not) for only sum queries, unlike \textsc{Obscure}'s result verification for all queries. \textsc{Obscure} verification methods neither involve the DB owner to verify the results nor require a trusted-third-party verifier.

\section{Preliminary}
\label{sec:Preliminary}
This section provides a description of entities, an adversarial model, and security properties for obliviously executing queries. 

\begin{table*}[!t]
\B
\centering
\scriptsize
\begin{tabular}{|l|l|}\hline
Simple aggregation queries & \texttt{select}  $\alpha$($A_i$) \texttt{from} $R$ \\\hline

Multi-dimensional aggregation queries & \texttt{select}  $\alpha$($A_i$) \texttt{from} $R$ \texttt{where} $A_k= v_k$ $\mathcal{OP}$ $A_l=v_l$ \\\hline

Multi-dimensional aggregation queries with group-by & \texttt{select} $A_i$, $\alpha$($A_j$) \texttt{from} $R$ \texttt{where} $A_k= v_k$ $\mathcal{OP}$ $A_l=v_l$ \texttt{group by} $A_i$ \\\hline

Multi-dimensional aggregation queries with group-by and having clause & \texttt{select} $A_i$, $\alpha$($A_j$) \texttt{from} $R$ \texttt{where} $A_k= v_k$ $\mathcal{OP}$ $A_l=v_l$ \texttt{group by} $A_i$ \texttt{having} $\gamma(A_j)=v_i$ \\\hline\hline

\multicolumn{2}{|p{17cm}|}{\textbf{Notations}: $\alpha$ and $\gamma$ are aggregation operators, such as \texttt{count}, \texttt{sum}, \texttt{avg}, \texttt{max}, and \texttt{min}. $\mathcal{OP}$ is a conjunctive or disjunctive operation. $A_i$, $A_j$, $A_k$, and $A_l$ are some attributes of a relation $R$} \\\hline


\end{tabular}
\BB
\caption{Query types supported by \textsc{Obscure}.}
\label{tab:Queries supported by Obscure}
\BBB\B
\end{table*}

\subsection{The Model}
\label{subsec:the model}
We assume the following three entities in our model.
\begin{enumerate}[noitemsep,nolistsep,leftmargin=0.1in]
\item
A set of $c>2$ \emph{non-communicating} servers. The servers do not exchange data with each other to compute any answer. The only possible data exchange of a server is with the user/querier or the database owner.

\item
The \emph{trusted} database (DB) owner, that creates $c$ secret-shares of the data and transfers the $i^{\mathit{th}}$ share to the $i^{\mathit{th}}$ server. The secret-shares are created by an algorithm that supports \emph{non-interactive} addition and multiplication of two shares, which is required to execute the private string-matching operation, at the server, as explained in \S\ref{sec:Building Blocks of the Algorithms}.\footnote{{\scriptsize The choice of the underlying non-interactive and string-matching-based secret-sharing mechanism does not change our proposed aggregation and verification algorithms.}}

\item
An (authenticated, authorized, and \emph{trusted}) user/querier, who executes queries on the secret-shared data at the servers. The query is sent to servers. The user fetches the partial outputs from the servers and performs a simple operation (polynomial interpolation using Lagrange polynomials~\cite{corless2013graduate}) to obtain the secret-value.
\end{enumerate}



\subsection{Adversarial Model}
\label{subsec:Adversarial Model}
We consider two adversarial models, in both of which the cloud servers (storing secret-shares) are not trustworthy. In the {\em honest but curious} model, the server correctly computes the assigned task without tampering with data or hiding answers. However, the server may exploit side information (\textit{e}.\textit{g}., query execution, background knowledge, and output size) to gain as much information as possible about the stored data. Such a model is considered widely in many cryptographic algorithms and in widely used in DaS~\cite{DBLP:conf/stoc/CanettiFGN96,DBLP:conf/sigmod/HacigumusILM02,DBLP:conf/icdcs/WangCLRL10,DBLP:conf/ccs/YuWRL10}. We also consider a malicious adversary that could deviate from the algorithm and delete tuples from the relation. Users and database owners, in contrast, are assumed to be not malicious.

Only authenticated users can request query on servers. Further, we follow the restriction of the standard SSS that the adversary cannot collude with all (or possibly the majority of) the servers. Thus, the adversary cannot generate/insert/update shares at the majority of the servers. Also, the adversary cannot eavesdrop on a majority of communication channels between the user and the servers. This can be achieved by either encrypting the traffic between user and servers, or by using anonymous routing~\cite{DBLP:journals/cacm/GoldschlagRS99}, in which case the adversary cannot gain knowledge of servers that store the secret-shares. Note that if the adversary could either collude with or successfully eavesdrop on the communication channels between the majority of servers and user, the secret-sharing technique will not apply.\footnote{{\scriptsize The DB owner/user can use anonymous routing to send their data to the servers, thereby preventing an adversary from  determining which user is connecting to which server. If the adversary knows the majority of the communication channels/servers, then it can construct the secret-shared query, outputs to the query, and the database.}} The validity of the assumptions behind secret-sharing has been extensively discussed in prior work~\cite{DBLP:conf/dev/RajanQABLV18,DBLP:conf/uss/FranklePSGW18,DBLP:books/cu/CDN2015,DBLP:conf/icassp/Orlandi11}. 
The adversary can be aware of the public information, such as the actual number of tuples and number of attributes in a relation, which will not affect the security of the proposed scheme, though such leakage can be prevented by adding fake tuples and attributes.\footnote{{\scriptsize The adversary cannot launch any attack against the DB owner. We do not consider cyber-attacks that can exfiltrate data from the DB owner directly, since defending against generic cyber-attacks is outside the scope of this paper.}}

\subsection{Security Properties}
\label{subsec:Security Properties}
In the above-mentioned adversarial model, an adversary wishes to learn the (entire/partial) data and query predicates. Hence, a secure algorithm must prevent an adversary to learn the data (\textit{i}) by just looking the cryptographically-secure data and deduce the frequency of each value (\textit{i}.\textit{e}., frequency-count attacks), and (\textit{ii}) when executing a query and deduce which tuples satisfy a query predicate (\textit{i}.\textit{e}., access-pattern attacks) and how many tuples satisfy a query predicate (\textit{i}.\textit{e}., output-size attacks). Thus, in order to prevent these attacks, our security definitions are identical to the standard security definition as in~\cite{DBLP:journals/joc/Canetti00,DBLP:conf/tcc/FreedmanIPR05,DBLP:conf/pkc/ChuT05}.
An algorithm is \emph{privacy-preserving} if it maintains the privacy of the querier (\textit{i}.\textit{e}., query privacy), the privacy of data from the servers, and performs identical operations, regardless of the user query.

\smallskip
\noindent\textbf{Query/Querier's privacy} requires that the user's query must be hidden from the server, the DB owner, and the communication channel. In addition, the server cannot distinguish between two or more queries of the \textit{same type} based on the output. Queries are of the same type based on their output size. For instance, all count queries are of the same type since they return almost an identical number of bits.

\noindent\textbf{Definition: User’s privacy.} \emph{For any probabilistic polynomial time adversarial server having a secret-shared relation $S(R)$ and any two input query predicates, say $p_1$ and $p_2$, the server cannot distinguish $p_1$ or $p_2$ based on the executed computations for either $p_1$ and $p_2$.}

\smallskip
\noindent\textbf{Privacy from the server} requires that the stored input data, intermediate data during a computation, and output data are not revealed to the server, and the secret value can only be reconstructed by the DB owner or an authorized user. In addition, two or more occurrences of a value in the relation must be different at the server to prevent frequency analysis while data at rest. Recall that due to secret-shared relations (by following the approach given in \S\ref{subsec:Building Blocks}), the server cannot learn the relations and frequency-analysis, and in addition, due to maintaining the query privacy, the server cannot learn the query and the output.

Here, we, also, must ensure that the server's behavior must be identical for a given query, and the servers provide an identical answer to the same query, regardless of the users (recall that user might be different compared to the data owner in our model). To show that we need to compare the real execution of the algorithm at the servers against the ideal execution of the algorithm at a trusted party having the same data and the same query predicate. An algorithm maintains the data privacy from the server if the real and ideal executions of the algorithm return an identical answer to the user.

\noindent\textbf{Definition: Privacy from the server.} \emph{For any given secret-shared relation $S(R)$ at a server, any query predicate $\mathit{qp}$, and any real user, say $U$, there exists a probabilistic polynomial time (PPT) user $U^{\prime}$ in the ideal execution, such that the outputs to $U$ and $U^{\prime}$ for the query predicate $\mathit{qp}$ on the relation $S(R)$ are identical.}

\smallskip
\noindent\textbf{Properties of verification.} We provide verification properties against malicious behaviors. A verification method must be oblivious and find any misbehavior of the servers when computing a query. We follow the verification properties from~\cite{murat08}, as follows: (\textit{i}) the verification method cannot be refuted by the majority of the malicious servers, and (\textit{ii}) the verification method should not leak any additional information.

\smallskip
\noindent\textbf{Algorithms' performance.} We analyze our oblivious aggregation algorithms on the following parameters, which are stated in Table~\ref{tab:Complexities of the operations}: (\textit{i}) \textit{Communication rounds}. The number of rounds that is required between the user and each server to obtain an answer to the query. (\textit{ii}) \textit{Scan cost at the server}. We measure scan cost at the server in terms of the number of the rounds that the server performs to \emph{read the entire dataset}. (\textit{iii}) \textit{Computational cost at the user}. The number of values/tuples that the user interpolates to know the final output.

\subsection{\textbf{\textsc{{\large Obscure}}} Overview}
\label{subsec:Obscure overview}
Let us introduce \textsc{Obscure} at a high-level. \textsc{Obscure} allows single-dimensional and multi-dimensional conjunctive/disjunctive equality queries. Note that the method of \textsc{Obscure} for handling these types of queries is different from SQL, since \textsc{Obscure} does not support query optimization and indexing\footnote{{\scriptsize For the class of queries considered (viz. aggregation with selection), the main optimization in standard databases is to push selections down and to determine whether an index-scan should be used or not. In secret-sharing, an index scan cannot be used (at least not in any obvious way), since sub-setting the data processed will reveal access-patterns, making the technique less secure. Hence, we avoid using any indexing structure.}} due to secret-shared data. Further, \textsc{Obscure} handles range-based queries by converting the range into equality queries. Executing a query on \textsc{Obscure} requires four phases, as follows:

\noindent\textsc{Phase 1:} \emph{Data upload by DB owner(s).} The DB owner uploads data to non-communicating servers using a secret-sharing mechanism that allows addition and multiplication (\textit{e}.\textit{g}.,~\cite{DBLP:conf/ccs/DolevGL15}) at the servers.

\noindent\textsc{Phase 2:} \emph{Query generation by the user.} The user generates a query, creates secret-shares of the query predicate, and sends them to the servers. For generating secret-shares of the query predicate, the user follows the strategies given in \S\ref{sec:Count Query} (count query), \S\ref{sec:Sum Queries} (sum queries), \S\ref{sec:maximum} (maximum/minimum), and \S\ref{subsec:Verification of Count Queries},\S\ref{subsec:Verification of Sum Queries} (verification).

\noindent\textsc{Phase 3:} \emph{Query processing by the servers.} The servers process an input query in an oblivious manner such that neither the query nor the results satisfying the query are revealed to the adversary. Finally, the servers transfer their outputs to the user.

\noindent\textsc{Phase 4:} \emph{Result construction by the user.} The user performs Lagrange interpolation on the received results, which provide an answer to the query. The user can also verify these results by following the methods given in \S\ref{subsec:Verification of Count Queries}, \S\ref{subsec:Verification of Sum Queries}, \S\ref{subsec:Verification of Maximum Query}.

\smallskip
{Table~\ref{tab:Queries supported by Obscure} shows queries supported by \textsc{Obscure}, where $\alpha$ and $\gamma$ are aggregation operators, such as count, sum, average, maximum, and minimum. In order to execute these operators, we provide algorithms in the following sections. As will become clear soon, the proposed algorithms may take at most three communication rounds between the servers and the user. Further, note that in \textsc{Obscure}, a group-by query requires us to know the name of groups, prior to query execution. For example, if the group-by operation is executed on \texttt{Department} attribute, then we need to know all unique department names.}

%


\section{Data Outsourcing}
\label{sec:Data Outsourcing}
This section provides details on creating and outsourcing a database of secret-shared form. The DB owner wishes to outsource a relation $R$ having attributes $A_1,A_2,\ldots,A_m$ and $n$ tuples, and creates the following two relations $R^1$ and $R^2$:

\noindent$\bullet$ \textbf{Relation $R^1$} that consists of all the attributes $A_1,A_2,\ldots,A_m$ along with two additional attributes, namely \texttt{TID} (tuple-id) and \texttt{Index}. As will become clear in \S\ref{sec:maximum}, the \texttt{TID} attribute will help in finding tuples having the maximum/minimum/top-k values, and the \texttt{Index} attribute will be used to know the tuples satisfying the query predicate. The $i^{\mathit{th}}$ values of the \texttt{TID} and \texttt{Index} attributes have the \emph{same} and \emph{unique} random number between 1 to $n$.

\noindent$\bullet$ \textbf{Relation $R^2$} that consists of three attributes \texttt{CTID} (cleartext tuple-id), \texttt{SSTID} (secret-shared tuple-id), and an attribute, say $A_c$, on which a comparison operator (minimum, maximum, and top-k) needs to be supported.\footnote{{\scriptsize If there are $x$ attributes on which comparison operators will be executed, then the DB owner will create $x$ relations, each with attributes \texttt{CTID}, \texttt{SSTID}, and one of the $x$ attributes.}}

The $i^{\mathit{th}}$ values of the attributes \texttt{CTID} and \texttt{SSTID} of the relation $R^2$ keep the $i^{\mathit{th}}$ value of the \texttt{TID} attribute of the relation $R^1$. The $i^{\mathit{th}}$ value of the attributes $A_c$ of the relation $R^2$ keeps the $i^{\mathit{th}}$ value of an attribute of the relation $R^1$ on which the user wants to execute a comparison operator. Further, the tuples of the relations $R^2$ are randomly permuted. The reason for doing permutation is that the adversary cannot relate any tuple of both the secret-shared relations, which will be clear soon by the example below.

\noindent\textbf{Note.} The relation $S(R^1)$ will be used to answer count and sum queries, while it will be clear in \S\ref{sec:maximum} how the user can use the two relations $S(R^1)$ and $S(R^2)$ together to fetch a tuple having maximum/minimum/top-k/reverse-top-k value in an attribute.
\bgroup
\def\arraystretch{0.86}
\begin{table}[!h]
\scriptsize
  \centering
  \begin{tabular}{|l|l|l|l|l|l|}
    \hline
      \texttt{EmpID} & \texttt{Name} & \texttt{Salary} & \texttt{Dept} \\ \hline\hline
    E101 & John &  1000 & Testing  \\ \hline
    E101 & John &  100000 & Security   \\ \hline
    E102 & Adam &  5000 & Testing  \\ \hline
    E103 & Eve  &  2000 & Design   \\ \hline
    E104 & Alice&  1500 & Design   \\ \hline
    E105 & Mike&  2000 & Design   \\ \hline
  \end{tabular}
  \caption{A relation: \texttt{Employee}.}
  \label{fig:database1}
\end{table}
\egroup

\noindent\emph{\underline{Example}}. Consider the \texttt{Employee} relation (see Table~\ref{fig:database1}). The DB owner creates $R^1=$ \texttt{Employee1} relation\footnote{{\scriptsize For verifying results of count and sum queries, we add two more attributes to this relation. However, we do not show here, since verification is not a mandatory step.}} (see Table~\ref{fig:employee1 relation}) with \texttt{TID} and \texttt{Index} attributes. Further, the DB owner creates $R^2=$ \texttt{Employee2} relation (see Table~\ref{fig:employee2 relation}) having three attributes \texttt{CTID}, \texttt{SSTID}, and \texttt{Salary}.

\bgroup
\def\arraystretch{.86}
\begin{table}[h]
\begin{center}
  \begin{minipage}[t]{.65\linewidth}
  \centering
 \scriptsize
\centering
\begin{tabular}{|p{0.6cm}|p{0.5cm}|p{0.68cm}|p{0.65cm}|p{0.25cm}|p{0.55cm}|}
    \hline
    \texttt{EmpID} & \texttt{Name} & \texttt{Salary} & \texttt{Dept} & \texttt{TID} &  \texttt{Index} \\ \hline\hline
    E101 & John &  1000 & Testing & 3&3 \\ \hline
    E101 & John &  100000 & Security &2&2  \\ \hline
    E102 & Adam &  5000 & Testing  & 5&5\\ \hline
    E103 & Eve  &  2000 & Design  &4&4 \\ \hline
    E104 & Alice&  1500 & Design  &1&1 \\ \hline
    E105 & Mike &  2000 & Design  &6&6 \\ \hline
  \end{tabular}
  \subcaption{$R^1=$ \texttt{Employee1} relation.}
\label{fig:employee1 relation}
  \end{minipage}
  \begin{minipage}[t]{.33\linewidth}
  \centering
  \scriptsize
\centering
\begin{tabular}{|p{0.30cm}|p{0.55cm}|p{0.68cm}|}
    \hline
    \texttt{CTID} & \texttt{SSTID} & \texttt{Salary} \\ \hline\hline
    1& 1 &  1500   \\ \hline
    5& 5 &  5000  \\ \hline
    3& 3 &  1000  \\ \hline
    6& 6 &  2000    \\ \hline
    2& 2 &  100000    \\ \hline
    4& 4 &  2000    \\ \hline
 \end{tabular}
 \subcaption{$R^2=$ \texttt{Employee2} relation.}
\label{fig:employee2 relation}
\end{minipage}
\end{center}
\BB\B
\caption{Two relations obtained from \texttt{Employee} relation.}
\label{fig:two realtions}
\end{table}
\egroup

\noindent\emph{Creating secret-shares.} Let $A_i[a_j]$ ($1\leq i\leq m+1$ and $1\leq j\leq n$) be the $j^{\mathit{th}}$ value of the attribute $A_i$. The DB owner creates $c$ secret-shares of each attribute value $A_i[a_j]$ of the relation $R^1$ using a secret-sharing mechanism that allows string-matching operations at the server (as specified in \S\ref{sec:Building Blocks of the Algorithms}). However, $c$ shares of the $j^{\mathit{th}}$ value of the attribute $A_{m+2}$ (\textit{i}.\textit{e}., \texttt{Index}) are obtained using SSS. This will result in $c$ relations: $S(R^1)_1$, $S(R^1)_2$, $\ldots$, $S(R^1)_c$, each having $m+2$ attributes. The notation $S(R^1)_k$ denotes the $k^{\mathit{th}}$ secret-shared relation of $R^1$ at the server $k$. We use the notation $A_i[S(a_j)]_k$ to indicate the $j^{\mathit{th}}$ secret-shared value of the $i^{\mathit{th}}$ attribute of a secret-shared relation at the server $k$.

Further, on the relation $R^2$, the DB owner creates $c$ secret-shares of each value of \texttt{SSTID} using a secret-sharing mechanism that allows string-matching operations on the servers and each value of $A_c$ using order-preserving secret-sharing~\cite{DBLP:conf/icde/EmekciAAG06,DBLP:conf/esorics/HadaviDJCG12,DBLP:journals/isci/EmekciMAA14}. The secret-shares of the relation $R^2$ are denoted by $S(R^2)_i$ ($1\leq i\leq c$). The attribute \texttt{CTID} is outsourced in cleartext with the shared relation $S(R^2)_i$. It is important to mention that \texttt{CTID} attribute allows fast search due to cleartext representation than \texttt{SSTID} attribute, which allows search over shares.

Note that the DB owner's objective is to hide any relationship between the two relations when creating shares of the relations $S(R^1)$ and $S(R^2)$, \textit{i}.\textit{e}., the adversary cannot know by just observing any two tuples of the two relations that whether these tuples share a common value in the attribute \texttt{TID}/\texttt{SSTID} and $A_c$ or not. Thus, shares of an $i^{\mathit{th}}$ ($1\leq i\leq n$) value of the attribute \texttt{TID} in the relation $S(R^1)_j$ and in the attribute \texttt{SSTID} of the relation $S(R^2)_j$ must be different at the $j^{\mathit{th}}$ server. Also, by default, the attribute $A_c$ have different shares in both the relations, due to using different secret-sharing mechanisms for different attributes. The DB owner outsources the relations $S(R^1)_i$ and $S(R^2)_i$ to the $i^{\mathit{th}}$ server.

\bgroup
\def\arraystretch{.91}
\begin{table}[t]
\begin{center}
\scriptsize
\begin{tabular}{|p{3cm}|p{1.55cm}| p{0.95cm}|p{.9cm}|p{1.1cm}|}\hline

Algorithms & Query conditions & Scan rounds at a server & Comm. rounds & Interpolated values at user \\ \hline\hline

\multirow{3}{*}{Count~\S\ref{sec:Count Query}} & 1D & 1 & 1 & 1 \\
    \hhline{~----}     & CE & 1 & 1  & 1  \\
    \hhline{~----}     & DE & 1 & 1  & $1$ \\\hline

\multirow{3}{*}{Sum~\S\ref{sec:Sum Queries}} & 1D & 1 & 1 & 1 \\
    \hhline{~----}     & CE & 1 & 1  & 1  \\
    \hhline{~----}     & DE & 1 & 1  & $1$  \\\hline

Unconditional max./min. (\texttt{SDBMax}~\S\ref{subsec: Unconditional Maximum Query}) & One occurrence with tuple & 1  & 1 & $m$ \\\hline

\multirow{2}{2.5cm}{Conditional maximum/minimum (\texttt{SDBMax}~\S\ref{subsec:conditional Maximum Query})} & Finding maximum & 1 & 2 & $n+1$ or $\mathcal{T}+1$ \\
    \hhline{~----}     & Tuple fetching & 2  & 2  & $n+m$ or $\mathcal{T}+m$ \\\hline


\multirow{2}{3cm}{Maximum/Minimum (\texttt{MDBMax}) One occurrences~\S\ref{app_subsec:Finding Maximum over Datasets Outsourced by Multiple DB Owners}} & Counting & $2n+1$ & 1 & 2 \\
\hhline{~----}     & Counting + tuple fetching & $2n+3$ & 3 & $2\mathcal{T}+\ell m$ \\\hline\hline


Group-by~\S\ref{subsec:Group-by Query} &  & 1 & 1 & $g$ \\\hline

Top-k or reverse top-k~\S\ref{app_sec:Minimum, Top-k, and Reverse Top-k}  &  Unique occurrence & 1 or $k$ & 2 or 1 & $k\times m$ \\\hline\hline

\multicolumn{5}{|p{8.8cm}|}{\textbf{Notations.} $m$: \# attributes. $n$: \# tuples. $D=n\times m$: the database. 1D: Single-dimensional equality query. CE: Conjunctive equality query. DE: Disjunctive equality query. $\mathcal{T}$: \# tuples satisfying a query predicate. $\ell$: \# tuples having the maximum/minimum in the desired attribute. $g$: \# groups. \textbf{Condition}: the above-mentioned rounds are given when we have $2l+1$ shares, where $l$ is the maximum length of a keyword.} \\\hline
\end{tabular}
 \end{center}
\BB
\caption{Complexities of the algorithms.}
\label{tab:Complexities of the operations}
\BBB
\end{table}
\egroup

\noindent\textbf{Note.} Naveed et al.~\cite{DBLP:conf/ccs/NaveedKW15} showed that a cryptographically secured database that is also an using order-preserving cryptographic technique (\textit{e}.\textit{g}., order-preserving encryption or OP-SS) may reveal the entire data when mixed with publicly known databases. Hence, in order to overcome such a vulnerability of order-preserving cryptographic techniques, we created two relations, and importantly,
the above-mentioned representation, even though it uses OP-SS does not suffer from attacks based on background knowledge, as mentioned in \S\ref{sec:Building Blocks of the Algorithms}. Of course, instead of using the two relations, the DB owner can outsource only a single relation without using OP-SS. In the case of a single relation, while we reduce the size of the outsourced dataset, we need to compare each pair of two shares, and it will result in increased communication cost, as well as, communication rounds, as shown in previous works~\cite{DBLP:conf/tcc/DamgardFKNT06,sepia}, which were developed to compare two shares.

\section{Count Query and Verification}
\label{sec:Count Query}
In this section, we develop techniques to support count queries over secret-shared dataset outsourced by a single or multiple DB owners. The query execution does not involve the DB owner or the querier to answer the query. Further, we develop a method to verify the count query results.

\smallskip
\noindent\textbf{Conjunctive count query.} Our conjunctive equality-based count query scans the entire relation only once for checking single/multiple conditions of the query predicate. For example, consider the following conjunctive count query: \texttt{select count(*) from R where $A_1 = v_1$ $\wedge$ $A_2 = v_2$ $\wedge$ $\ldots$ $\wedge$ $A_{m} = v_m$}.

The user transforms the query predicates to $c$ secret-shares that result in the following query at the $j^{\mathit{th}}$ server: \texttt{select count(*) from $S(R^1)_j$ where $A_1 = S(v_1)_j$ $\wedge$ $A_2 = S(v_2)_j$ $\wedge$ $\ldots$ $\wedge$ $A_{m} = S(v_m)_j$}. Note that the single-dimensional query will have only one condition. Each server $j$ performs the following operations:

\centerline{$\mathit{Output}=\textstyle\sum_{k=1}^{k=n} \prod_{i=1}^{i=m} (A_i[S(a_k)]_j \otimes S(v_i)_j)$}

\noindent$\otimes$ shows a string-matching operation that depends on the underlying text representation. For example, if the text is represented as a unary vector, as explained in \S\ref{sec:Building Blocks of the Algorithms}, $\otimes$ is a bit-wise multiplication and
addition over a vector's elements, whose results will be $0$ or $1$ of secret-share form. Each server $j$ compares the query predicate value $S(v_i)$ against $k^{\mathit{th}}$ value ($1\leq k\leq n$) of the attribute $A_i$, multiplies all the resulting comparison for each of the attributes for the $k^{\mathit{th}}$ tuple. This will result in a single value for the $k^{\mathit{th}}$ tuple, and finally, the server adds all those values. Since secret-sharing \emph{allows the addition of two shares}, the sum of all $n$ resultant shares provides the occurrences of tuples that satisfy the query predicate of secret-share form in the relation $S(R^1)$ at the $j^{\mathit{th}}$ server. On receiving the values from the servers, the user performs Lagrange interpolation~\cite{corless2013graduate} to get the final answer in cleartext.

\smallskip
\noindent\emph{Correctness}. The occurrence of $k^{\mathit{th}}$ tuple will only be included when the multiplication of $m$ comparisons results in $1$ of secret-share form. Having only a single $0$ as a comparison resultant over an attribute of $k^{\mathit{th}}$ tuple produce $0$ of secret-share form; thus, the $k^{\mathit{th}}$ tuple will not be included. Thus, the correct occurrences over all tuples are included that satisfy the query's \texttt{where} clause.

\smallskip
\noindent\emph{\underline{Example.}} We explain the above conjunctive count query method using the following query on the \texttt{Employee} relation (refer to Table~\ref{fig:database1}): \texttt{select count(*) from Employee where Name = `John' and Salary = `1000'}. Table~\ref{fig:conjunctive count} shows the result of the private string-matching on the attribute \texttt{Name}, denoted by $o_1$, and on the attribute \texttt{Salary}, denoted by $o_2$. Finally, the last column shows the result of the query for each row and the final count answer for all the tuples. Note that for the purpose of explanation, we use cleartext values; however, the server will perform all operations over secret-shares. For the first tuple, when the servers check the first value of \texttt{Name} attribute against the query predicate \texttt{John} and the first value of \texttt{Salary} attribute against the query predicate \texttt{1000}, the multiplication of both the results of string-matching becomes 1. For the second tuple, when the server checks the second value of \texttt{Name} and \texttt{Salary} attributes against the query predicate \texttt{John} and \texttt{1000}, respectively, the multiplication of both the results become 0. All the other tuples are processed in the same way.

\bgroup
\def\arraystretch{.86}
\begin{table}[!t]
\scriptsize
  \centering
  \begin{tabular}{|l||l|l|l|l|l|l|}
    \hline
    \texttt{Name} & $o_1$ & \texttt{Salary} & $o_2$ & $o_1\times o_2$\\ \hline\hline
     John &  1 & 1000   & 1 & 1   \\ \hline
     John &  1 & 100000 & 0 & 0   \\ \hline
     Adam &  0 & 5000   & 0 & 0   \\ \hline
     Eve  &  0 & 2000   & 0 & 0   \\ \hline
     Alice&  0 & 1500   & 0 & 0   \\ \hline
     Mike &  0 & 2000   & 0 & 0   \\ \hline\hline
     ~ &  && & 1\\\hline
  \end{tabular}
  \B
  \caption{An execution of the conjunctive count query.}
  \label{fig:conjunctive count}
\BB\B
\end{table}
\egroup


\medskip
\noindent\textbf{Disjunctive count query.} Our disjunctive count query also scans the entire relation only once for checking multiple conditions of the query predicate, like the conjunctive count query. Consider, for example, the following disjunctive count query: \texttt{select count(*) from R where $A_1 = v_1$ $\vee$ $A_2=v_2$ $\vee$ $\ldots$ $\vee$ $A_{m} = v_m$}

The user transforms the query predicates to $c$ secret-shares that results in the following query at the $j^{\mathit{th}}$ server: \texttt{select count(*) from $S(R^1)_j$ where $A_1 = S(v_1)_j$ $\vee$ $\ldots$ $\vee$ $A_{m} = S(v_m)_j$} The server $j$ performs the following operation:

\centerline{$\mathit{Result}_i^k= A_i[S(a_k)]_j \otimes S(v_i)_j, 1\leq i\leq m$}
\centerline{$\mathit{Output} = \textstyle\sum_{k=1}^{k=n}(((\mathit{Result}_1^k\textnormal{ \texttt{OR} } \mathit{Result}_2^k)\textnormal{ \texttt{OR} }\mathit{Result}_3^k)\ldots$}
\centerline{$\textnormal{ \texttt{OR} }\mathit{Result}_m^k)$}
To capture the \texttt{OR} operation for each tuple $k$, the server generates $m$ different results either 0 or 1 of secret-share form, denoted by $\mathit{Result}_i$ ($1\leq i\leq m$), each of which corresponds to the comparison for one attribute. To compute the final result of the \texttt{OR} operation for each tuple $k$, one can perform binary-tree style computation. However, for simplicity, we used an iterative \texttt{OR} operation, as follows:
\centerline{$\mathit{temp}_1^k=\mathit{Result}_1^k + \mathit{Result}_2^k - \mathit{Result}_1^k \times \mathit{Result}_2^k$}
\centerline{$\mathit{temp}_2^k=\mathit{temp}_1^k + \mathit{Result}_3^k - \mathit{temp}_1^k \times \mathit{Result}_3^k$}
\centerline{$\vdots$}
\centerline{$Output^k=\mathit{temp}_{m-1}^k + \mathit{Result}_m^k - \mathit{temp}_{m-1}^k \times \mathit{Result}_m^k$}
After performing the same operation on each tuple, finally, the server adds all the resultant of the \texttt{OR} operation ($\textstyle\sum_{k=1}^{k=n} \mathit{Output}^k$) and sends to the user. The user performs an interpolation on the received values that is the answer to the disjunctive count query.

\noindent\emph{Correctness}. The disjunctive counting operation counts only those tuples that satisfy one of the query predicates. Thus, by performing \texttt{OR} operation over string-matching resultants for an $i^{\mathit{th}}$ tuple results in 1 of secret-share form, if the tuple satisfied one of the query predicates. Thus, the sum of the \texttt{OR} operation resultant surely provides an answer to the query.

\smallskip\noindent\textbf{Information leakage discussion.} The user sends query predicates of secret-share form, and the string-matching operation is executed on all the values of the desired attribute. Hence, access-patterns are hidden from the adversary, so that the server cannot distinguish any query predicate in the count queries. The output of any count query is of secret-share form and contains an identical number of bits. Thus, based on the output size, the adversary cannot know the exact count, as well as, differentiate two count queries. However, the adversary can know whether the count query is single-dimensional, conjunctive or disjunctive count query.


\subsection{Verifying Count Query Results}
\label{subsec:Verification of Count Queries}

In this section, we describe how results of count query can be verified. Note that we explain the algorithms only for a single-dimensional query predicate. Conjunctive and disjunctive predicates can be handled in the same way. 

Here, our objective is to verify that (\textit{i}) all tuples of the databases are checked against the count query predicates, and (\textit{ii}) all answers to the query predicate ($0$ or $1$ of secret-share form) are included in the answer. In order to verify both the conditions, the server performs two functions, $f_1$ and $f_2$, as follows:

\centerline{$\textstyle \mathit{op}_1 = f_1(x)= \sum_{i=1}^{i=n} (S(x_i) \otimes o_i)$}
\centerline{$\textstyle \mathit{op}_2 = \mathit{op}_1 + f_2(y)=  \mathit{op}_1 +\sum_{i=1}^{i=n}f_2 (S(y_i) \otimes (1-o_i))$}

\noindent\textit{i}.\textit{e}., the server executes the functions $f_1$ and $f_2$ on $n$ secret-shared values each (of two newly added attributes $A_x$ and $A_y$, outsourced by the DB owner, described below). In the above equations $o_i$ is the output of the string-matching operation carried on the $i^{\mathit{th}}$ value of an attribute, say $A_j$, on which the user wants to execute the count query. The server sends the outputs of the function $f_1$, denoted by $\mathit{op}_1$, and the sum of the outputs of $f_1$ and $f_2$, denoted by $\mathit{op}_2$, to the user. The outputs $\mathit{op}_1$ and $\mathit{op}_2$ ensure the count result verification and that the server has checked each tuple, respectively. The verification method for a count query works as follows:

\smallskip
\noindent\textbf{The DB owner.} For enabling a count query result verification over any attribute, the DB owner adds two attributes, say $A_x$ and $A_y$, having initialized with one, to the relation $R^1$. The values of the attributes $A_x$ and $A_y$ are also outsourced of SSS form (not unary representations) to the servers.

\noindent\textbf{Server.} Each server $k$ executes the count query, as mentioned in \S\ref{sec:Count Query}, \textit{i}.\textit{e}., it executes the private string-matching operation on the $i^{\mathit{th}}$ ($1\leq i\leq n$) value of the attribute $A_j$ against the query predicate and adds all the resultant values. In addition, each server $k$ executes the functions $f_1$ and $f_2$. The function $f_1$ (and $f_2$) multiplies the $i^{\mathit{th}}$ value of the $A_x$ (and $A_y$) attribute by the $i^{\mathit{th}}$ string-matching resultant (and by the complement of the $i^{\mathit{th}}$ string-matching resultant). The server $k$ sends the following three things: (\textit{i}) the sum of the string-matching operation over the attribute $A_j$, as a result, say $\langle\mathit{result}\rangle_k$, of the count query, (\textit{ii}) the outputs of the function $f_1$: $\langle\mathit{op}_1\rangle_k$, and (\textit{iii}) the sum of outputs of the function $f_1$ and $f_2$: $\langle\mathit{op}_2\rangle_k$, to the user.

\noindent\textbf{User-side.} The user interpolates the received three values from each server, which result in $\mathit{Iresult}$, $\mathit{Iop}_1$, and $\mathit{Iop}_2$. If the server followed the algorithm, the user will obtain: $\mathit{Iresult}=\mathit{Iop}_1$ and $\mathit{Iop}_2=n$, where $n$ is the number of tuples in the relation, and it is known to the user.

\smallskip
\noindent\textit{\underline{Example}.} In Appendix~\ref{app_sec:Count Query Verification over Secret-Shared Values}, we provide an example of count query verification over secret-shares. However, here, we explain the above method using the following query on the \texttt{Employee} relation (refer to Table~\ref{fig:database1}): \texttt{select count(*) from Employee where Name = `John'}. Table~\ref{fig:verify count} shows the result of the private string-matching, functions $f_1$ and $f_2$ at a server. Note that for the purpose of explanation, we use cleartext values; however, the server will perform all operations over secret-shares. For the first tuple, when the servers check the first value of \texttt{Name} attribute against the query predicate, the result of string-matching becomes 1 that is multiplied by the first value of the attribute $A_x$, and results in 1. The complement of the resultant is multiplied by the first value of the attribute $A_y$, and results in 0. All the other tuples are processed in the same way. Note that for this query, $\mathit{result}=\mathit{op}_1= 2$ and $\mathit{op}_2=6$, if server performs each operation correctly.

\bgroup
\def\arraystretch{.86}
\begin{table}[!t]
\scriptsize
  \centering
  \begin{tabular}{|l||l|l|l|l|l|l|}
    \hline
    \texttt{Name} & String-matching results & $f_1$ & $f_2$\\ \hline\hline
     John &  1 & 1 & 0   \\ \hline
     John &  1 & 1 & 0   \\ \hline
     Adam &  0 & 0 & 1   \\ \hline
     Eve  &  0 & 0 & 1   \\ \hline
     Alice&  0 & 0 & 1   \\ \hline
     Mike &  0 & 0 & 1   \\ \hline\hline
     ~ & 2 &2& 4\\\hline
  \end{tabular}
  \B
  \caption{An execution of the count query verification.}
  \label{fig:verify count}
\BBB
\end{table}
\egroup

\noindent\emph{Correctness}. 
Consider two cases: (\textit{i}) all servers discard an entire identical tuple for processing, or (\textit{ii}) all servers correctly process each value of the attribute $A_j$, $\mathit{op}_1$, and $\mathit{op}_2$; however, they do not add an identical resultant, $o_i$ ($1\leq i\leq n$), of the string-matching operation. In the first case, the user finds $\mathit{Iresult}=\mathit{Iop}_1$ to be true. However, the second condition ($\mathit{Iop}_2=n$) will never be true, since discarding one tuple will result in $\mathit{Iop}_2=n-1$. In the second case, the servers will send the wrong $\mathit{result}$ by discarding an $i^{\mathit{th}}$ count query resultant, and they will also discard the $i^{\mathit{th}}$ value of the attribute $A_x$ to produce $\mathit{Iresult}=\mathit{Iop}_1$ at the user-side. Here, the user, however, finds the second condition $\mathit{Iop}_2=n$ to be false.

Thus, the above verification method correctly verifies the count query result, always, under the assumption of SSS that an adversary cannot collude all (or the majority of) the servers, as given in \S\ref{subsec:Adversarial Model}. 

\section{Sum and Average Queries}
\label{sec:Sum Queries}
The sum and average queries are based on the search operation as mentioned above in the case of conjunctive/disjunctive count queries. In this section, we briefly present sum and average queries on a secret-shared database outsourced by single or multiple DB owners. Then, we develop a result verification approach for sum queries.

\smallskip
\noindent\textbf{Conjunctive sum query.} Consider the following query: \texttt{select sum($A_\ell$) from $R$ where $A_1= v_1 \wedge A_2= v_2\wedge \ldots \wedge A_m=v_m$}.

In the secret-sharing setting, the user transforms the above query into the following query at the $j^{\mathit{th}}$ server: \texttt{select sum($A_\ell$) from $S(R^1)_j$ where $A_1= S(v_1)_j\wedge A_2= S(v_2)_j \wedge \ldots \wedge A_m=S(v_m)_j$}. This query will be executed in a similar manner as conjunctive count query except for the difference that the $i^{\mathit{th}}$ resultant of matching the query predicate is multiplied by the $i^{\mathit{th}}$ values of the attribute $A_\ell$. The $j^{\mathit{th}}$ server performs the following operation on each attribute on which the user wants to compute the sum, \textit{i}.\textit{e}., $A_\ell$ and $A_q$:

\centerline{$\textstyle\sum_{k=1}^{k=n} A_\ell[S(a_k)]_j \times (\prod_{i=1}^{i=m} (A_i[S(a_k)]_j \otimes S(v_i)_j))$}

\noindent\emph{Correctness}. The correctness of conjunctive sum queries is similar to the argument for the correctness of conjunctive count queries.

\smallskip
\noindent\textbf{Disjunctive sum query.} Consider the following query: \texttt{select sum($A_\ell$) from $R$ where $A_1= v_1\vee A_2=v_2 \vee \ldots \vee A_m=v_m$}. The user transforms the query predicates to $c$ secret-shares that results in the following query at the $j^{\mathit{th}}$ server:

\centerline{\texttt{select sum($A_\ell$) from $S(R^1)_j$}}
 \centerline{\texttt{where $A_1 = S(v_1)_j$ $\vee$ $A_2 = S(v_2)_j$ $\vee$ $\ldots$ $\vee$ $A_{m} = S(v_m)_j$}}

The server $j$ executes the following computation:

\centerline{$\mathit{Result}_i^k= A_i[S(a_k)]_j \otimes S(v_i)_j, 1\leq i\leq m, 1\leq k\leq n$}
\centerline{$\displaystyle\mathit{Output} = \textstyle\sum_{k=1}^{k=n} A_\ell[S(a_k)]_j \times (((\mathit{Result}_1^k\textnormal{ \texttt{OR} } \mathit{Result}_2^k)\textnormal{ \texttt{OR} }$}
\centerline{$\mathit{Result}_3^k)\ldots \textnormal{ \texttt{OR} }\mathit{Result}_n^k)$}

The server multiplies the $k^{\mathit{th}}$ comparison resultant by the $k^{\mathit{th}}$ value of the attribute, on which the user wants to execute the sum operation (\textit{e}.\textit{g}., $A_{\ell}$), and then, adds all values of the attribute $A_{\ell}$.

\noindent\emph{Correctness}. The correctness of a disjunctive sum query is similar to the correctness of a disjunctive count query.

\smallskip
\noindent\textbf{Average queries.} In our settings, computing the average query is a combination of the counting and the sum queries. The user requests the server to send the count and the sum of the desired values, and the user computes the average at their end.

\noindent\textbf{Information leakage discussion.} Sum queries work identically to count queries. Sum queries, like count queries, hide the facts which tuples are included in the sum operation, and the sum of the values.

\subsection{Result Verification of Sum Queries}
\label{subsec:Verification of Sum Queries}
Now, we develop a result verification approach for a single-dimensional sum query. The approach can be extended for conjunctive and disjunctive sum queries. Let $A_\ell$ be an attribute whose values will be included by the following sum query. \texttt{select sum($A_\ell$) from $R$ where $A_q = v$}.

Here, our objective is to verify that (\textit{i}) all tuples of the databases are checked against the sum query predicates, $A_q=v$, and (\textit{ii}) only all qualified values of the attribute $A_{\ell}$ are included as an answer to the sum query. The verification of a sum query first verifies the occurrences of the tuples that qualify the query predicate, using the mechanism for count query verification (\S\ref{subsec:Verification of Count Queries}). Further, the server computes two functions, $f_1$ and $f_2$, to verify both the conditions of sum-query verification in an oblivious manner, as follows:

\centerline{$\textstyle \mathit{op}_1 = f_1(x) = \sum_{i=1}^{i=n} o_i(x_i+a_i+o_i)$}
\centerline{$\textstyle \mathit{op}_2 = f_1(x) = \sum_{i=1}^{i=n}o_i(y_i+a_i+o_i)$}

\noindent\textit{i}.\textit{e}., the server executes the functions $f_1$ and $f_2$ on $n$ values, described below. In the above equations, $o_i$ is the output of the string-matching operation carried on the $i^{\mathit{th}}$ value of the attribute $A_q$, and $a_i$ be the $i^\mathit{th}$ ($1\leq i \leq n$) value of the attribute $A_\ell$. The server sends the sum of the outputs of the function $f_1$, denoted by $\mathit{op}_1$, and the outputs of $f_2$, denoted by $\mathit{op}_2$, to the user. Particularly, the verification method for a sum query works as follows:

\smallskip\noindent\textbf{The DB owner.} Analogous with the count verification method, if the data owner wants to provide verification for sum queries, new attributes should be added. Thus, the DB owner adds two attributes, say $A_x$ and $A_y$, to the relation $R^1$. The $i^{\mathit{th}}$ values of the attributes $A_x$ and $A_y$ are any two random numbers whose difference equals to $-a_i$, where $a_i$ is the $i^\mathit{th}$ value of the attribute $A_\ell$. The values of the attributes $A_x$ and $A_y$ are also secret-shared using SSS. For example, in Table~\ref{fig:verify sum}, boldface numbers show these random numbers of the attribute $A_x$ and $A_y$ in cleartext.

\noindent\textbf{Servers.} The servers execute the above-mentioned sum query, \textit{i}.\textit{e}., each server $k$ executes the private string-matching operation on the $i^{\mathit{th}}$ ($1\leq i\leq n$) value of the attribute $A_q$ against the query predicate $v$ and multiplies the resultant value by the $i^{\mathit{th}}$ value of the attribute $A_\ell$. The server $k$ adds all the resultant values of the attributes $A_\ell$.

\noindent\textit{Verification stage}. The server $k$ executes the functions $f_1$ and $f_2$ on each value $x_i$ and $y_i$ of the attributes $A_x$ and $A_y$, by following the above-mentioned equations. Finally, the server $k$ sends the following three things to the user: (\textit{i}) the sum of the resultant values of the attributes $A_\ell$, say $\langle\mathit{sum}_{\ell}\rangle_k$, (\textit{ii}) the sum of the output of the string-matching operations carried on the attribute $A_q$, say $\langle\mathit{sum}_{q}\rangle_k$,\footnote{{\scriptsize If users are interested, they can also verify this result using the method given in \S\ref{subsec:Verification of Count Queries}.}} against the query predicate, and (\textit{iii}) the sum of outputs of the functions $f_1$ and $f_2$, say $\langle\mathit{sum}_{f1f2}\rangle_k$.

\noindent\textbf{User-side.} The user interpolates the received three values from each server, which results in $\mathit{Isum}_{\ell}$, $\mathit{Isum}_q$, and $\mathit{Isum}_{f1f2}$. The user checks the value of $\mathit{Isum}_{f1f2}-2\times \mathit{Isum}_q$ and $\mathit{Isum}_{\ell}$, and if it finds equal, then it implies that the server has correctly executed the sum query.

\smallskip
\noindent\textit{\underline{Example}.} We explain the above method using the following query on the \texttt{Employee} relation (refer to Table~\ref{fig:database1}): \texttt{select sum(Salary) from Employee where Dept = `Testing'}. Table~\ref{fig:verify sum} shows the result of the private string-matching ($o$), the values of the attributes $A_x$ and $A_y$ in boldface, and the execution of the functions $f_1$ and $f_2$ at a server. Note that for the purpose of explanation, we show the verification operation in cleartext; however, the server will perform all operations over secret-shares.

For the first tuple, when the server checks the first value of \texttt{Dept} attribute against the query predicate, the string-matching resultant, $o_1$, becomes 1 that is multiplied by the first value of the attribute \texttt{Salary}. Also, the server adds the salary of the first tuple to the first values of the attributes $A_x$ and $A_y$ with $o_1$. Then, the server multiplies the summation outputs by $o_1$.

For the second tuple, the servers perform the same operations, as did on the first tuple; however, the string-matching resultant $o_2$ becomes $0$, which results in the second values of the attributes $A_x$ and $A_y$ to be $0$. The servers perform the same operations on the remaining tuples. Finally, the servers send the summation of $o_i$ (\textit{i}.\textit{e}., $2$), the sum of the salaries of qualified tuples (\textit{i}.\textit{e}., $6000$), and the sum of outputs of the functions $f_1$ and $f_2$ (\textit{i}.\textit{e}., $6004$), to the user. Note that for this query,
$\mathit{Isum}_{f1f2}-2\times \mathit{Isum}_q= \mathit{Isum}_{\ell}$, \textit{i}.\textit{e}., $6004-2\times 2= 6000$.

\noindent\emph{Correctness}. The occurrences of qualified tuples against a query predicates can be verified using the method given in \S\ref{subsec:Verification of Count Queries}. Consider two cases: (\textit{i}) all servers discard an entire identical tuple for processing, or (\textit{ii}) all servers correctly process the query predicate, but they discard the $i^{\mathit{th}}$ values of the attributes $A_\ell$, $A_x$, and $A_y$.

\bgroup
\def\arraystretch{.86}
\begin{table}[!t]
\scriptsize
  \centering
  \begin{tabular}{|p{.6cm}|p{.6cm}||p{.6cm}|p{2.2cm}|p{2.6cm}|}
    \hline
    \texttt{Dept} & Salary & $o$ values & $A_x$ and $f_1$ & $A_y$ and $f_2$\\ \hline\hline
     Testing &  1000 & 1 &  1(\textbf{200}+1000+1)$=$1201  & 1(\textbf{$-$1200}+1000+1)$=-$199  \\ \hline

     Security &100000 & 0 &  0(\textbf{1000}+100000+0)$=$0 & 0(\textbf{$-$101000}+100000+0)$=$0   \\ \hline

     Testing &  5000 & 1 & 1(\textbf{$-$5900}+5000+1)$=-$899 & 1(\textbf{900}+5000+1)$=$5901  \\ \hline

     Design  &  2000 & 0   & 0(\textbf{2000}+2000+0)$=$0 & 0(\textbf{$-$4000}+2000+1)$=$0 \\ \hline

     Design&  1500 & 0     & 0(\textbf{500}+1500+0)$=$0 & 0(\textbf{$-$2000}+1500+0)$=$0\\ \hline

     Design &  2000 & 0     & 0(\textbf{$-$2100}+2000+0)$=$0 & 0(\textbf{100}+2000+0)$=$0\\ \hline\hline
     & & 2 & $\sum f_1=$302 & $\sum f_2=$5702 \\\hline
  \end{tabular}
  \B
  \caption{An execution of the sum query verification.}
  \label{fig:verify sum}
\BBB
\end{table}
\egroup

The first case is easy to deal with, since the count query verification will inform the user that an identical tuple is discarded by the server for any processing. In the second case, the user finds
$\mathit{Isum}_{f1f2}-2\times \mathit{Isum}_q\neq \mathit{Isum}_{\ell}$, since an adversary cannot provide a wrong value of $\mathit{Isum}_q$, which is detected by count query verification. In order to hold the equation $\mathit{Isum}_{f1f2}-2\times \mathit{Isum}_q = \mathit{Isum}_{\ell}$, the adversary needs to generate shares such that $\mathit{Isum}_{f1f2}-\mathit{Isum}_{\ell}=2\times \mathit{Isum}_q$, but an adversary cannot generate any share, as per the assumption of SSS that an adversary cannot produce a share, since it requires to collude all (or the majority of) the servers, which is impossible due to the assumption of SSS, as mentioned in \S\ref{subsec:Adversarial Model}.

\section{Maximum Query}
\label{sec:maximum}
This section provides methods for finding the maximum value and retrieving the corresponding tuples for the two types of queries, where the first type of query (\texttt{QMax1}) does not have any query condition, while another (\texttt{QMax2}) is a conditional query, as follows:
\begin{center}
\texttt{\textbf{QMax1}. select * from Employee where Salary in (select max(Salary) from Employee)}

\texttt{\textbf{QMax2}. select * from Employee as E1 where E1.Dept = 'Testing' and Salary in (select max(salary) from Employee as E2 where E2.Dept = 'Testing')}\footnote{{\scriptsize Note that we considered only a single-dimensional condition in \texttt{QMax2} query. Our proposed algorithms (without any modification) can find maximum/minimum while satisfying conjunctive and disjunctive conditions.}}
\end{center}
Note that the string-matching secret-sharing algorithms (as explained in \S\ref{sec:Building Blocks of the Algorithms}) cannot find the maximum value, as these algorithms provide only equality checking mechanisms, not comparing mechanisms to compare between values. For answering maximum queries, we provide two methods: The first method, called \texttt{SDBMax} is applicable for the case when only a single DB owner outsources the database. It will be clear soon that \texttt{SDBMax} takes only one communication round when answering an unconditional query (like \texttt{QMax1}) and at most two communication rounds for answering a conditional query (like \texttt{QMax2}). The second method, called \texttt{MDBMax} is applicable to the scenario when multiple DB owners outsource their data to the servers.

\smallskip
\noindent\textbf{\texttt{SDBMax}.} In this section, we assume that $A_c$ be an attribute of the relation $S(R^1)$ on which the user wishes to execute maximum queries. Our idea is based on a combination of OP-SS~\cite{DBLP:conf/icde/EmekciAAG06,DBLP:conf/esorics/HadaviDJCG12} and SSS~\cite{DBLP:journals/cacm/Shamir79,DBLP:conf/ccs/DolevGL15} techniques. Specifically, for answering maximum queries, \texttt{SDBMax} uses the two relations $S(R^1)$ and $S(R^2)$, which are secured using secret-shared and OP-SS, respectively, as explained in \S\ref{subsec:the model}. In particular, according to our data model (\S\ref{subsec:the model}), the attribute $A_c$ will exist in the relations $S(R^1)_i$ and $S(R^2)_i$ at the server $i$. The strategy is to jointly execute a query on the relations $S(R^1)_i$ and $S(R^2)_i$ and obliviously retrieve the entire tuple from $S(R^1)_i$. In this paper, due to space restrictions, we develop \texttt{SDBMax} for the case when only a single tuple has the maximum value; for example, in \texttt{Employee} relation (see Table~\ref{fig:database1}), the maximum salary over all employees is unique.

\subsection{Unconditional Maximum Query}
\label{subsec: Unconditional Maximum Query}

Recall that by observing the shares of the attribute $A_c$ of the relation $S(R^1)$, the server cannot find the maximum value of the attribute $A_c$. However, the server can find the maximum value of the attribute $A_c$ using the relation $S(R^2)$, which is secret-shared using OP-SS. Thus, to retrieve a tuple having the maximum value in the attribute $A_c$ of the relation $S(R^1)_i$, the $i^{\mathit{th}}$ server executes the following steps:
\begin{enumerate}[noitemsep,nolistsep,leftmargin=0.01in]
  \item \emph{On the relation $S(R^2)_i$}. Since the secret-shared values of the attribute $A_c$ of the relation $S(R^2)_i$ are comparable, the server $i$ finds a tuple $\langle S(t_k), S(\mathit{value})\rangle_i$ having the maximum value in the attribute $A_c$, where $S(t_k)_i$ is the $k^{\mathit{th}}$ secret-shared tuple-id (in the attribute \texttt{SSTID}) and $S(\mathit{value})_i$ is the secret-shared value of the $A_c$ attribute in the $k^{\mathit{th}}$ tuple.

  \item \emph{On the relation $S(R^1)_i$}. Now, the server $i$ performs the join of the tuple $\langle S(t_k), S(\mathit{value})\rangle_i$ with all the tuples of the relation $S(R^1)_i$ by comparing the tuple-ids (\texttt{TID} attribute's values) of the relation $S(R^1)_i$ with $S(t_k)_i$, as follows:
      \centerline{$\textstyle \sum_{k=1}^{k=n} A_{p}[S(a_k)]_i \times ({\textnormal{\texttt{TID}}}[S(a_k)]_i \otimes S(t_k)_i)$}
      Where $p$ ($1\leq p \leq m$) is the number of attributes in the relation $R$ and \texttt{TID} is the tuple-id attribute of $S(R^1)_i$. The server $i$ compares the tuple-id $\langle S(t_k)\rangle_i$ with each $k^{\mathit{th}}$ value of the attribute \texttt{TID} of $S(R^1)_i$ and multiplies the resultant by the first $m$ attribute values of the tuple $k$. Finally, the server $i$ adds all the values of each $m$ attribute.
\end{enumerate}

\noindent\textit{Correctness}. The server $i$ can find the tuple having the maximum value in the attribute $A_c$ of the relation $S(R^2)_i$. Afterward, the comparison of the tuple-id $S(t_k)_i$ with all the values of the \texttt{TID} attribute of the relation $S(R^1)_i$ results in $n-1$ zeros (when the tuple-ids do not match) and only one (when the tuple-ids match) of secret-share form. Further, the multiplication of the resultant ($0$ or $1$ of secret-share form) by the entire tuple will leave only one tuple in the relation $S(R^1)_i$, which satisfies the query.

\smallskip\noindent\textbf{Information leakage discussion.} The adversary will know only the order of the values, due to OP-SS implemented on the relation $S(R^2)$. However, revealing only the order is not threatening, since the adversary may know the domain of the values, for example, the domain of age or salary.

Recall that, as mentioned in \S\ref{subsec:the model}, the relations $S(R^1)$ and $S(R^2)$ share attributes: \texttt{TID}/\texttt{SSTID} and $A_c$ (the attribute on which a comparison operation will be carried). However, by just observing these two relations, the adversary cannot know any relationship between them, as well as, which tuple of the relation $S(R^1)$ has the maximum value in the attribute $A_c$, due to different representations of common \texttt{TID}/\texttt{SSTID} and $A_c$ values between the relations. Furthermore, after the above-mentioned maximum query (\texttt{QMax1}) execution, the adversary cannot learn which tuple of the relation $S(R^1)$ has the maximum value in the attribute $A_c$, due to executing an identical operation on each tuple of $S(R^1)$ when joining with a single tuple of $S(R^2)$.


\subsection{Conditional Maximum Query}
\label{subsec:conditional Maximum Query}
The maximum value of the attribute $A_c$ may be different from the $A_c$'s maximum value of the tuple satisfying the \texttt{where} clause of a query. For example, in \texttt{Employee} relation, the maximum salary of the testing department is 2000, while the maximum salary of the employees is 100000. Thus, the method given for answering unconditional maximum queries is not applicable here. In the following, we provide a method to answer maximum queries that have conditional predicates (like \texttt{QMax2}), and that uses \emph{two} communication rounds between the user and the servers, as follows:

\smallskip
\noindent\emph{Round 1}. The user obliviously knows the indexes of the relation $S(R^1)$ satisfying the \texttt{where} clause of the query (the method for obliviously finding the indexes is given below).

\smallskip
\noindent\emph{Round 2}. The user interpolates the received indexes and sends the desired indexes in cleartext to the servers. Each server $i$ finds the maximum value of the attribute $A_c$ in the requested indexes by looking into the attribute \texttt{CTID} of the relation $S(R^2)_i$ and results in a tuple, say $\langle S(t_k), S(\mathit{value})\rangle_i$, where $S(t_k)_i$ shows the secret-shared tuple-id (from \texttt{SSTID} attribute) and $S(\mathit{value})_i$ shows the secret-shared maximum value. Now, the server $i$ performs a join operation between all the tuples of $S(R^1)_i$ and $\langle S(t_k), S(\mathit{value})\rangle_i$, as performed when answering unconditional maximum (\texttt{QMax1}) queries; see \S\ref{subsec: Unconditional Maximum Query}. This operation results in a tuple that satisfies the conditional maximum query.

\smallskip\noindent\textit{Note.} The difference between the methods for answering unconditional and conditional maximum queries is that first we need to know the desired indexes of $S(R^1)$ relation satisfying the \texttt{where} clause of a query in the case of conditional maximum queries.

\noindent\textit{Correctness.} The correctness of the above method can be argued in a similar way as the method for answering unconditional maximum queries.

\smallskip\noindent\textbf{Information leakage discussion.} In round 1, due to obliviously retrieving indexes of $S(R^1)$, the adversary cannot know which tuples satisfy the query predicate. In round 2, the user sends only the desired indexes in cleartext to fasten the lookup of the maximum salary. Note that by sending indexes, the adversary learns
the number of tuples that satisfies the query predicate;\footnote{{\scriptsize The adversary may already know the classification of tuples based on some criteria, due to her background knowledge. For example, the number of employees working in a department or the number of employees of certain names/age. Hence, revealing the number of tuples satisfying a query does not matter a lot; however, revealing that which tuples satisfy a query may jeopardize the data security/privacy.}} however, the adversary cannot learn which tuples of the relation $S(R^1)$ have those indexes. Due to OP-SS, the adversary also knows only the order of values of $A_c$ attribute in the requested indexes. However, joining the tuple of $S(R^2)$, which has the maximum value in $A_c$ attribute, with all tuples of $S(R^1)$ will not reveal which tuple satisfies the query predicate, as well as, have the maximum value in $A_c$.


\smallskip\noindent\textbf{Aside: Hiding frequency-analysis in round 2 used for conditional maximum queries}. In the above-mentioned round 2, the user reveals the number of tuples satisfying a query predicate. Now, below, we provide a method to hide frequency-count information:

\noindent\textit{User-side.} The user interpolates the received indexes (after round 1) and sends the desired indexes with some fake indexes, which do not satisfy the query predicate in the round 1, in cleartext to the servers. Let $x=r+f$ be the indexes that are transmitted to the servers, where $r$ and $f$ be the real and fake indexes, respectively. Note that the maximum value of the attribute $A_c$ over $x$ tuples may be more than the maximum value over $r$ tuples. Hence, the user does the following computation to appropriately send the indexes: The user arranges the $x$ indexes in a $\sqrt{x}\times \sqrt{x}$ matrix, where all $r$ real indexes appear before $f$ fake indexes. Then, the user creates $\sqrt{x}$ groups of tuples ids, say $g_1,g_2,\ldots,g_{\sqrt{x}}$, where all tuples ids in an $i^{\mathit{th}}$ row of the matrix become a part of the group $g_i$. Note that in this case only one of the groups, say $g_{\mathit{mix}}$, may contain both the real and fake indexes. Now, the user asks the server to find the maximum value of the attribute $A_c$ in each group except for the group $g_{\mathit{mix}}$ and to fetch all $\sqrt{x}$ tuples of the group $g_{\mathit{mix}}$.

\noindent\textit{Server.} For each group, $g_j$, except the group $g_{\mathit{mix}}$, each server $i$ finds the maximum value of the attribute $A_c$ by looking into the attribute \texttt{CTID} of the relation $S(R^2)_i$ and results in a tuple, say $\langle S(t_k), S(\mathit{value})\rangle_i$. Further, the server $i$ fetches all $\sqrt{x}$ tuples of the group $g_{\mathit{mix}}$. Then, the server $i$ performs a join operation (based on the attribute \texttt{TID} and \texttt{SSTID}, as performed in the second step for answering unconditional maximum queries; see \S\ref{subsec: Unconditional Maximum Query}) between all the tuples of $S(R^1)_i$ and $2\sqrt{x}-1$ tuples obtained from the relation $S(R^2)$, and returns $2\sqrt{x}-1$ tuples to the user. The user finds the maximum value over the $r$ real tuples. Note that $2\sqrt{x}-1$ tuples must satisfy a conditional maximum query; however, due to space restrictions, we do not prove this claim here.

Note that this method, on one hand, hides the frequency-count; on the other hand, it requires the servers and the user process more tuples than the method that reveals the frequency-count.

\smallskip
\noindent\textbf{Obliviously finding the indexes.} For finding the indexes, each server $k$ executes the following operation:
${\textnormal{\texttt{Index}}}[i]_k \times (A_p[i]_k\otimes S(v)_k)$, \textit{i}.\textit{e}., the server executes string-matching operations on each value of the desired attribute, say $A_p$, of the relation $S(R^1)$ and checks the occurrence of the query predicate $v$. Then, the server $k$ multiplies the $i^{\mathit{th}}$ resultant of the string-matching operation by the $i^{\mathit{th}}$ value of \texttt{Index} attribute of the relation $S(R^1)$. Finally, the server sends all the $n$ values of the attribute \texttt{Index} to the user, where $n$ is the number of tuples in the relation. The user interpolates the received values and knows the desired indexes.\footnote{{\scriptsize The servers can also check conjunctive and/or disjunctive conditions, like one-dimensional condition (see \S\ref{sec:Count Query} to recall the method of evaluating conjunctive and/or disjunctive conditions). Here, the server multiplies the $i^{\mathit{th}}$ resultant of conjunctive and/or disjunctive conditions matching by the $i^{\mathit{th}}$ value of \texttt{Index} attribute of the relation $S(R^1)$, and then, sends all the $n$ values of the attribute \texttt{Index} to the user.}}

\bgroup
\def\arraystretch{.86}
\begin{table}[!t]
\B
\scriptsize
  \centering
  \begin{tabular}{|p{.65cm}|p{.55cm}|p{.85cm}|p{.55cm}|p{.3cm}|p{.10cm}|p{1.8cm}|l|l|l|}
    \hline
      \texttt{EmpID}$^{\prime}$ & \texttt{Name}$^{\prime}$ & \texttt{Salary}$^{\prime}$ & \texttt{Dept}$^{\prime}$ &  \texttt{TID} & $o$ & $A_x$ & $A_y$\\ \hline\hline
    106 & 47 &  1000 & 80    &  3 & 1 & 1(\textbf{500}+1233)=1733 & 1(\textbf{-733}+1233)=500\\ \hline
    106 & 47 &  100000 & 120 &  2 & 0 & 0(\textbf{400}+100273)=0 &
    0(\textbf{-99873}+100273)=0 \\ \hline

    107 & 19 &  5000 & 80    &  5 & 0 & 0(\textbf{200}+5211)=0 &
    0(\textbf{-5011}+5211)=0 \\ \hline

    108 & 32 &  2000 & 51    &  4 & 0 & 0(\textbf{600}+2195)=0 &
    0(\textbf{-1595}+2195)=0 \\ \hline

    109 & 30 &  1500 & 51    &  1 & 0 & 0(\textbf{300}+1690)=0 &
    0(\textbf{-1390}+1690)=0 \\ \hline

    110 & 38 &  2000 & 51    &  6 & 0 & 0(\textbf{100}+2199)=0 &
    0(\textbf{-2099}+2199)=0 \\ \hline\hline
     & &   &    &   &  & $op_1=1733$ & $op_2=500$\\ \hline

  \end{tabular}
  \B
  \caption{An execution of the tuple retrieval verification.}
  \label{fig:verification tuple}
  \BBB
\end{table}
\egroup

\subsection{Verification of Maximum Query}
\label{subsec:Verification of Maximum Query}
This section provides a method to verify the tuple having maximum value in an attribute, $A_c$. Note that verifying only the maximum value of the tuple is trivial, since $\langle S(value)\rangle_i$ of $S(R^2)_i$ is also a part of the attribute of $A_c$ of $S(R^1)_i$, and servers send a joined output of the relations (see step 2 in \S\ref{subsec: Unconditional Maximum Query}). Thus, servers cannot alter the maximum value. However, servers can alter other attribute values of the tuple. Thus, we provide a method to verify the received tuple.


\smallskip
\noindent\textbf{Verification of retrieved tuple.}
This method is an extension of the sum verification method (as given in \S\ref{subsec:Verification of Sum Queries}). The server computes two functions, $f_1$ and $f_2$, in an oblivious manner, as follows:

\centerline{$\textstyle \mathit{op}_1 = f_1(x) = \sum_{i=1}^{i=n} o_i(x_i+s_{ij})$}
\centerline{$\textstyle \mathit{op}_2 = f_1(x) = \sum_{i=1}^{i=n}o_i(y_i+s_{ij})$}

\noindent\textit{i}.\textit{e}., the server executes the functions $f_1$ and $f_2$ on $n$ values, described below. In the above equations, $o_i$ is the output of the string-matching operation carried on the $i^{\mathit{th}}$ value of the \texttt{TID} attribute, and $s_{i,j}$ be the $i^\mathit{th}$ ($1\leq i \leq n$) value of the attribute $j$, where $1\leq j\leq m$. The server sends the difference of the outputs of the functions $f_1$ and $f_2$ to the user. Particularly, the tuple verification method works as follows:

\noindent\textbf{The DB owner.} The DB owner adds one value to each of the attribute values of a tuple along with new attributes, say $A_x$ and $A_y$. 

Let $A_1$ be an attribute having only numbers. For $A_1$ attribute, the newly added $i^{\mathit{th}}$ value in cleartext is same as the existing $i^{\mathit{th}}$ value in $A_1$ attribute. Let $A_2$ be an attribute having English alphabets, say attribute \texttt{Name} in Employee relation in Table~\ref{fig:database1}. The new value is the sum of the positions of each appeared alphabet in English letters; for example, the first value in the attribute \texttt{Name} is \texttt{John}, the DB owner adds 47 (10+15+8+14). When creating shares of the two values at the $i^{\mathit{th}}$ position of the attribute $A_1$ or $A_2$, the first value's shares are created using the mechanism that supports string-matching at the server, as mentioned in \S\ref{subsec:Building Blocks}, and the second value's shares are created using SSS.

The $i^{\mathit{th}}$ values of the attributes $A_x$ and $A_y$ are two random numbers whose difference equals to $-a_i$, where $a_i$ is the $i^\mathit{th}$ value obtained after summing all the newly added values to each attribute of the $i^\mathit{th}$ tuple. The values of the attributes $A_x$ and $A_y$ are secret-shared using SSS. E.g., in Table~\ref{fig:verification tuple}, numbers show newly added values to attributes \texttt{Name}$^{\prime}$, \texttt{Dept}$^{\prime}$, and random numbers (in bold-face) of the attributes $A_x$ and $A_y$ in cleartext (a prime ($^{\prime}$) symbol is used to distinguish these values from the original attribute values).

\noindent\textbf{Servers.} Each server $k$ executes the method for tuple retrieval as given in step 2 in \S\ref{subsec: Unconditional Maximum Query}. Then, the server $k$ executes functions $f_1$ and $f_2$, \textit{i}.\textit{e}., adds all the $m$ newly added values (one in each attribute) to $x_i$ and $y_i$ of the attributes $A_x$ and $A_y$, respectively, and then, multiply the resultant of the string-matching operation carried on \texttt{TID} attribute of the relation $S(R^1)_k$. Finally, the server $k$ sends the following two things to the user: (\textit{i}) the tuple having the maximum value in the attribute $A_c$ of the relation $S(R^1)_k$; and (\textit{ii}) the difference of outputs of the functions $f_1$ and $f_2$, say $\langle\mathit{diff}_{f1f2}\rangle_k$.

\noindent\textbf{User.} After interpolation, the user obtains the desired tuple and a value, say $\mathit{Idiff}_{f1f2}$. Like the DB owner, the user generates a value for each of the attribute values of the received tuple (see the first step above for generating values), compares against $\mathit{Isum}_{f1f2}$, and if it finds equal, then it implies that the server has correctly sent the tuple.

\noindent\textit{\underline{Example}.} Table~\ref{fig:verification tuple} shows the verification process for the first tuple-id of employee relation; see Table~\ref{fig:database1}. Note that the values and computation are shown in the cleartext; however, the values are of secret-share form and the computation will be carried on shares at servers.



\section{Other Operations}
\label{app_sec:Other Operations Related to a Maximum Query}
This section considers two more cases of a maximum query, where the maximum value can occur in multiple tuples (\S\ref{subsec:multi maxes}) and find the maximum value (or retrieve the tuple having the maximum value) over a dataset outsourced by multiple DB owners (\S\ref{app_subsec:Finding Maximum over Datasets Outsourced by Multiple DB Owners}). Further, we present an algorithm for a group-by query.

\subsection{Multiple Occurrences of the Maximum Value}
\label{subsec:multi maxes}
In practical applications, more than one tuple may have the maximum value in an attribute, \textit{e}.\textit{g}., two employees (E103 and E015) earn the maximum salary in design department; see Table~\ref{fig:database1}. However, the above-mentioned methods (for \texttt{QMax1} or \texttt{QMax2}) cannot fetch all those tuples from the relation $S(R^1)$ in one round. The reason is that since the server $i$ uses OP-SS values of the attribute $A_c$ in the relation $S(R^2)_i$ for finding the maximum value, where more than one occurrences of a value have different representations, the sever $i$ cannot find all the tuples of $S(R^2)$ having the identical maximum value, by looking OP-SS values. 

In this subsection, we, thus, provide a simple two-communication-round method for solving unconditional maximum queries. This method can be easily extended to conditional maximum queries.

\noindent\textbf{Data outsourcing.} The DB owner outsources the relation $S(R^1)$ as mentioned in \S\ref{subsec:the model}. However, the DB owner outsources the relation $S(R^2)$ with four columns: \texttt{CTID}, \texttt{SSTID}, \texttt{OP-SS-}$A_c$, and \texttt{SS-}$A_c$. The first three columns are created in the same way as mentioned in \S\ref{subsec:the model}. The $i^{\mathit{th}}$ value of  \texttt{SS-}$A_c$ attribute has the same value as the $i^{\mathit{th}}$ value of  \texttt{OP-SS-}$A_c$ attribute. However, this value is secret-shared using the unary representation, as the column $A_c$ of the relation $S(R^1)$ has, and the DB owner uses different polynomials over the $i^{\mathit{th}}$ value of the attribute $A_c$ of $S(R^1)$ and the attribute \texttt{SS-}$A_c$ of $S(R^2)$; thus, the adversary cannot observe that which two values are identical in  the two relations.

\noindent\textbf{Query execution.} The method uses two communication rounds as follows:

\noindent\textit{Round 1.} In round 1, the server $i$ finds a tuple $\langle S(t_k), S(\mathit{value}_1), S(\mathit{value}_2)\rangle_i$ having the maximum value (denoted by
$\langle S(\mathit{value})_1\rangle_i$) in the attribute $A_c$, where $S(t_k)_i$ is the $k^{\mathit{th}}$ secret-shared tuple-id (in the attribute \texttt{SSTID}) and
$\langle S(\mathit{value})_2\rangle_i$ is the secret-shared value of the \texttt{SS-}$A_c$ attribute in the $k^{\mathit{th}}$ tuple. Afterward, the server $i$ performs the following:
\centerline{$\textstyle {\textnormal{\texttt{Index}}}[k] \times (A_C[S(k)]_i \otimes S(\mathit{value}_2)_i), 1\leq k\leq n$}
\textit{i}.\textit{e}., the server compares $\langle S(\mathit{value}_2)\rangle_i$ with each $k^{\mathit{th}}$ value of the attribute $A_c$ of the relation $S(R^1)$ and multiplies the resultant by the $k^{\mathit{th}}$ index values. The server $i$ provides a list of $n$ numbers to the user.

\noindent\textit{Round 2.} After interpolating $n$ numbers, the user gets a list of $n$ numbers having 0 and \texttt{Index} values, where the desired maximum value of the attribute $A_c$ exists. Then, the user fetches all the tuples having the maximum values based on the received \texttt{Index} value. In particular, the user creates new secret-shares of the matching indexes in a way that the server can perform searching operation on \texttt{TID} attribute. The server executes the following computation to retrieve all the tuples, say $\mathcal{T}$, having the maximum value in the attribute $A_c$:
\centerline{$\textstyle \sum_{k=1}^{k=n} A_{p}[S(a_k)]_i \times ({\textnormal{\texttt{TID}}}[S(a_k)]_i \otimes S(t_j)_i)$}

Where $1\leq p\leq m$, $1\leq j\leq \mathcal{T}$ and $1\leq k\leq n$, \textit{i}.\textit{e}., the server $i$ compares each received tuple-id $\mathcal{T}$ with each tuple-id of the relation $S(R^1)_i$ and multiplies the resultant to the first $m$ attributes of the relation $S(R^1)_i$. Finally, the server $i$ adds all the attribute values for each tuple-id $\mathcal{T}$.

\noindent\textit{Complexities}. As mentioned, fetching all tuples having the maximum value in the attribute $A_c$ requires two communication rounds when answering an unconditional query. Further, each server scans the entire relation $S(R^1)$ twice. However, finding the maximum number over the attribute \texttt{OP-SS-}$A_c$ can be done using an index.

\noindent\textit{Information leakage discussion}. The adversary learns the order of the values. The adversary will not learn which tuple has the maximum value in the attribute $A_c$. But, the adversary may learn how many tuples have the maximum value. This can be prevented by asking queries for fake tuples in round 2 by generating random \texttt{TID} values, which should be larger than $n$ (the number of tuples in the relation).

\noindent\textit{Aside.} We can prevent having to outsource $S(R^2)$, by adding one additional communication round between the user and the server. In that case, the server provides a tuple having the maximum value in the attribute $A_c$, and then, the user finds occurrences of the maximum value in the relation $S(R^2)$ by using one additional round.


\noindent\textbf{Note. Answering conditional maximum query.} The above mechanism can easily be extended to support conditional maximum queries.  For answering a conditional maximum query, the user includes the above-mentioned two steps to the method given in \S\ref{subsec:conditional Maximum Query}. Thus, fetching all tuples having the maximum value in the attribute $A_c$ requires three communication rounds, and each server scans the entire relation $S(R^1)$ three times. In particular, in the first round, the server $i$ provides \texttt{Index} values to the user. In the second round, the server $i$ finds the tuple having the maximum value in the attribute $A_c$ from the requested tuple-ids, implements the above-mentioned method given in round 1, and provides a list of $n$ numbers. In the last round, the user fetches all the desired tuples.

\subsection{Group-by Query}
\label{subsec:Group-by Query}
A group-by query in combination with aggregation (viz., count/sum), can be executed similar to the aggregation query as mentioned in \S\ref{sec:Count Query} and \S\ref{sec:Sum Queries}, if the set of possible values -- for the attribute on which the group-by query will be executed -- is known to the user in advance. For example, consider the following group-by query on \texttt{Employee} relation, shown in Table~\ref{fig:database1}:

\centerline{\texttt{select Dept, count(Dept) from Employee}}
\centerline{\texttt{group by Dept}}
In this query, the user needs to know the name of departments, \textit{i}.\textit{e}, \texttt{Testing}, \texttt{Security}, and \texttt{Design}, and then, the user can execute the query at the servers for each department. Below, we briefly summarize, the execution of a group-by query with count/sum aggregation operation.

\noindent\textit{Group-by query with count}. Consider the following group-by query: \texttt{select $A_i$, count($A_i$) from $R$ group by $A_i$}. For answering this group-by query, the server $j$ executes the following computation on each tuple of the relation $R$ for each group ($1$ to $g$):

\centerline{$\mathit{Output}_l=\textstyle\sum_{k=1}^{k=n} (A_i[S(a_k)]_j \otimes S(v_l)_j)$}

\noindent Where $1\leq l\leq g$, $v_l$ is the name of each group, $\otimes$ shows a string-matching operation, and $\mathit{Output}_l$ is the answer to the group-by query. The server $j$ will return $\langle S(v_l)_j,\mathit{Output}_l\rangle$, where $1\leq l\leq g$. The user interpolates the received answers from the server to obtain the final answer to the query. Note that since the user will receive each group name, the user will know the correct answer to group-by queries for each group.



\noindent\textit{Group-by query with sum}. Consider the following group-by query involving sum operation: \texttt{select $A_i$, sum($A_{\ell}$) from $R$ group by $A_i$}. For answering this group-by query, the server $j$ executes the following computation on each tuple of the relation $R$ for each group ($1$ to $g$):

\centerline{$\textstyle\mathit{Sum}_l = \sum_{k=1}^{k=n} A_\ell[S(a_k)]_j \times (A_i[S(a_k)]_j \otimes S(v_l)_j)$}

\noindent Where $1\leq l\leq g$, $v_l$ is the name of each group, $\otimes$ shows a string-matching operation, and $\mathit{Sum}_l$ is the answer to the group-by query. The server $j$ will return $\langle S(v_l)_j,\mathit{Sum}_l\rangle$, where $1\leq l\leq g$.

\smallskip
\noindent\textbf{Information leakage discussion.} In executing the following query: {\texttt{select Dept, count(Dept) from Employee group by Dept}}, the adversary may learn the number of groups in an attribute, by receiving only three values, one value for each department. However, the user may also hide such information, by asking queries  for additional fake groups. For example, the user may ask the count query for fake groups such as \texttt{Sale} and \texttt{Production}, including the three real groups (\texttt{Testing}, \texttt{Security}, and \texttt{Design}). Since the stored data and query predicates are secret-shared, the adversary cannot learn how many unique values exist in an attribute. In this case, the count query answer for real groups, after interpolation at the user-side, will produce the desired answers, but for the fake group, the user will obtain zero as the answer. Since the user knows the real and fake groups, the user can distinguish the results.

Note that since the proposed  algorithms for group-by queries produce the result of secret-shared form, it prevents the adversary to know the frequency-count of each group. In addition, since the proposed algorithms check each group name against the desired attribute's values of each tuple, it hides access-patterns and prevents the adversary to know which group name is real or fake. Further, note that different attributes in a relation may have a different number of unique values, and hence, group-by queries over different attributes will produce a different number of answers, (depending on the unique values in attributes). We can also hide this fact by executing a group-by query for fake groups. While such a method will prevent information leakages based on the number of groups across different groups, it will incur computational cost and communication cost.


\subsection{Bucketization-based Range Queries}
\label{app_sec:Bucketization-based Range Queries}
As we mentioned, we convert a range query into several point queries that cover the entire range. However, as per Exp 8 (Figure~\ref{fig:range_query_without_bucket}), as the range increases, the computation time also increases. In order to reduce the computation time, we propose a new method that creates bins over the domain of attribute values and organizes these bins into a $k$-way tree, where $k$ is the number of child nodes of a node or the number of values in each node at the lowest level. The bucketization-based range queries works as follows:

\noindent\textbf{DB owner.} Assume that the domain of values in an attribute has $1,2,\ldots,n$ numbers. The DB owner first creates a $k$-way tree, by creating $n/k$ nodes at the 0$^{\mathit{th}}$-level by placing $1,2,\ldots, k$ numbers in the first node, $k+1,k+2,\ldots, 2k$ numbers in the second node, and so on. The first level node has $\lceil n/k^2\rceil$ nodes, where the first node of the first level becomes the parent of the first $k$ nodes of 0$^{\mathit{th}}$ node. The second node of the first level becomes a parent of $k+1, k+2,\ldots, 2k$ nodes of 0$^{\mathit{th}}$-level. In this way, the DB owner constructs a $k$-way tree of height $\lceil\log_k (n/k)+1\rceil$. Now for each level, except the root node and the leaf level, the DB owner adds one attribute in the relation $R$. An $i^{\mathit{th}}$ value of the attribute corresponding to a level, say $j$,
is set to be the node id of the $j^{\mathit{th}}$ level's node that covers the $i^{\mathit{th}}$ value at the leaf level (\textit{i}.\textit{e}., level 0).

\begin{figure}[!t]
    	   \centering
	   \includegraphics[scale=0.37]{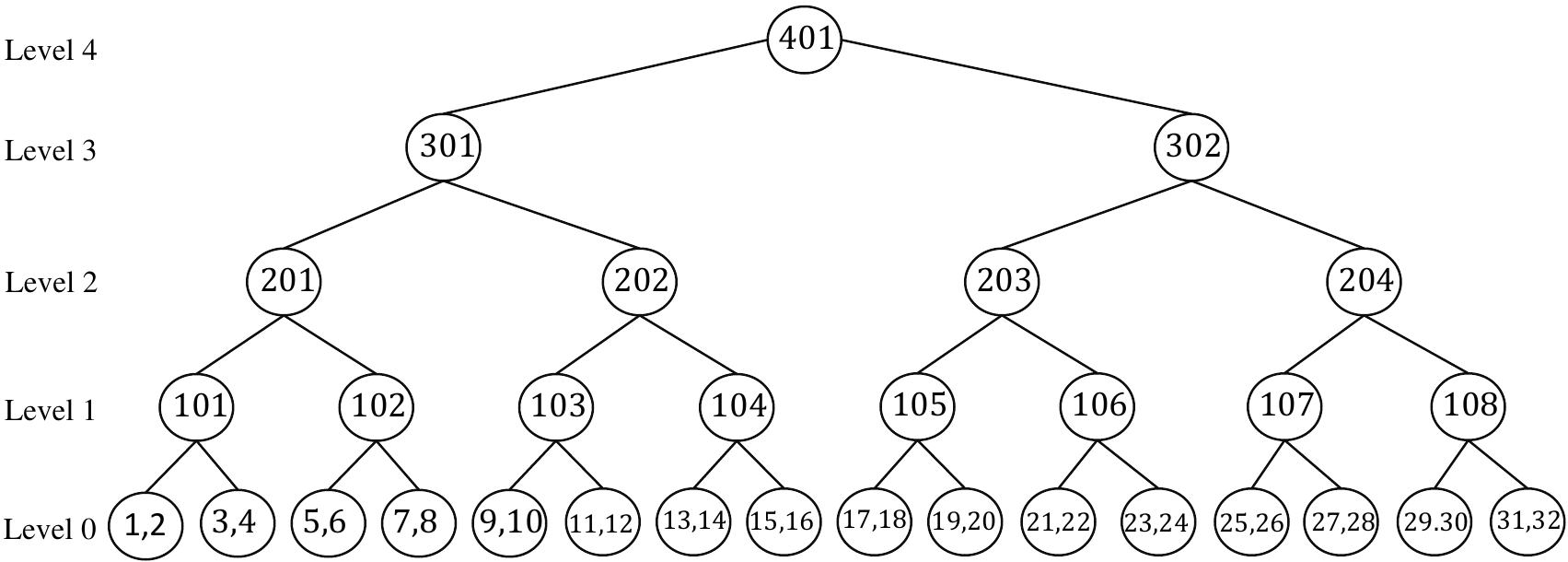}
	   \caption{2-way tree for 32 values.}
        \label{fig:2-way tree}
	   \BBB
\end{figure}

Assume that an attribute $A$ of a relation $R$ has 32 numbers ($1,2,\ldots, 32$).\footnote{{\scriptsize For simplicity of presentation, we assume that the attribute has 32 continuous numbers. Having any 32 numbers will not affect the algorithm. In the case of any 32 number,  we will create $k$-way tree for the minimum and maximum value in the domain, so that the resulting tree will have many empty nodes.}} Here, we show how does the DB owner create a $2$-way tree and three additional columns. Figure~\ref{fig:2-way tree} shows a $2$-way tree for 32 numbers. In a $2$-way tree, the 0$^{\mathit{th}}$ level has $n/k=16$ nodes, each with two numbers. The tree height is $\log_k (n/k)+1=5$. Here, the DB owner adds three columns, say $A_1$, $A_2$, and $A_3$, in the relation for levels 1, 2, and 3 of the tree; see Table~\ref{tab:new table range}. Note that, for example, $9^{\mathit{th}}$ value of the attributes $A_1$, $A_2$, and $A_3$ contains node-ids of the respective levels that cover $9^{\mathit{th}}$ value of the level 0. Thus, the attribute $A_1$ contains 103, since \texttt{Node 103} covers the value 9, the attribute $A_2$ contains 201, since \texttt{Node 201} covers the value 9, and the attribute $A_3$ contains 301, since \texttt{Node 301} covers the value 9.

\noindent\textit{Creating secret-shares of the relation.} The DB owner $c$ secret-shares of each attribute value $A_i[a_j]$ of the relation $R$ using a secret-sharing mechanism that allows string-matching operations
at the server (as specified in \S\ref{sec:Building Blocks of the Algorithms}).

\noindent\textbf{User.} We assume that the user is aware of the $k$ value used in the $k$-way tree creation. For a given range, the user first finds the minimal set of nodes that cover the range, and then, creates secret-shares of those node values. We follow a least-match method for searching node values. Assume a query for counting the number of tuples having values between \texttt{1} and \texttt{13}. The best-match method will find only \texttt{Node 301} that satisfies this query. However, it will cover some other values too, resulting in a wrong answer to the query. Thus, using a minimal set of nodes that cover the range, the user breaks the range into sub-ranges such as \texttt{Node 201}, \texttt{Node 103}, and value \texttt{13}. Note that by breaking the range from 1-13 into point queries requires searching 13 different values. However, in the modified representation of ranges using $2$-way tree, the server will search only for three values.

Finally, the user creates secret-shares of these three values (\texttt{Node 201}, \texttt{Node 103}, and value 13) and sends them to the servers with the information of the desired attribute on which the server should search for a value.

\noindent\textbf{Server.} The server executes the count query as mentioned in \S\ref{sec:Count Query}. Particularly, in this example, each server searches for \texttt{Node 201} in the attribute $A_2$, for \texttt{Node 103} in the attribute $A_1$, and for the value \texttt{13} in the attribute $A$. Finally, the server adds the outputs of all three individual searches, which produce the final answer to the count query.

\medskip
\noindent\textbf{Note.} By following the same idea of breaking a range into sub-range, one can execute conjunctive and disjunctive count/sum queries.

\section{Experiments}
\label{sec:Experiments}

This section evaluates the scalability of \textsc{Obscure} and compares it against other SSS- and MPC-based systems. We used a 16GB RAM machine as a DB owner, as well as, a user that communicates with AWS servers. For our experiments, we used two types of AWS servers -- a relatively weaker 32 GB, 2.5 GHz, Intel Xeon CPU (Exp 2, 5, 6), and a powerful 144GB RAM, 3.0GHz Intel Xeon CPU with 72 cores to study the impact of multi-threaded processing (Exp 3, 8).


\bgroup
\def\arraystretch{.86}
\begin{table}[!t]
\centering
\scriptsize
\begin{tabular}{|l|l|l|l|}\hline
$A$  & $A_1$  & $A_2$  & $A_3$  \\\hline
1  & 101 & 201 & 301 \\\hline
2  & 101 & 201 & 301 \\\hline
3  & 101 & 201 & 301 \\\hline
4  & 101 & 201 & 301 \\\hline
5  & 102 & 201 & 301 \\\hline
6  & 102 & 201 & 301 \\\hline
7  & 102 & 201 & 301 \\\hline
8  & 102 & 201 & 301 \\\hline
9  & 103 & 202 & 301 \\\hline
10 & 103 & 202 & 301 \\\hline
11 & 103 & 202 & 301 \\\hline
12 & 103 & 202 & 301 \\\hline
13 & 104 & 202 & 301 \\\hline
14 & 104 & 202 & 301 \\\hline
15 & 104 & 202 & 301 \\\hline
16 & 104 & 202 & 301 \\\hline
17 & 105 & 203 & 302 \\\hline
18 & 105 & 203 & 302 \\\hline
19 & 105 & 203 & 302 \\\hline
20 & 105 & 203 & 302 \\\hline
21 & 106 & 203 & 302 \\\hline
22 & 106 & 203 & 302 \\\hline
23 & 106 & 203 & 302 \\\hline
24 & 106 & 203 & 302 \\\hline
25 & 107 & 204 & 302 \\\hline
26 & 107 & 204 & 302 \\\hline
27 & 107 & 204 & 302 \\\hline
28 & 107 & 204 & 302 \\\hline
29 & 108 & 204 & 302 \\\hline
30 & 108 & 204 & 302 \\\hline
31 & 108 & 204 & 302 \\\hline
32 & 108 & 204 & 302 \\\hline
\end{tabular}
\caption{A relation $R$ having three new attributes, $A_1$, $A_2$, and $A_3$, based on bucketization of range values.}
\label{tab:new table range}
\BB\B
\end{table}
\egroup

\B
\subsection{\textbf{\textsc{{\large Obscure}}} Evaluation}
\noindent\textbf{Secret-share (SS) dataset generation.} We used four columns (Orderkey (OK), Partkey (PK), Linenumber (LN), and Suppkey(SK)) of LineItem table of TPC-H benchmark to generate 1M and 6M rows. To the best of our knowledge, this is the first such experiment of SSS-based approaches to such large datasets. We next explain the method followed to generate SS data for 1M rows. A similar method was used for generating SS data for 6M rows.

The four columns of LineItem table only contain numbers: OK: 1 to 300,000 (1,500,000 in 6M), PK: 1 to 40,000 (200,000 in 6M), LN: 1 to 7, and SK: 1 to 2000 (200,000 in 6M). The following steps are required to generate SS of the four columns in 1M rows:

\begin{enumerate}[noitemsep,nolistsep,leftmargin=0.01in]
  \item The first step was to pad each number of each column with zeros. Hence, all numbers in a column contain identical digits to prevent an adversary to know the distribution of values. For example, after padding 1 of OK was 000,001. Similarly, values of PK and SK were padded. We did not pad LN values, since they took only one digit.

  \item The second step was representing each digit into a set of ten numbers, as mentioned in \S\ref{subsec:Building Blocks}, having only 0s or 1s. For example, 000,001 (one value of OK attribute) was converted into 60 numbers, having all zeros except positions 1, 11, 21, 31, 41, and 52. Here, a group of the first ten numbers shows the first digit, \textit{i}.\textit{e}., 0, a group of 11th to 20th number shows the second digit, \textit{i}.\textit{e}., 0, and so on.\footnote{{\scriptsize
      One may use binary representation for representing secret-shares, since it is compact as compared to unary representation. However, in binary representation, the polynomial degree increases significantly, when we perform string-matching operations. For example, consider a decimal number, say $n$ ($=400$), having $l_d$ ($=3$) digits in decimal, and takes $l_b$ ($=9$) digits in binary ($110010000$). Here, representing 400 using unary and binary representations will take 30 and 9 numbers, respectively. However, when the user wishes to perform the minimum computation by interpolating only the desired answer, we need at least $2\times l_d+1$ and $2\times l_b+1$ servers for string-matching, when using unary and binary representation, respectively.}} Similarly, each value of PK, SK, and LN was converted. We also added columns for TID, Index, count, sum, and maximum verification, and it resulted in the relation $R^1$. Further, we created another relation, $R^2$, with three attributes CTID, SSTID, and OK, as mentioned in \S\ref{sec:Data Outsourcing}.

  \item The third step was creating SS of these numbers. We selected a polynomial $f(x)= \mathit{secret\_value}+a_1x$, where $a_1$ was selected randomly between 1 to 10M for each number, the modulus is chosen as 15,000,017, and $x$ was varied from one to fifteen to obtain fifteen shares of each value. On $R^2$, we implemented OP-SS on OK attribute, and also generated fifteen shares of SSTID. Thus, we got $S(R^1)_i$ and $S(R^2)_i$, $1\leq i\leq 15$. (Exp 5 will discuss in detail why are we generating fifteen shares.) For sum and tuple retrieval queries' time minimization, we add four more attributes corresponding to each of the four attributes in LineItem table. A value of each of the four attributes has only one secret-shared value, created using SSS (not after padding). But, one can also implement the same query on secret-shared values obtained after step 2.

  \item Lastly, we placed $i^{\mathit{th}}$ share of $S(R^1)$ and $S(R^2)$ to $i^{\mathit{th}}$ AWS server.
\end{enumerate}

\medskip
\noindent\textbf{Exp 1. Data generation time.} Table~\ref{tab:Data generation} shows the time to generate secret-shared LineItem table of size 1M and 6M rows, at the DB owner machine. Note that due to unary representation, the size of the data is large; however, the data generation time of \textsc{Obscure} is significantly less than an MPC system, which will be discussed in \S\ref{subsec:Comparing with Other Works}. 

\bgroup
\def\arraystretch{0.87}
\begin{table}[!h]
\scriptsize
  \centering
    \begin{tabular}{|l|l|l|}
    \hline
    Tuples          & Time & Size (in GB) \\ \hline
    1M & $\approx$ 10 mins & $|S(R^1)|=1.3$, $|S(R^2)|=0.3$  \\ \hline
    6M & $\approx$ 1.4 hours & $|S(R^1)|=14$, $|S(R^2)|=3$  \\ \hline
    \end{tabular}
    \B
    \caption{Exp. 1. Average time and size for shared data generation using \emph{single-threaded implementation} at the DB owner.}
    \label{tab:Data generation}
    \BB
\end{table}
\egroup

\begin{figure}[!t]
		\begin{center}
			\begin{minipage}{.95\linewidth}
				\centering
				\includegraphics[scale=0.54]{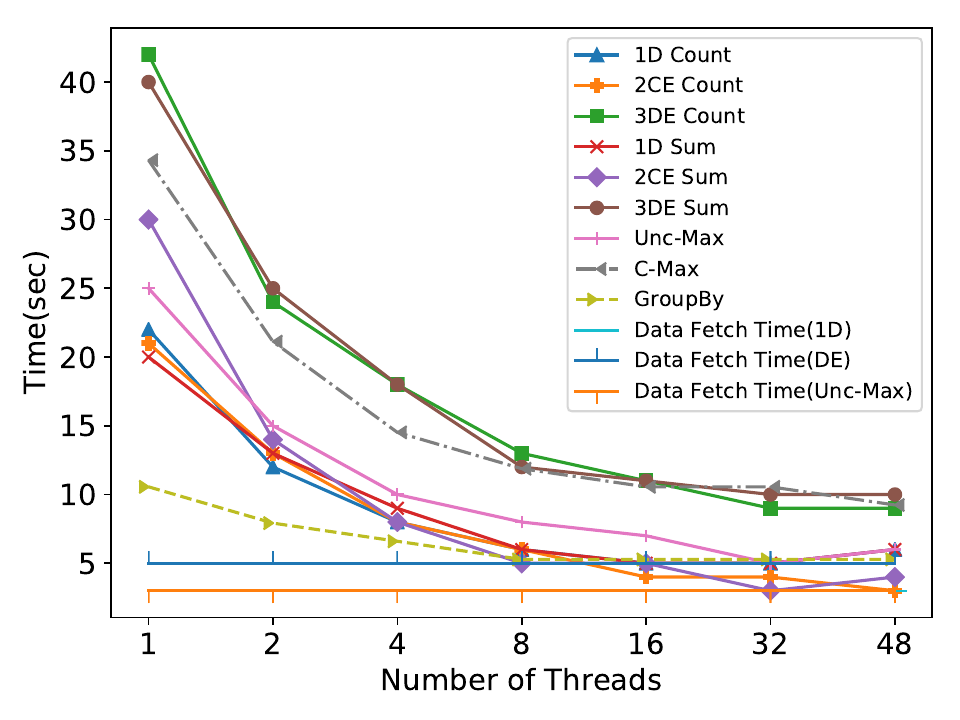}\B
				\subcaption{1M rows.}
				\label{fig:1M rows parallelism}
			\end{minipage}

			\begin{minipage}{.95\linewidth}
				\centering
				\includegraphics[scale=0.54]{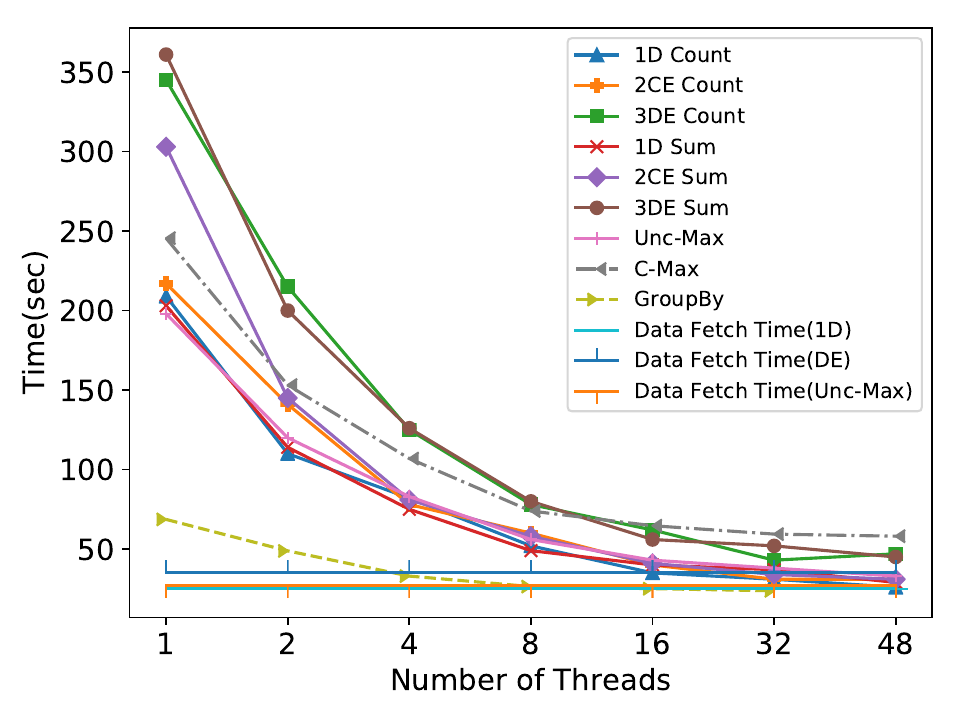}\B
				\subcaption{6M rows.}
				\label{fig:6M rows parallelism}
			\end{minipage}			
			  \end{center}
		\BB
		\caption{Exp 3. Impact of parallelism, evaluated using \textbf{AWS servers with 144GB RAM, 3.0GHz Intel Xeon CPU with 72 cores.}}
		\label{fig:impact of parallelism}
		\BBB
	\end{figure}

\medskip\noindent\textbf{Exp 2. \textsc{Obscure} performance.} In \textsc{Obscure}, we used multiple cores of CPU by writing parallel programs for one-dimensional (1D) count/sum, two-dimensional conjunctive-equality (2CE) count/sum, three-dimensional disjunctive-equality (3DE) count/sum, unconditional maximum (Unc-Max), conditional maximum (C-Max), and group-by queries on the LineItem table having 1M and 6M rows. A parallel program divides rows into blocks with one thread processing one block, and then, the intermediate results (generated by each thread) are reduced by the master thread to produce the final result.

For this experiment having fifteen shares, we used \textit{AWS servers with 144GB RAM, 3.0GHz Intel Xeon CPU with 72 cores}, and varied the \textit{degree of parallelism up to 48} (number of parallel threads). Increasing more threads did not provide speed-up, since the execution time reached close to the time spent in the sequential part of the program (Amdahl's law); furthermore, the execution time increases due to thread maintenance overheads. Figure~\ref{fig:impact of parallelism} shows as the number of threads increases, the computation time decreases. Also, observe that the data fetch time from the database remains (almost) same and less than the processing time. Further, the computation time reduces significantly due to using many threads on powerful servers (Figure~\ref{fig:impact of parallelism}). Also, note that as the size of data increases, the time increases slightly more than linearly. This is due to the unary representation that requires 10 more numbers (for the 6M rows table) to cover one new additional digit in all attribute values (except LN attribute). This increase results in additional multiplications during string-matching. An important observation is that executing any query took at most 13seconds on 1M rows and 75seconds on 6M rows.

\noindent\textit{Count and sum queries}. Figure~\ref{fig:impact of parallelism} shows the time taken by 1D, 2CE conjunctive-equality, and three-dimensional disjunctive-equality (3DE) count and sum queries. CE queries were executed on OK and LN, and DE queries involved OK, PK, and LN attributes. Observe that as the number of predicates increases, the computation time also increases, due to an increasing number of multiplications. The time difference between computations on 1M and 6M rows is about 6-7.4\%.

\noindent\textit{Maximum queries}. Fetching the tuple having the maximum value in an unconditional maximum query was very efficient, due to OP-SS, and took at most 9seconds on 1M rows and at most 50seconds on 6M rows; see Figure~\ref{fig:impact of parallelism}. We executed 1D conditional maximum query (C-Max). C-Max requires to know the tuple-ids that satisfy the condition in relation $S(R^1)$, and then, determining the maximum value from $S(R^2)$. Note that in both UnC-Max and C-Max, we achieve the maximum efficiently, due to OP-SS, (while also preventing background-knowledge-based attacks on OP-SS). The time difference between fetching a tuple having the maximum value from 1M and 6M data is about 5.5-6.6\%.

\noindent\textit{Group-by queries}. A group-by query works in a similar manner to 1D count/sum query. Figure~\ref{fig:impact of parallelism} shows the time taken by a group-by query when the number of groups was seven (due to LN attribute that has seven values), where we counted the number of OK values corresponding to each LN value.

\medskip\noindent\textbf{Exp 3. Impact of local processing at a resource-constrained user.} To show the practicality of \textsc{Obscure}, we did an experiment, where a resource-constrained user downloads the entire encrypted data and executes the computation at their end after decrypting the data and loading into a database system. We restricted the user to have a machine with 1GB RAM and single-core 1.35GHz CPU using docker, unlike multicore servers used in Exp 3, and executed the same queries that we executed in Exp 3. With this setup, decryption time at the user side was 54s and 259s for 1M and 6M rows, respectively. Further, loading decrypted data into a database system (MySQL) at the user-side took 20s and 120s for 1M and 6M rows, respectively. All queries used in Exp 2 were executed in 1-5s for both 1M and 6M rows. Note that the user computation time is significantly higher compared to the computation time of queries in Exp 3. For example, end-to-end 1D count query execution in Exp 3 over 6M secret-shared rows took 26s (see Figure~\ref{fig:6M rows parallelism}), while the same query took 385s when decrypting and loading the data into MySQL at the resource-constrained user.

\begin{wrapfigure}{r}{5.4cm}
    \begin{minipage}{.9\linewidth}
	   \centering \includegraphics[scale=0.35]{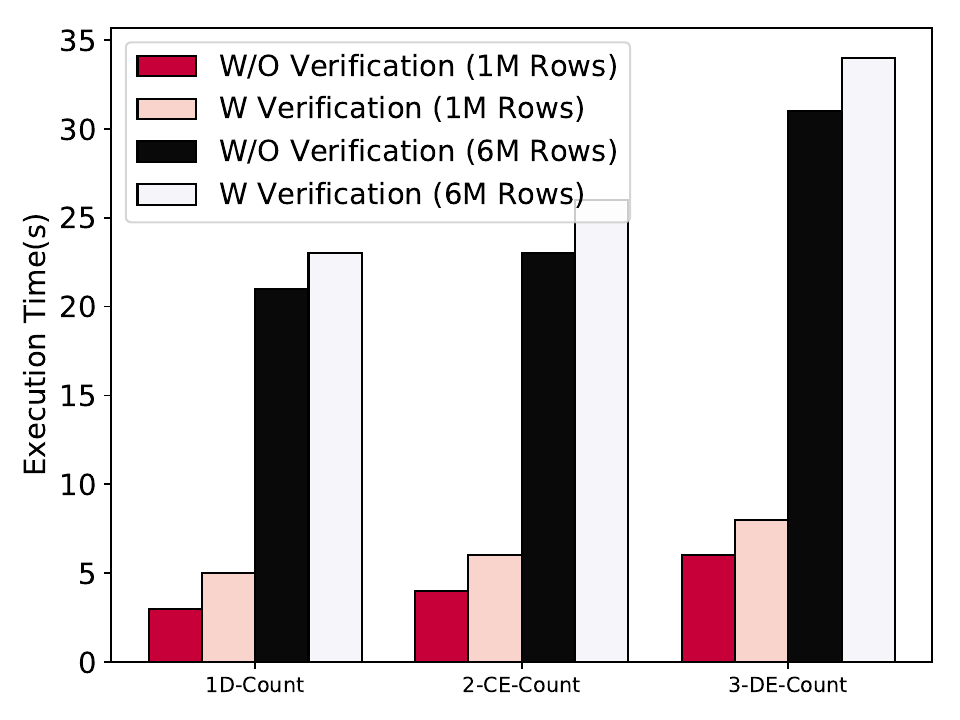}\B
	   \subcaption{Count query.}
	   \label{fig:verification Count query}
	\end{minipage}

    \begin{minipage}{.9\linewidth}
	   \centering
	\includegraphics[scale=0.35]{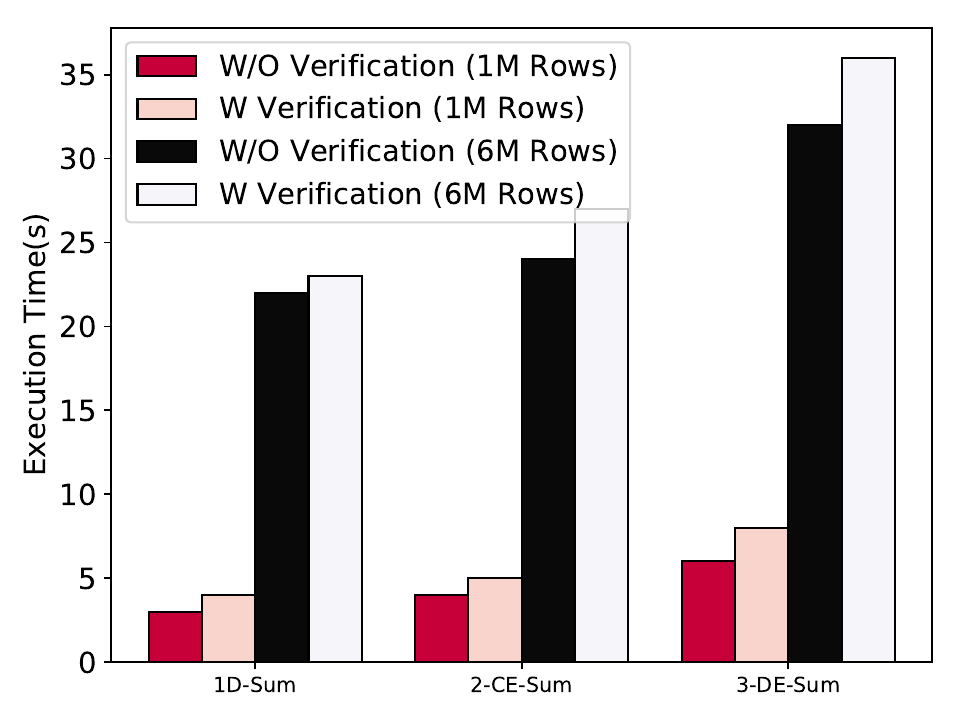}\B
	   \subcaption{Sum query.}
	   \label{fig:verification sum query}
	\end{minipage}

    \begin{minipage}{.9\linewidth}
	   \centering \includegraphics[scale=0.35]{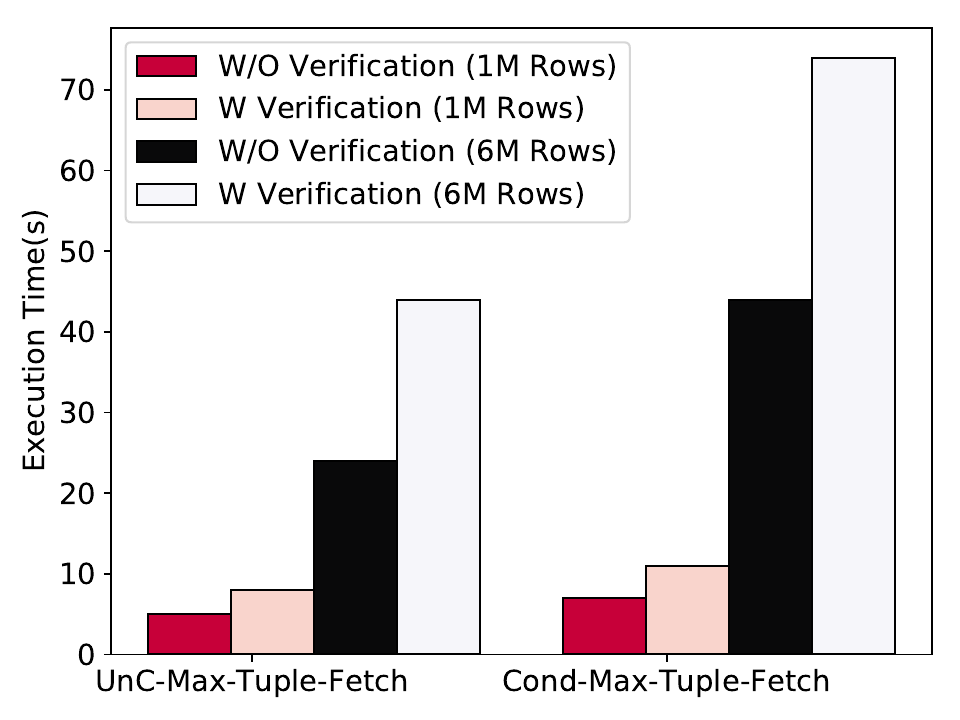}\B
	   \subcaption{Maximum tuple fetch query.}
	   \label{fig:Tuple fetch query}
	\end{minipage}	
\caption{Exp 4. Result verification.}
\label{fig:result verification}
\end{wrapfigure}

\medskip
\noindent\textbf{Exp 4. Overheads of result verification.} This experiment finds the overheads of the result verification approaches. Figure~\ref{fig:verification Count query} shows that count result verification steps do not incur a significant cost at the servers, since executing result verification requires only two more multiplications and modulo on each row's $A_x$ and $A_y$ values (see \S\ref{subsec:Verification of Count Queries}). However, in the case of a sum query, the cost increases, due to first verifying count query results, and then, sum query results. If one drops count query result verification, the cost decreases significantly; see Figure~\ref{fig:verification sum query}. Figure~\ref{fig:Tuple fetch query} shows the time comparison between fetching a tuple having the maximum value in an attribute and verifying that tuple. Here, in the case of UnC-Max-Tuple-Fetch, this step does not involve any condition checking. However, in the case of Cond-Max-Tuple-Fetch, we need to first apply count query verification method to verify that query predicate(s) are evaluated correctly. As mentioned previously, we are evaluating conditional maximum query for 1D predicate; hence, this step increases the time of verification by 304 and 790 seconds (s), in the case of 1M and 6M rows, respectively.

\setlength{\columnsep}{6pt}

\medskip
\noindent\textbf{Exp 5. Impact of number of shares.} In this experiment, we study the  impact of the number of shares on the performance of \textsc{Obscure}. For this experiment, we used four different setups with data, secret shared between 3, 5, 11, and 15 servers. Due to space restrictions, we show results for 1M rows only. Figure~\ref{fig:Impact of the number of shares} shows computation time at the server and user side, with a different number of shares.

The results demonstrate two tradeoffs, first between the number of shares and computation time at the user, and second between the number of shares and the amount of data transferred from each server to the user. As the number of shares decreases, the computation time at the user increases; since the string-matching operation results in the degree of polynomials to be doubled, and if servers do not have enough shares, they cannot compute the final answer and may require more than one round of communication with the user to compute the SS aggregate value. Thus, the communication cost also increases with a decreasing number of shares.

From Figure~\ref{fig:Impact of the number of shares}, it is clear that as the number of shares increases, the computation time at the user decreases and at the server increases, while the overall query execution time decreases, generally. In Appendix~\ref{app_sec:Impact of Number of Shares}, we discuss the processing of each query under a different number of shares. 


\begin{figure}[!t]
    	   \centering
	   \includegraphics[scale=0.34]{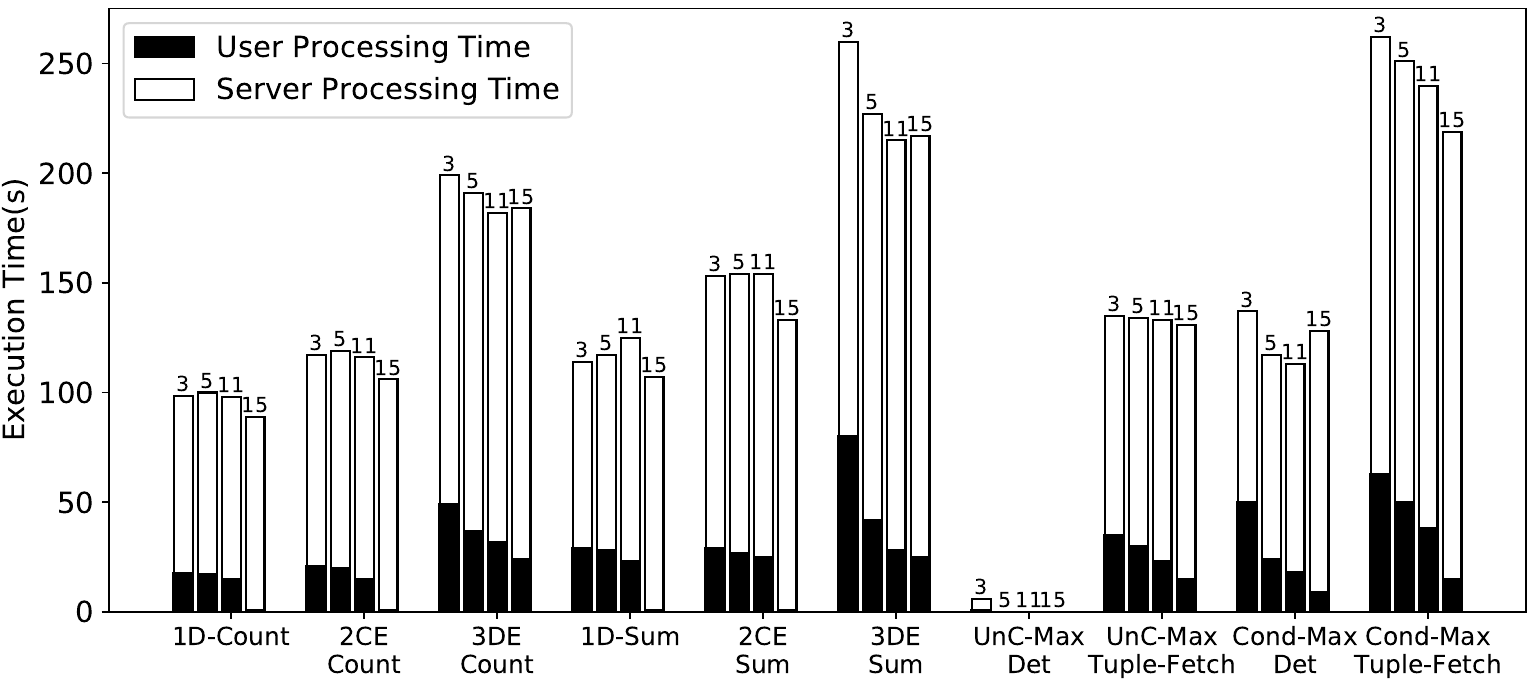}
	   \caption{Exp 5. Impact of the number of shares, using a single threaded implementation on 32GB RAM, 2.5GHz Intel Xeon CPU.}
	   \label{fig:Impact of the number of shares}
	\BB
\end{figure}

\begin{figure*}[!t]
		\B
		\begin{center}
			\begin{minipage}{.32\linewidth}
				\centering
	   \includegraphics[scale=0.37]{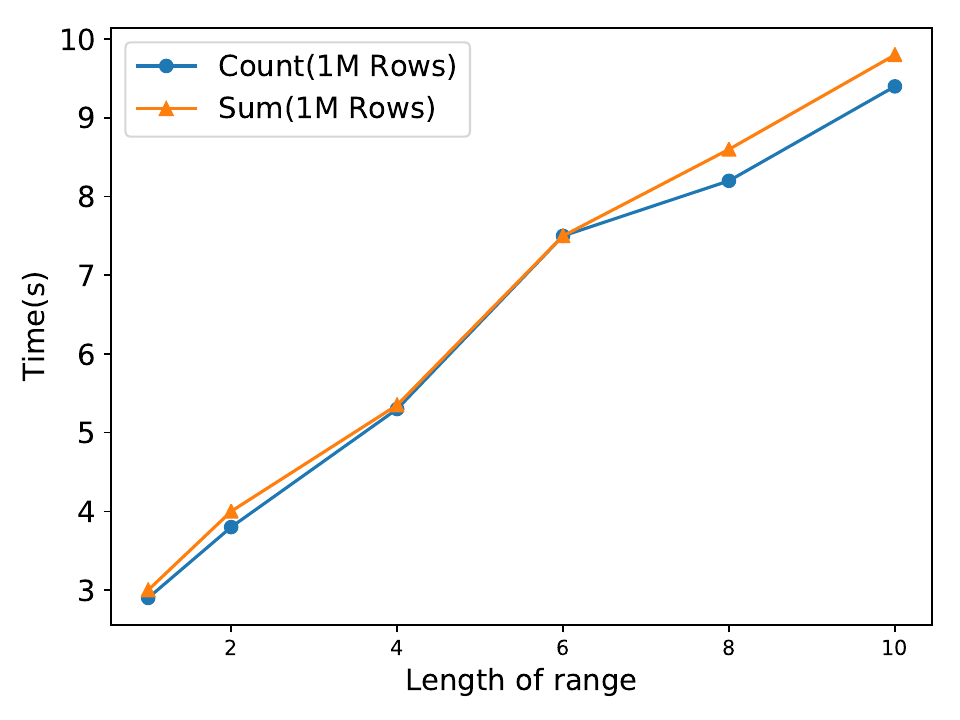}
\BB
	   \caption{Exp 7: Impact of executing range queries.}
        \label{fig:range_query_without_bucket}
			\end{minipage}
			\begin{minipage}{.32\linewidth}
				\centering
		   \includegraphics[scale=0.35]{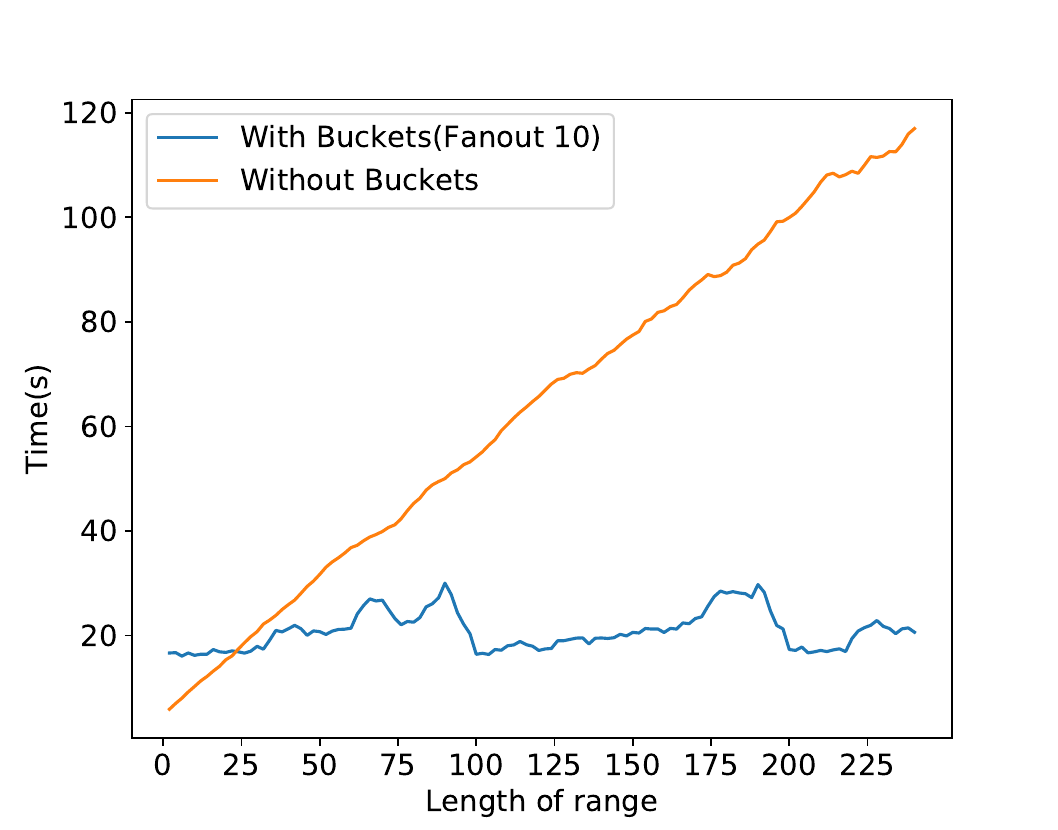}
\BB
	   \caption{Exp 8: Impact of executing range queries using bucketization.}
        \label{fig:range_with_bucket}
        			\end{minipage}			
\begin{minipage}{.32\linewidth}
				\centering
\includegraphics[scale=0.3]{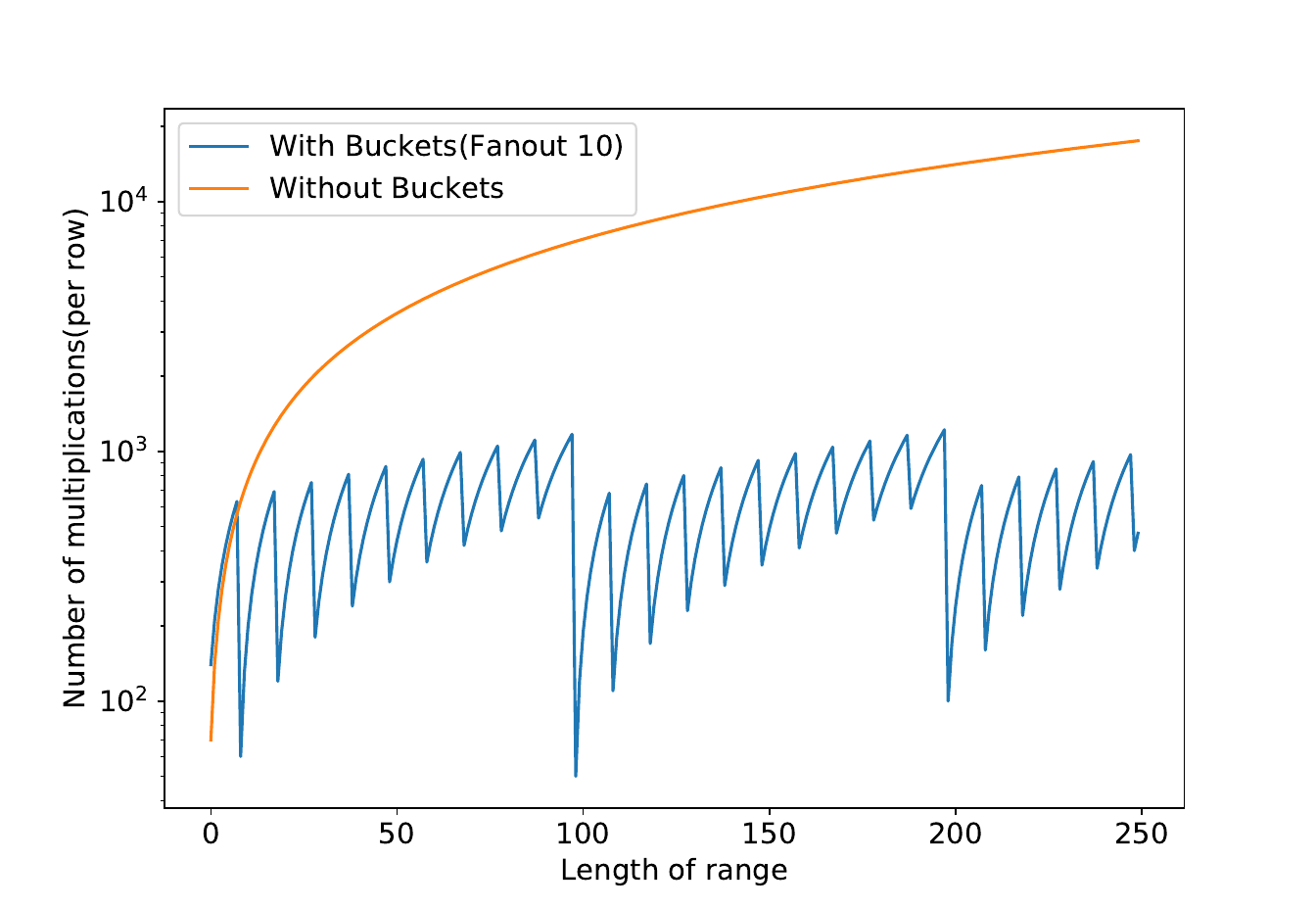}
\BB
	   \caption{Exp 8: Number of multiplication operations used in string-matching operation in range queries.}
        \label{fig:range_multi}		
        	\end{minipage}
		\end{center}
		\BBB\BBB
	\end{figure*}

\begin{figure*}[!bh]
\begin{center}
    \centering
	\includegraphics[scale=0.37]{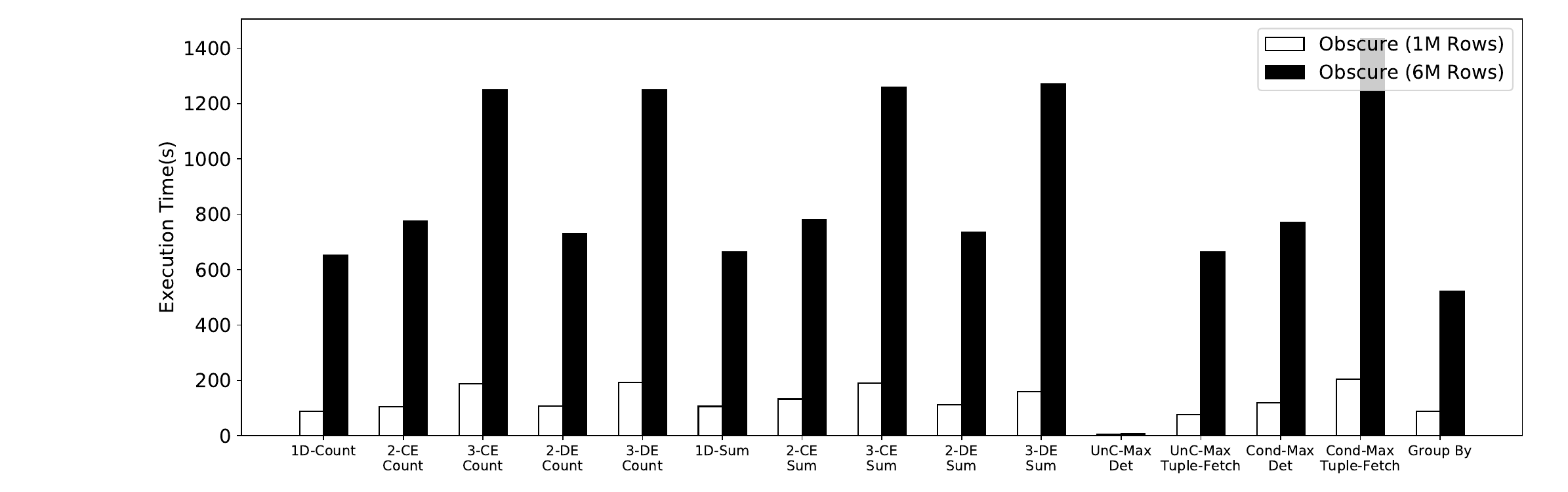}\B
\end{center}
\BB
\caption{Exp 9. \textsc{Obscure} performance using \textbf{a single-threaded implementation on 32GB RAM}, 2.5GHz Intel Xeon CPU.}
\label{fig:fig_scalability}
\BBB
\end{figure*}

\medskip
\noindent\textbf{Exp 6. Impact of communication cost.} An interesting point was the impact of the communication cost. Since servers send data to the user over the network, it may affect the overall performance. As mentioned in Exp 4., using 3 servers, the communication cost increases as compared to 15 servers. For instance, in executing DE count/sum queries over PK, LN, and OK attributes took the highest amount of data transfer when using 3 servers. Since the number of digits of the three predicates was 12 in 1M rows and 14 in 6M rows, each server sends 12 files (each of size 7MB) in the case of 1M rows and 14 files (each of size 48MB).

Hence, the server to user communication was 84MB/server in the case of 1M rows and 672MB/server in the case of 6M rows. However, in the case of 15 servers, the server to user communication was 7MB/server in the case of 1M rows and 48MB/server in the case of 6M rows. When using slow (100MB/s), medium (500MB/s), and fast (1GB/s) speed of data transmission, the data transmission time in the case of 15 servers was negligible. However, in the case of 6M, it took 7s, 1s, less than 1s per server, respectively, on slow, medium, and fast transmission speed.

Observe that the computation time at the server was at least 40s in any query on 6M rows (when using 72 core servers; Figure~\ref{fig:6M rows parallelism}) that was significantly more than the communication time between user and servers. Thus, \emph{the communication time does not affect the servers's computation time, which was the bottleneck}.

\medskip
\noindent\textbf{Exp 7. Range queries.} We evaluated range queries for 1D-count and 1D-sum operations. Given a range query involving $k$ continuous values, we converted it into $k$ 1D-count/sum queries (one per value in the range). However, this may require scanning the secret-shared relation $k$-times at the server. In order to reduce the number of scans, we processed (as per 1D-count or sum query) all the $k$-values in the range on each tuple, before processing the next tuple. As a result, we got $k$ values (as per the 1D-count or sum query) after processing the entire relation. Finally, the server adds all $k$ values and sends them to the user. We implemented a range query involving 1D-count/sum operations, using 48 threads on AWS servers with 144GB RAM, 3.0GHz Intel Xeon CPU with 72 cores. Figure~\ref{fig:range_query_without_bucket} shows that as the length of range increases the computation time also increases. In Appendix~\ref{app_sec:Bucketization-based Range Queries}, we provide a bucketization-based approach to reduce the computation time while increasing the range values.


%
%

\medskip
\noindent\textbf{Exp 8. Bucketization-based range queries.} We pre-computed the range information with $k=10$ for OrderKey values (with domain of 1-150000) of LineItem table of TPC-H. We outsourced the secret-shared version of this range information along with the original data. We executed 1D-count query on OrderKey values. Figure~\ref{fig:range_with_bucket} shows that with pre-computed range information in the form of buckets significantly decreases the computation time (\textit{i}.\textit{e}., the amount of time spent in range-based filtering over secret-shared data). Note that the na\"{\i}ve range implementation scales linearly with the length of range, whereas the bucketized pre-computed range information along with each tuple, takes almost constant time even after increasing the range length. Furthermore, it sometimes drops as fewer buckets are able to cover the entire range. For example, a range of 1-99 requires 19 searches (9 for 9 buckets covering values from 1 to 90 and extra 9 searches for values 91 to 99), whereas for a range from 1 to 100 we only need 1 bucket to represent it, therefore the time required to execute a range query for 1 to 100 decreases. Figure~\ref{fig:range_multi} shows how the number of multiplications per row varies with the increase in the range length. We can see that the na\"{\i}ve implementation requires many multiplications as compared to the bucketization-based range algorithm. However, since we store range information along with each tuple, the size of the database increases, requiring more time to scan the table. Figure~\ref{fig:range_with_bucket} shows that for small-sized ranges (length $<10$), the na\"{\i}ve algorithm performs better as the number of multiplication require by both the algorithms are equal, but the scanning time for the na\"{\i}ve algorithm is smaller than the bucketization-based algorithm.

\medskip\noindent\textbf{Exp 9. \textsc{Obscure} performance on a weaker machine.} In this experiment, we explored \textsc{Obscure} on a relatively weaker single-threaded machine with 32GB. \emph{We chose this machine since (as will be clear in~\S\ref{subsec:Comparing with Other Works}) the MPC system, we used, can only work on a local single-threaded machine. To be able to compare against that we also execute \textsc{Obscure} on 32GB AWS servers.} Note that single-threaded implementation of \textsc{Obscure} incurs time overheads, which are significantly reduced when using many threads on powerful servers; see Exp 2. Likewise Exp 2, we executed count, sum, unconditional and conditional maximum, and group-by queries on the LineItem table having 1M and 6M rows using fifteen shares; see Figure~\ref{fig:fig_scalability}. Note that Figure~\ref{fig:fig_scalability} shows that determining only the maximum value is efficient due to OP-SS, in the case of unconditional maximum queries (UnC-Max-Det, \texttt{QMax1}, see \S\ref{sec:maximum}).

\subsection{Comparing with Other Works}
\label{subsec:Comparing with Other Works}
The previous works on SSS-based techniques either did not report any experiments~\cite{DBLP:conf/icde/EmekciAAG06,DBLP:conf/ccs/DolevGL15}
or scaled to only a very small dataset, or used techniques that, while efficient, were insecure~\cite{DBLP:journals/isci/EmekciMAA14,DBLP:journals/isci/XiangLCGY16}. For instance,~\cite{DBLP:journals/isci/EmekciMAA14,DBLP:journals/isci/XiangLCGY16} are both vulnerable to access-pattern attacks.
Furthermore, these approaches achieve efficient query processing times (\textit{e}.\textit{g}., 90 ms for aggregation queries on databases of size 150K) by executing queries on SS data identically to that on cleartext, which requires user sides to retain polynomials, which were used to generate SS-data. Thus, as mentioned in \S\ref{subsec:Comparison with Existing Work}, the DB owner keeps
$n\times m$ polynomials, where $n$ and $m$ are the number of rows and
columns in a database, respectively.

MPC-based methods, \textit{e}.\textit{g}.,~\cite{sepia,DBLP:conf/ccs/BonawitzIKMMPRS17,DBLP:journals/iacr/ArcherBLKNPSW18,Sharemind}, are secure, they also do not scale to large datasets due to high overhead of share creation and/or query execution. For example, MPC-based Sepia~\cite{sepia} used 65K values for only count operation without any condition with the help of three to nine servers, and recent Bonawitz et al.~\cite{DBLP:conf/ccs/BonawitzIKMMPRS17} (appeared in CCS 2017) used only 500K values for count and sum of the numbers. Note that Sepia~\cite{sepia} and Bonawitz et al.~\cite{DBLP:conf/ccs/BonawitzIKMMPRS17} do not support conjunctive/disjunctive count/sum queries.

We evaluated one of the state-of-the-art industrial MPC-based systems that we refer to system Z to get a better sense of its performance compared to \textsc{Obscure}, whose performance is given in Figure~\ref{fig:fig_scalability}. However, we note that the MPC systems, as mentioned in \S\ref{sec:introduction}, are not available as either open source, and, often, not even available for purchase, except in the context of a contract. We were able to gain access to System Z, due to our ongoing collaboration with the team under the anonymity understanding. We installed system Z (having three SS of LineItem) on the local machine, since it was not allowed to install it on AWS. Also, note that we cannot directly compare system Z and \textsc{Obscure}, since system Z uses a single machine to keep all three shares. Inserting 1M rows in system Z took 9 hours, while the size of SS data was 1GB. We executed the same queries using the system Z, which took the following time: 532s for 1D count, 808s for CE count, 1099s for DE count, 531s for 1D sum, 801s for CE sum, 1073s for DE sum, 2205s for UnC-Max-Tuple-Fetch, and 2304s for Cond-Max-Tuple-Fetch.

\section{Conclusion}
\label{sec:Conclusion}
We proposed \textsc{Obscure} that is a information-theoretically secure, oblivious, and communication efficient system for answering aggregation queries (count, sum, and maximum having single-dimensional, conjunctive, or disjunctive query predicates) on a secret-shared dataset outsourced by either a single DB or multiple DB owners. \textsc{Obscure} also supports efficient result verification algorithms to protect against malicious adversarial cloud servers that deviate from the algorithm, due to software/hardware bugs. Our experimental results on 1M rows and 6M secret-shared rows using AWS servers show better performance as compared a simple strategy of downloading encrypted data, decrypting, and then, executing the query at a resource-constrained user. Further, we showed a tradeoff between the number of shares and performance.

\noindent
\textbf{Future directions.} While \textsc{Obscure} supports a wide range of aggregation queries, there are some issues that we plan to extend in the future, listed below:
\begin{enumerate}[nolistsep,noitemsep,leftmargin=0.01in]
  \item Reducing the number of communication rounds between the user and the server to one for any aggregation query.
  \item Designing an algorithm for group-by queries without knowing the unique values of the attribute on which the group-by query will be executed; recall that the existing algorithm for group-by queries requires to know the unique values in an attribute.
  \item Dealing with multiple aggregation operators in a query. For example, \textsc{Obscure} can execute the following query in one communication round between the user and the server: \texttt{SELECT avg(age), max(age) FORM Employee} by creating two sub-queries, one for average and another for maximum. One may consider how to execute such a query only in one communication round, without creating two sub-queries.
\item Extending this work on GPU-based efficient join and nested queries, since the proposed algorithms use multiplication and addition operations, which can be supported by GPU very efficiently.
\end{enumerate}

\section*{Acknowledgment}

{{ This material is based on research sponsored by DARPA under agreement number FA8750-16-2-0021. The U.S. Government is authorized to reproduce and distribute reprints for Governmental purposes notwithstanding any copyright notation thereon. The views and conclusions contained herein are those of the authors and should not be interpreted as necessarily representing the official policies or endorsements, either expressed or implied, of DARPA or the U.S. Government. This work is partially supported by NSF grants 1527536 and 1545071. This work of Y. Li is supported by National Natural Science Foundation of China (Grant no. 61402393, 61601396).}}


\BBB\BBB
\begin{IEEEbiography}[{\includegraphics[width=1in,height=1.0in,clip,keepaspectratio]{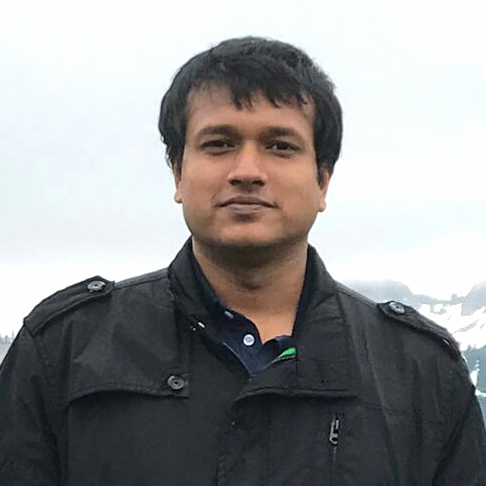}}]{Peeyush Gupta} is a Ph.D. student, advised by Prof. Sharad Mehrotra, at University of California, Irvine, USA. He obtained his Master of Technology  degree in Computer Science from Indian Institute of Technology, Bombay, India, in 2013. His research interests include IoT data management, time series database systems, and data security and privacy.
\end{IEEEbiography}
\BBB\BBB

\begin{IEEEbiography}[{\includegraphics[width=1in,height=1.0in,clip,keepaspectratio]{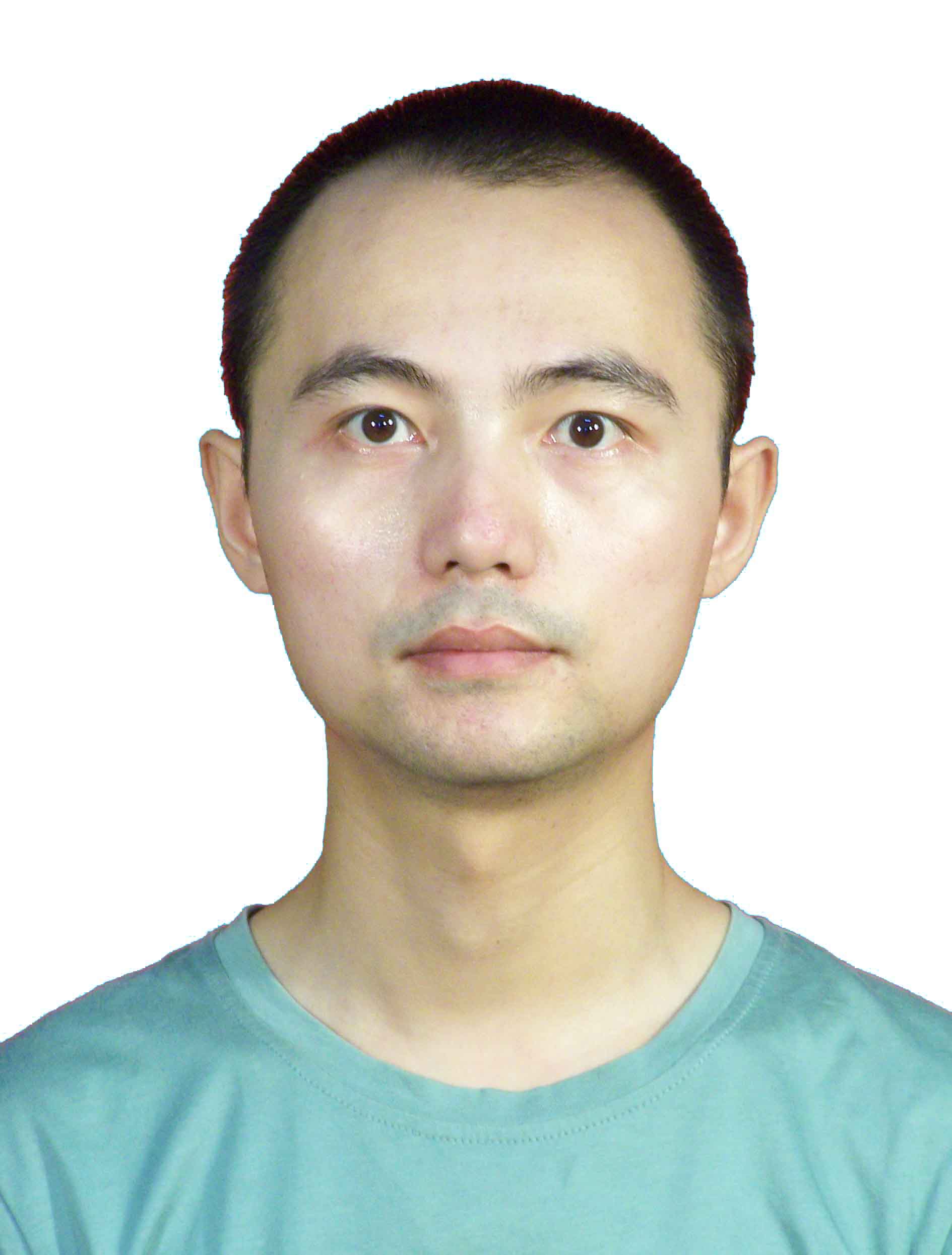}}]{Yin Li} is an associate professor in the School of Cyberspace Security, Dongguan University of Technology, China. Previously, he was an associate professor in Department of Computer Science and Technology, Xinyang Normal University, China. He received his Ph.D. degree in Computer Science from Shanghai Jiaotong University (SJTU), Shanghai in 2011. He received~ his~ B.Sc. degree and M.Sc. degree from Information Engineering University, Zhenzhou, in 2004 and 2007.  He was a Post Doc at Ben-Gurion University of the Negev, Israel, assisted by Prof. Shlomi Dolev. His current research interests include algorithm and architectures for computation in finite field, computer algebra, and secure cloud computing.
\end{IEEEbiography}
\BBB\BBB

\begin{IEEEbiography}[{\includegraphics[width=1in,height=1.0in,clip,keepaspectratio]{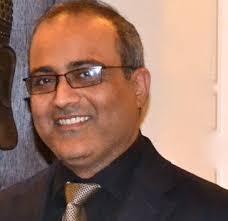}}]{Sharad Mehrotra} received the PhD degree in computer science from the University of Texas, Austin, in 1993. He is currently a professor in Department of Computer Science, University of California, Irvine. Previously, he was a professor with the University of Illinois at Urbana Champaign. He has received numerous awards and honors, including the 2011 SIGMOD Best Paper Award, 2007 DASFAA Best Paper Award, SIGMOD test of time award, 2012, DASFAA ten year best paper awards for 2013 and 2014, 1998 CAREER Award from the US National Science Foundation (NSF), and ACM ICMR best paper award for 2013. His primary research interests include the area of database management, distributed systems, secure databases, and Internet of Things.
\end{IEEEbiography}
\BBB\BBB
\begin{IEEEbiography}[{\includegraphics[width=1in,height=1.0in,clip,keepaspectratio]{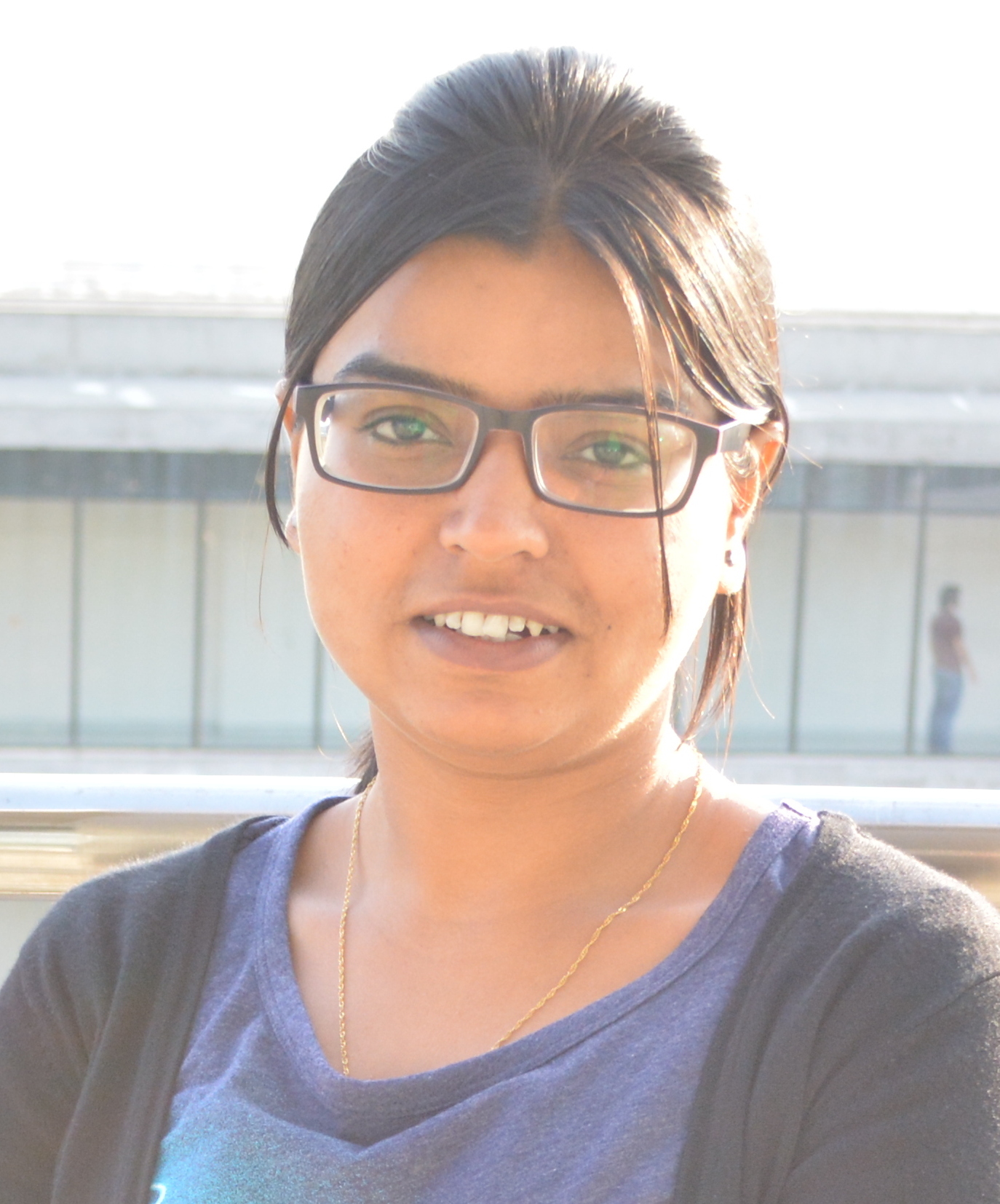}}]{Nisha Panwar} is an assistant professor at Augusta University, Georgia. She obtained her Ph.D. in Computer Science from Ben-Gurion University, Israel, in 2016, where he worked with Prof. Shlomi Dolev and Prof. Michael Segal.
She received her Master of Technology (M.Tech.) degree in Computer Engineering from National Institute of Technology, Kurukshetra, India in 2011. She was a Post Doc at University of California, Irvine, USA. Her research interests include security and privacy issues in IoT systems, as well as, in vehicular networks, computer network and communication, and distributed algorithms.
\end{IEEEbiography}
\BBB\BBB

\begin{IEEEbiography}[{\includegraphics[width=1in,height=1.0in,clip,keepaspectratio]{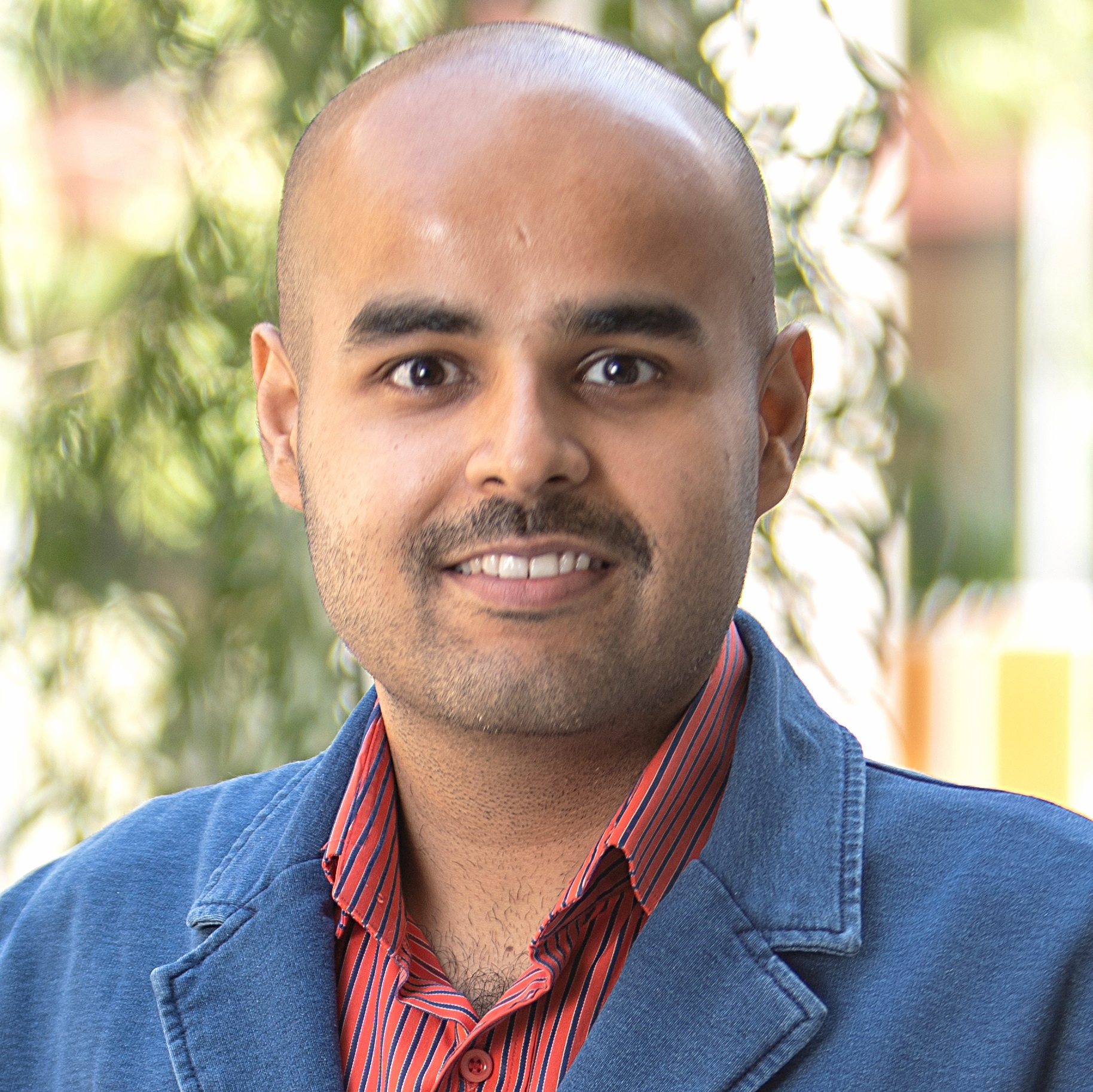}}]{Shantanu Sharma} received his Ph.D. in Computer Science in 2016 from Ben-Gurion University, Israel. During his Ph.D., he worked with Prof. Shlomi Dolev and Prof. Jeffrey Ullman. He obtained his Master of Technology (M.Tech.) degree in Computer Science from National Institute of Technology, Kurukshetra, India, in 2011. He was awarded a gold medal for the first position in his M.Tech. degree. Currently, he is pursuing his Post Doc at University of California, Irvine, USA, assisted by Prof. Sharad Mehrotra. His research interests include data security and privacy, building secure and privacy-preserving systems on sensor data for smart buildings,
designing models for MapReduce computations, distributed algorithms, mobile computing, and wireless communication.
\end{IEEEbiography}
\BBB\BBB


%
%

\newpage

\appendices
\section{Count Query Verification over Secret-Shared Values}
\label{app_sec:Count Query Verification over Secret-Shared Values}
This section shows an example for count query verification over a secret-shared relation.

\noindent\textbf{\emph{\underline{Example.}}} Assume that the domain of symbols has only two symbols, namely A and B. Thus, A can be represented as $\langle 1,0\rangle$, and B can be represented as $\langle 0,1\rangle$.

\noindent\emph{DB owner side}. Suppose that the DB owner wants to outsource three rows having A, B, A, respectively. The DB owner adds two attributes, $A_x$ and $A_y$, initialized with one, to the relation; see Table~\ref{tab:Non-secret-shared relation at the DB owner}.
\bgroup
\def\arraystretch{.86}
\begin{table}[h]
\centering
  \scriptsize
  \begin{tabular}{|l|l|l|}\hline
  Values & $A_x$ & $A_y$ \\\hline
  A & 1 & 0  \\\hline
  B & 1 & 0  \\\hline
  A & 1 & 0  \\\hline
\end{tabular}
\caption{Non-secret-shared relation at the DB owner.}
\label{tab:Non-secret-shared relation at the DB owner}
\end{table}
\egroup

The DB owner uses any polynomials of an identical degree, as shown in Table~\ref{tab:count_Secret-shares by DB owner}, to create four shares. Further, the $i^{\mathit{th}}$ share is placed to the $i^{\mathit{th}}$ server.

\bgroup
\def\arraystretch{.86}
\begin{table}[h]
\centering
  \scriptsize
  \begin{tabular}{|p{0.5cm}|p{.65cm}|l|p{0.8cm}|p{0.8cm}|p{0.8cm}|p{0.9cm}|}\hline

  Values & Vector values & Polynomials & First shares & Second shares & Third shares & Fourth shares\\\hline

  \multirow{2}{*}{A} & 1 & $x+1$ & 2 & 3 & 4 & 5\\
  \hhline{~------}   & 0 &$3x+0$ & 3 & 6 & 9 &10\\ \hline

  \multirow{2}{*}{B} & 0 &$4x+0$ & 4 & 8 & 12 & 16\\
  \hhline{~------}   & 1 &$2x+1$ & 3 & 5 &  7 &  9\\ \hline

  \multirow{2}{*}{A} & 1 &$5x+1$ & 6 & 11& 16 &21\\
  \hhline{~------}   & 0 &$4x+0$ & 4 & 8 & 12 &16\\ \hline

\multirow{3}{*}{$A_x$} & 1 & $x+1$ & 2 & 3 & 4 & 5\\
  \hhline{~------}     & 1 &$2x+1$ & 3 & 5 & 7 & 9\\
  \hhline{~------}     & 1 &$4x+1$ & 5 & 9 &13 &17\\ \hline

\multirow{3}{*}{$A_y$} & 1 &$3x+1$ & 4 & 7 & 10 & 13\\
  \hhline{~------}     & 1 &$5x+1$ & 6 & 11& 16 & 21\\
  \hhline{~------}     & 1 &$2x+1$ & 3 & 5 & 7  & 9 \\ \hline
\end{tabular}
\caption{Secret-shares of a relation shown in Table~\ref{tab:Non-secret-shared relation at the DB owner}.}
\label{tab:count_Secret-shares by DB owner}
\end{table}
\egroup

\noindent\emph{User-side}. Suppose that the user wants to search for a symbol B. The user will first represent B as a unary vector, $\langle0,1\rangle$, and then, create secret-shares of B, as shown in Table~\ref{tab:count_Secret-shares by user}.

\bgroup
\def\arraystretch{.86}
\begin{table}[h]
  \centering
  \scriptsize
\begin{tabular}{|p{.65cm}|l|p{0.8cm}|p{0.8cm}|p{0.8cm}|p{0.9cm}|}\hline
  Vector values & Polynomials & First shares & Second shares & Third shares & Fourth shares \\\hline
  0 & $2x+0$ & 2 & 4 & 6 & 8  \\\hline
  1 & $x+1$  & 2 & 3 & 4 & 5  \\\hline
\end{tabular}
\caption{Secret-shares of a vector $\langle 0,1\rangle$, created by the user/querier.}
\label{tab:count_Secret-shares by user}
\end{table}
\egroup

\noindent\emph{Server-side}. Each server executes the count query, as mentioned in \S\ref{sec:Count Query}, and the functions $f_1$ and $f_2$.

\centerline{$\textstyle \mathit{op}_1 = f_1(x)= \sum_{i=1}^{i=n} (S(x_i) \otimes o_i)$}
\centerline{$\textstyle \mathit{op}_2 = \mathit{op}_1 + f_2(y)=  \mathit{op}_1 +\sum_{i=1}^{i=n}f_2 (S(y_i) \otimes (1-o_i))$}

The function $f_1$ (and $f_2$) multiplies the $i^{\mathit{th}}$ value of the $A_x$ (and $A_y$) attribute by the $i^{\mathit{th}}$ string-matching resultant (and by the complement of the $i^{\mathit{th}}$ string-matching resultant). Each server $i$ ($1\leq i\leq 4$) sends the following three things: (\textit{i}) the result of the count query $\langle\mathit{result}\rangle_i$, (\textit{ii}) the outputs of the function $f_1$: $\langle\mathit{op}_1\rangle_i$, and (\textit{iii}) the sum of outputs of the function $f_1$ and $f_2$: $\langle\mathit{op}_2\rangle_i$, to the user. Tables~\ref{tab:server1} -~\ref{tab:server4} show the working of servers over secret-shares.

\bgroup
\def\arraystretch{.86}
\begin{table}[h]
  \centering
  \scriptsize
  \begin{tabular}{|l|p{1.3cm}|p{1.6cm}|p{1.5cm}|p{1.6cm}|l|l|}\hline

Value & SMR ($o$)        & Function $f_1$  & $1-o$    & Function $f_2$      \\\hline
A     & $(2,3)\otimes (2,2)=10$ & $2\times 10=20$ & $1-10=-9$ &  $4\times -9 =-36$ \\\hline
B     & $(4,3)\otimes (2,2)=14$ & $3\times 14=42$ & $1-14=-13$ & $6\times -13 =-78$ \\\hline
A     & $(6,4)\otimes (2,2)=20$ & $5\times 20=100$ & $1-20=-19$ & $3\times -19 =-57$ \\\hline
      & $\mathit{result}_1=44$  & $\langle\mathit{op}_1\rangle_1=162$ &   & $\langle\mathit{op}_2\rangle_1=-9$\\\hline
\end{tabular}
\caption{Server 1 execution.}
\label{tab:server1}
\end{table}
\egroup

\bgroup
\def\arraystretch{.86}
\begin{table}[h]
  \centering
  \scriptsize
  \begin{tabular}{|l|p{1.3cm}|p{1.6cm}|p{1.5cm}|p{1.6cm}|l|l|}\hline

Value & SMR ($o$)        & Function $f_1$  & $1-o$    & Function $f_2$      \\\hline
A     & $(3,6)\otimes (4,3)=30$ & $3\times 30=90$ & $1-30=-29$ &  $7\times -29 =-203$ \\\hline
B     & $(8,5)\otimes (4,3)=47$ & $5\times 47=235$ & $1-47=-46$ & $11\times -46 =-506$ \\\hline
A     & $(11,8)\otimes(4,3)=68$ & $9\times 68=612$ & $1-68=-67$ & $5\times -67 =-335$ \\\hline
      & $\mathit{result}_2=145$  & $\langle\mathit{op}_1\rangle_2=937$ &   & $\langle\mathit{op}_2\rangle_2=-107$\\\hline
\end{tabular}
\caption{Server 2 execution.}
\label{tab:server2}
\end{table}
\egroup

\bgroup
\def\arraystretch{.86}
\begin{table}[h]
  \centering
  \scriptsize
  \begin{tabular}{|l|p{1.5cm}|p{1.6cm}|p{1.5cm}|p{1.6cm}|l|l|}\hline

Value & SMR ($o$)        & Function $f_1$  & $1-o$    & Function $f_2$      \\\hline
A     & $(4,9)\otimes  (6,4)=60$  & $4\times 60=240$ & $1-60=-59$ &  $10\times -59 =-590$ \\\hline
B     & $(12,7)\otimes (6,4)=100$ & $7\times100=600$ & $1-100=-99$ & $16\times -99 =-1584$ \\\hline
A     & $(16,12)\otimes(6,4)=144$ & $13\times 44=1872$ & $1-144=-143$ & $7\times -143 =-1001$ \\\hline
      & $\mathit{result}_3=304$  & $\langle\mathit{op}_1\rangle_3=2812$ &   & $\langle\mathit{op}_2\rangle_3=-3175$\\\hline
\end{tabular}
\caption{Server 3 execution.}
\label{tab:server3}
\end{table}
\egroup

\bgroup
\def\arraystretch{.86}
\begin{table}[!h]
  \centering
  \scriptsize
  \begin{tabular}{|l|p{1.5cm}|p{1.6cm}|p{1.5cm}|p{1.6cm}|l|l|}\hline

Value & SMR ($o$)        & Function $f_1$  & $1-o$    & Function $f_2$      \\\hline
A     & $(5,12)\otimes  (8,5)=100$  & $5\times 100=500$ & $1-100=-99$ &  $13\times -99 =-1287$ \\\hline
B     & $(16,9)\otimes (8,5)=173$ & $9\times173=1557$ & $1-173=-172$ & $21\times -172 =-3612$ \\\hline
A     & $(21,16)\otimes(8,5)=248$ & $17\times248=4216$ & $1-248=-247$ & $9\times -247 =-2223$ \\\hline
      & $\mathit{result}_4=521$  & $\langle\mathit{op}_1\rangle_4=6273$ &   & $\langle\mathit{op}_2\rangle_4=-7122$\\\hline
\end{tabular}
\caption{Server 4 execution.}
\label{tab:server4}
\end{table}
\egroup


\noindent\emph{User-side}. The user interpolates the received values from each server, which result in $\mathit{Iresult}$, $\mathit{Iop}_1$, and $\mathit{Iop}_2$, as follows:

\centerline{\scriptsize
$\mathit{Iresult}=
\frac{(x-2)(x-3)(x-4)}{(1-2)(1-3)(1-4)}\times 44 +
\frac{(x-1)(x-3)(x-4)}{(2-1)(2-3)(2-4)}\times 145 +$}
\centerline{\scriptsize
$\frac{(x-1)(x-2)(x-4)}{(3-1)(3-2)(3-4)}\times 304 +
\frac{(x-1)(x-2)(x-3)}{(4-1)(4-2)(4-3)}\times 521 =1
$}

\centerline{\scriptsize
$\mathit{Iop}_1=
\frac{(x-2)(x-3)(x-4)}{(1-2)(1-3)(1-4)}\times 162 +
\frac{(x-1)(x-3)(x-4)}{(2-1)(2-3)(2-4)}\times 937 +$}
\centerline{\scriptsize
$\frac{(x-1)(x-2)(x-4)}{(3-1)(3-2)(3-4)}\times 2812 +
\frac{(x-1)(x-2)(x-3)}{(4-1)(4-2)(4-3)}\times 6273 =1
$}

\centerline{\scriptsize
$\mathit{Iop}_2=
\frac{(x-2)(x-3)(x-4)}{(1-2)(1-3)(1-4)}\times -9 +
\frac{(x-1)(x-3)(x-4)}{(2-1)(2-3)(2-4)}\times -107 +$}
\centerline{\scriptsize
$\frac{(x-1)(x-2)(x-4)}{(3-1)(3-2)(3-4)}\times -363 +
\frac{(x-1)(x-2)(x-3)}{(4-1)(4-2)(4-3)}\times -849 =3
$}


Note that the user obtains: $\mathit{Iresult}=\mathit{Iop}_1$ and $\mathit{Iop}_2=n$, where $n$ is the number of tuples in the relation, and it is known to the user. Thus, it is proved that the servers followed the count query algorithm.

\section{Finding Maximum over Datasets Outsourced by Multiple DB Owners}
\label{app_subsec:Finding Maximum over Datasets Outsourced by Multiple DB Owners}
In this section, we explain a method, named \texttt{MDBMax} for the case when multiple DB owners outsource their data to servers, \textit{e}.\textit{g}., smart meters. Note that for the case of multiple DB owners, \texttt{SDBMax} method cannot work, as different DB owners do not share any information for creating OP-SS. We describe \texttt{MDBMax} for a list, say $A_c$, having $n$ numbers outsourced by $k$ DB owners/devices, where $k\leq n$.

\noindent\textit{Data outsourcing.} Consider that an $i^{\mathit{th}}$ DB owner wishes to outsource a number, say $v$. The $i^{\mathit{th}}$ DB owner creates shares of $v$ using a secret-sharing mechanism that allows string-matching operations at the server and sends to the $c$ non-communicating servers, as described in \S\ref{subsec:the model}. However, note that, here, we do not outsource numbers using the unary representation, which was used for other queries in previous sections. In this case, the DB owner first creates a binary representation of the number and then creates the shares. Binary representation allows us to execute 2's complement-based signbit computation, as follows:

\noindent\textit{Query execution.} \texttt{MDBMax} uses 2's complement-based signbit computation for each pair of shares at a server. The server $j$ considers an $i^{\mathit{th}}$ ($1\leq i\leq n$) share as the maximum value and compares the $i^{\mathit{th}}$ share against the remaining $n-1$ shares.

Thus, for each number at the $i^{\mathit{th}}$ position, say $V_i$, the server $j$ computes the signbit with all the other numbers using 2's complement-based subtraction, \textit{i}.\textit{e}., $\mathit{signbit}(V_i-V_x)$, $x\neq i$, and $1\leq x, i\leq n$. Recall that the signbit results in $1$ of secret-share form, if $V_i<V_x$; otherwise, $0$. Then, the server $j$ adds all $n-1$ signbit values computed for the $i^{\mathit{th}}$ share of the list $A_c$. Therefore, after comparing each pair of inputs and adding corresponding $n-1$ signbit values, the server $j$ has a vector, say $\mathit{vec}$, of $n$ shares. The user asks the count query (\S\ref{sec:Count Query}) to find the occurrences of $0$ in $\mathit{vec}$ (it will be clear soon why the user is asking for counting $0$) and the sum of the values of $A_c$ for which the count query resulted in 1 of secret-shared form.

\noindent\emph{\underline{Example.}} The following table shows how does the server find the maximum value without using OP-SS. Note that for the purpose of explanation, we use cleartext values and computations; however, the server will perform all operations over secret-shared
numbers. The list $A_c$ contains five numbers: 10, 20, 90, 50, and 90. Note that the sum of signbit for the maximum value is 0. The server executes the count query for the value of 0, multiplies the $i^{\mathit{th}}$ resultant to the $i^{\mathit{th}}$ value of $A_c$, and sends the sum of the count query results and the sum of values of $A_c$ after multiplication. The user receives 2 and 180 as the output of the count and sum queries, respectively, and so that the user knows the maximum value is 90.
\begin{center}
\scriptsize
\begin{tabular}{|l|l|l|l|l|l|l|l|l|}
\hline

  \multirow{2}{*}{$A_c$} & \multicolumn{5}{|c|}{Signbits}  & \multirow{2}{.8cm}{Sum of signbits} & \multirow{2}{1.8cm}{String-matching result} & \multirow{2}{.9cm}{Maximum value} \\
    \hhline{~-----~~~} & 10 & 20 & 90 & 50 & 90 & & & \\\hline

10 & 0 & 1  &  1 &  1 &  1 &  4 & 0 & 0 \\\hline
20 & 0 & 0  &  1 &  1 &  1 &  3 & 0 & 0 \\\hline
90 & 0 & 0  &  0 &  0 &  0 &  0 & 1 & 90 \\\hline
50 & 0 & 0  &  1 &  0 &  1 &  2 & 0 & 0 \\\hline
90 & 0 & 0  &  0 &  0 &  0 &  0 & 1 & 90 \\\hline\hline
\multicolumn{7}{|c|}{Answers to the count and sum queries} & 2 & 180 \\\hline
\end{tabular}
\end{center}
\noindent\emph{Complexities.} \texttt{MDBMax} requires $n^2$ comparisons and $2n+1$ scan rounds of the list $A_c$, where the first $n$ rounds are used in comparing each pair of numbers, other $n$ rounds are used for adding $n-1$ signbits for each number, and one additional round for executing count and sum queries.

\smallskip\noindent\textbf{Minimum queries over numbers outsourced by multiple DB owners.} Here, we also compare each pair of numbers. However, for each number at the $i^{\mathit{th}}$ position, say $V_i$, we compute the signbit with all the other numbers using 2's complement-based subtraction, as follows: $\mathit{signbit}(V_x-V_i)$, $x\neq i$, and $1\leq x, i\leq n$. As a result, after adding $n-1$ signbits for each number, the minimum values has 0, and the user asks for the count query for 0 and the sum
of the values of $A_c$ for which the count query resulted in 1 of
secret-shared form.

\section{Minimum and Top-k}
\label{app_sec:Minimum, Top-k, and Reverse Top-k}
In this section, we focus on the minimum and top-k/reverse-top-k finding algorithms on an attribute, say $A_c$. Further, we assume that any value in the attribute $A_c$ appears only once.

\medskip\noindent\textbf{Minimum.} Consider the following two queries \texttt{QMin1} (unconditional minimum) and \texttt{QMin2} (conditional minimum).
\begin{center}
\texttt{\textbf{QMin1}. select * from Employee where Salary in (select min(Salary) from Employee)}

\texttt{\textbf{QMin2}. select * from Employee as E1 where E1.Dept = 'Testing' and Salary in (select min(salary) from Employee as E2 where E2.Dept = 'Testing')}
\end{center}
Here, in short, we explain how to execute these queries on the relations $S(R^1)$ and $S(R^2)$, since these queries are similar to maximum queries \S\ref{sec:maximum}. To execute an unconditional minimum query, the user follows the same strategy for solving \texttt{QMax1} (\S\ref{subsec: Unconditional Maximum Query}); however, the user asks for the minimum value from the relation $S(R^1)$. First, each server $i$ finds a tuple, say $\langle S(t_k), S(\mathit{value})\rangle_i$, where $S(t_k)_i$ is the $k^{\mathit{th}}$ secret-shared tuple-id (in the attribute \texttt{SSTID}) and $S(\mathit{value})_i$ is the secret-shared minimum value of the $A_c$ attribute in the $k^{\mathit{th}}$ tuple. Finally, the server $i$ compares the tuple-id $\langle S(t_k)\rangle_i$ with each $k^{\mathit{th}}$ value of the attribute \texttt{TID} of $S(R^1)_i$ and multiplies the resultant by the first $m$ attribute values of the tuple $k$. Finally, the server $i$ adds all the values of each $m$ attribute.

To execute a conditional minimum query, the user operates in two rounds, like a conditional maximum query; see \S\ref{subsec:conditional Maximum Query}. In the round 1, the user obliviously knows the tuple-ids of the relation $S(R^1)$ satisfying query predicate. In round 2, the user interpolates the received tuple-ids and sends the desired tuple-ids in cleartext to the servers. Each server $i$ finds the minimum value of the attribute $A_c$ in the requested tuple-ids by looking into the attribute \texttt{CTID} of the relation $S(R^2)_i$ and results in a tuple, say $\langle S(t_k), S(\mathit{value})\rangle_i$, where $S(t_k)_i$ shows the secret-shared tuple-id (from \texttt{SSTID} attribute) and $S(\mathit{value})_i$ shows the secret-shared minimum value. Finally, the server $i$ performs a join operation between all the tuples of $S(R^1)_i$ and $\langle S(t_k), S(\mathit{value})\rangle_i$, as performed when answering unconditional maximum (\texttt{QMax1}) queries; see \S\ref{subsec: Unconditional Maximum Query}.

\noindent\textit{Correctness and information leakage}. The correctness arguments and information leakage of a minimum query is similar to maximum queries.

\smallskip
\noindent\textbf{Top-k.} We again consider unconditional and conditional queries in the case of a top-k query. In both the cases, the user follows a similar approach, like maximum queries; see \S\ref{sec:maximum}; however, the user asks for top-k values instead of the maximum value.

\noindent\textit{Unconditional top-k query}. To retrieve tuples having the top-k values in the attribute $A_c$ of the relation $S(R^1)_i$, the $i^{\mathit{th}}$ server executes the following steps:
\begin{enumerate}[noitemsep,nolistsep,leftmargin=0.1in]
  \item \emph{On the relation $S(R^2)_i$}. Since the secret-shared values of the attribute $A_c$ of the relation $S(R^2)_i$ are comparable, the server $i$ finds a set of $k$ tuples, where $k$ tuples have the top-k values in the attribute $A_c$. One of the $k$ tuples is denoted by $\langle S(t_\ell), S(\mathit{value})\rangle_i$, where $S(t_\ell)_i$ is the $\ell^{\mathit{th}}$ secret-shared tuple-id (in the attribute \texttt{SSTID}) and $S(\mathit{value})_i$ is the secret-shared value of the $A_c$ attribute in the $j^{\mathit{th}}$ tuple.

\begin{table*}[t]
\scriptsize
\begin{tabular}{|l|l|l|l|l|l|}\hline

TID ($r$) & Dept & SM result ($o$) & Count ($a$)& $x=1$ & $x=2$ \\ \hline\hline

3& Testing & 1 & 1 &
$r\times o[1-(\mathit{signbit}(x-1)+\mathit{signbit}(1-x))] =3$ &
$r\times o[1-(\mathit{signbit}(x-1)+\mathit{signbit}(1-x))]=0$  \\\hline

2& Security & 0 &1&
$r\times o[1-(\mathit{signbit}(x-1)+\mathit{signbit}(1-x))]=0$ &
$r\times o[1-(\mathit{signbit}(x-1)+\mathit{signbit}(1-x))]=0$  \\\hline

5& Testing & 1 &2&
$r\times o[1-(\mathit{signbit}(x-2)+\mathit{signbit}(2-x))]=0$ &
$r\times o[1-(\mathit{signbit}(x-2)+\mathit{signbit}(2-x))]=5$   \\\hline

4& Design & 0 &2&
$r\times o[1-(\mathit{signbit}(x-2)+\mathit{signbit}(2-x))]=0$ &
$r\times o[1-(\mathit{signbit}(x-2)+\mathit{signbit}(2-x))]=0$  \\\hline

1& Design & 0 &2&
$r\times o[1-(\mathit{signbit}(x-2)+\mathit{signbit}(2-x))]=0$ &
$r\times o[1-(\mathit{signbit}(x-2)+\mathit{signbit}(2-x))]=0$ \\\hline\hline

6& Design & 0 &2&
$r\times o[1-(\mathit{signbit}(x-2)+\mathit{signbit}(2-x))]=0$ &
$r\times o[1-(\mathit{signbit}(x-2)+\mathit{signbit}(2-x))]=0$ \\\hline\hline

\multicolumn{4}{|l|}{Tuple-ids after adding values of the columns}  & 3&5 \\\hline

\end{tabular}
\caption{Knowing tuple-ids of employees working in testing department.}
\label{tab:Knowing tuple ids}
\end{table*}

  \item \emph{On the relation $S(R^1)_i$}. Now, the server $i$ performs the join of all the top-$k$ tuples with all the tuples of the relation $S(R^1)_i$ by comparing the tuple-ids (\texttt{TID} attribute's values) of the relation $S(R^1)_i$:
      \centerline{$\textstyle \sum_{j=1}^{j=n} A_{p}[S(a_j)]_i \times ({\textnormal{\texttt{TID}}}[S(a_j)]_i \otimes S(t_\ell)_i)$}
      Where $1\leq \ell\leq  k$ and $p$ ($1\leq p \leq m$) is the number of attributes in the relation $R$ and \texttt{TID} is the tuple-id attribute of $S(R^1)_i$.
      To say, the server $i$ compares each tuple-id $\langle S(t_\ell)\rangle_i$ with each $j^{\mathit{th}}$ value of the attribute \texttt{TID} of $S(R^1)_i$ and multiplies the resultant by the first $m$ attribute values of the tuple $j$. Finally, the server $i$ adds all the values of each $m$ attribute.
\end{enumerate}

\noindent\textit{Conditional top-k query}. Answering conditional top-k queries require when all the values of the attribute $A_c$ are unique requires two communication rounds between the user and the servers, like a conditional maximum query, see \S\ref{subsec:conditional Maximum Query}, as follows:

\noindent\emph{Round 1}. The user obliviously knows the tuple-ids of the relation $S(R^1)$ satisfying the query predicate.

\noindent\emph{Round 2}. The user interpolates the received tuple-ids and sends the desired tuple-ids in cleartext to the servers. Each server $i$ finds the top-k values of the attribute $A_c$ in the requested tuple-ids by looking into the attribute \texttt{CTID} of the relation $S(R^2)_i$ and results in a set of $k$ tuples. Now, the server $i$ performs a join operation between all the tuples of $S(R^1)_i$ and each of the $k$ tuples of the relation $S(R^2)$, as performed above in answering an unconditional top-k query.

\noindent\textbf{Note.} A reverse-top-k query can also be executed in the same manner as top-k queries; however, the user asks for the minimum-k values.

\section{Methods for Finding Tuple-Ids}
\label{app_sec:Methods for Finding Tuple Ids}
A trivial solution for knowing the tuple-ids satisfying a query predicate is given in \S\ref{subsec:conditional Maximum Query} that transmits $n$ numbers from each server to the user. In the following method, we allow the adversary to know an upper bound on the number of tuples, say $\mathcal{T}$, satisfy the query predicate. The method executes $\mathcal{T}$ computations on each tuple and maintains $\mathcal{T}$ variables for each tuple. Thus, the server performs significant computations, when $\mathcal{T}$ is large.

\medskip\noindent\textbf{The method.} The server creates $\mathcal{T}$ columns,\footnote{The user either provides an upper bound on the number of tuples that can satisfy the query predicate or knows the occurrences of the query predicate by executing the count query.} one for each tuple-id that satisfies the query predicate, say $v$. Note that actually we do not need to create any column during implementation, we need to have $\mathcal{T}$ variables. For the purpose of explanation, we show $\mathcal{T}$ columns. Each column has allocated one of the values from 1 to $\mathcal{T}$ of secret-share form (provided by the user). After an oblivious computation over each tuple, if there are $\mathcal{T}$ occurrences of $v$, then each of the $\mathcal{T}$ columns will have one of the exact tuple-id where $v$ occurs. The server executes the following operation:
\centerline{$r\times o[1-(\mathit{signbit}(x-a)+\mathit{signbit}(a-x))]$}
Where $r$ is the tuple-id; $o= A_{\ell}[S(a_i)]\otimes S(v)$, $1\leq i\leq n$, \textit{i}.\textit{e}., the resultant output of matching the predicate $v$ with each value of the attribute $A_{\ell}$; $a= \textstyle\sum_{i=1}^{i=n} A_{\ell}[S(a_i)]\otimes S(v)$, \textit{i}.\textit{e}., the accumulated counting of the predicate $v$ in the attribute $A_{\ell}$; and $x$ ($1\leq x\leq \mathcal{T}$) is a value of the column, created for storing the tuple-id.

\noindent\emph{Details}. For $r^{\mathit{th}}$ ($1\leq r\leq n$) value of the attribute $A_{\ell}$, the server executes counting operations for finding the occurrences of $v$ in $A_{\ell}$. The occurrences of $v$ in the above-mentioned equation is denoted by $a$. For each resultant $a$, the server compares $a$ against each of the $\mathcal{T}$ values using 2's complement method (as given in \S\ref{sec:Building Blocks of the Algorithms}). The occurrence of $v$ matches with only one of the $\mathcal{T}$ values, and thus, $\mathit{signbit}(x-a)+\mathit{signbit}(a-x)$ results in $0$, \textit{i}.\textit{e}., the difference of signbits of comparing two identical numbers is $0$. For all the other subtraction, it will be either $1$ or $-1$ of secret-share form. Note that for all the values of $\mathcal{T}$ that do not match with $a$, the above-mentioned equation will be 0 of secret-share form.

Since for the occurrence of $v$ matching with one of the values of $\mathcal{T}$, $\mathit{signbit}(x-a)+\mathit{signbit}(a-x)$ results in $0$, we subtract it from $1$ to keep $1$ on which we can multiply the tuple-id $r$. Thus, if the tuple $r$ has $v$ in the attribute $A_{\ell}$, the server keeps $r$ to one of the $\mathcal{T}$ columns. It is important to note that if the $r^{\mathit{th}}$ tuple has $v$ in attribute $A_{\ell}$ and $(r+1)^{\mathit{th}}$ tuple do not have $v$ in attribute $A_{\ell}$, the value of accumulated count, $a$, will be same for the tuples $r^{\mathit{th}}$ and $(r+1)^{\mathit{th}}$. Hence, the server may also keep the $(r+1)^{\mathit{th}}$ tuple-id in the same column where it has kept $r^{\mathit{th}}$ tuple-id. In order to prevent this, we also multiply the result of the string-matching operation (denoted by $o$, see the above equation. Thus, the $(r+1)^{\mathit{th}}$ tuple-id will not be stored. Finally, the server performs the addition operations on each $\mathcal{T}$ column and sends the final sum of each column to the user.

\noindent\textit{\underline{Example}.} Table~\ref{tab:Knowing tuple ids} shows an implementation of tuple-id finding method in cleartext to know the tuple-ids that have \texttt{Dept = Testing}; see Figure~\ref{fig:employee1 relation} for Employee relation. Note that for each row, we perform string-matching operations whose results are stored in the variable $o$, and all the occurrences of the query predicate are stored in the variable $a$. The user asks the server to create two columns ($\mathcal{T}=2$) for keeping tuple-ids.

For the first tuple, the string-matching operation results in $o=1$ and $a=1$, since the occurrence of the query predicate (\texttt{Testing}) matches with the department of the first tuple. The server computes the signbit (by placing $x=1$ and $a=1$) that results in $0$, and subtracts it from $1$ before multiplying by $r=3$ and $o=1$. Hence, the first column keeps the tuple-id 3. The second column of the first row has 0, since $\mathit{signbit}(2-1)+\mathit{signbit}(1-2)$ $=$ $0+1=1$. Note that when processing the second row, the server finds the signbit of $a$ equals to the value of the first column, while the second tuple does not have \texttt{Testing} department. The multiplication of the resultant of the signbit comparison by $o$ makes the values of the first column $0$, while the second column has $0$ too. The server processes the third tuple like the first tuple. Here, the second column keeps the tuple-id, since for the second column the current value of accumulated count $a$ matches with the column number, while the first column stores $0$, due to $1-(\mathit{signbit}(1-2)+\mathit{signbit}(2-1)$ $=$ $1-(1+0)$. The server processes the remaining tuples in a similar manner.

\section{Impact of Number of Shares}
\label{app_sec:Impact of Number of Shares}

In this section, we discuss the impact of number of shares on query execution (Experiment 5).

\smallskip
\noindent\textbf{\textit{1D-count query.}} Consider a 1D-count on OK attribute (see Figure~\ref{fig:Impact of the number of shares}). Each OK value needs six digits, which we denote as: $\langle d_1,d_2,d_3,d_4,d_5,d_6 \rangle$. In order to evaluate a query predicate over OK attribute in a 1D-count query, using one round of communication between the user and the servers, we need at most thirteen servers/shares, as mentioned in \S\ref{sec:Building Blocks of the Algorithms}.

When using three servers, the computation time at each server and the user was 81s and 17s, respectively. Here, the servers can only compare individual digits of each OK value against the query predicate; they cannot evaluate the entire query predicate by comparing the entire OK value. Thus, the server sends partial results corresponding to each digit of OK value to the user. For each tuple in the result, the user, then, interpolates  string-matching resultant of each digit, multiplies them, and finally, adds 1M values, resulting in an answer to the count query.

In the case of five shares, each server checks two digits of each OK value against the query predicate, \textit{i}.\textit{e}., the server checks $\langle d_1,d_2\rangle$, $\langle d_3,d_4 \rangle$, and $\langle d_5,d_6 \rangle$, and sends the partial results to the user. Note that checking two digits of each OK value requires more multiplication and modulus operations than using three shares, and thus, the server computation time increases to 83s. Here, the user receives a smaller set of partial results, and thus, the user's interpolation task reduces. 
Note that the total time when using five shares is higher than three shares, since the server performs more computations. In the case of eleven shares, the computation time at the server is higher as compared to three and five shares, since the server is able to check five digits ($\langle d_1,d_2,d_3,d_4,d_5\rangle$) of the query predicate. In the case of fifteen servers, each server checks the entire predicate, and hence more computation time is required at the server, due to more multiplication and modulo operation. However, in the case of fifteen shares, the user pays only for interpolating one value, which is the answer to the count query.

\smallskip
\noindent\textbf{\textit{2CE-count.}} Now, consider a 2CE-count query on OK and LN attributes, where the number of digits in OK and LN were 6 and 1 respectively. In order to execute a 2CE-count query, we need at most fifteen shares. A 2CE-count query execution time follows a trend similar to 1D-count query (see Figure~\ref{fig:Impact of the number of shares}) except that there are more number of digits involved. Hence, we skip details here.

\smallskip
\noindent\textbf{\textit{3DE-count.}} 3DE-count queries were executed on OK, PK, and LN attributes, which have in total 12 digits, hence, we need at least 25 shares to compute the answer of a 3DE-count query in one round. Since we use at most fifteen shares, the servers send the partial results of string-matching, as 1D-count query. The user interpolates them and obtains the answer at their end. This query follows a similar trend like 1D-count query; hence, we omit details here.

\begin{figure}[t]
    	   \centering
	   \includegraphics[scale=0.34]{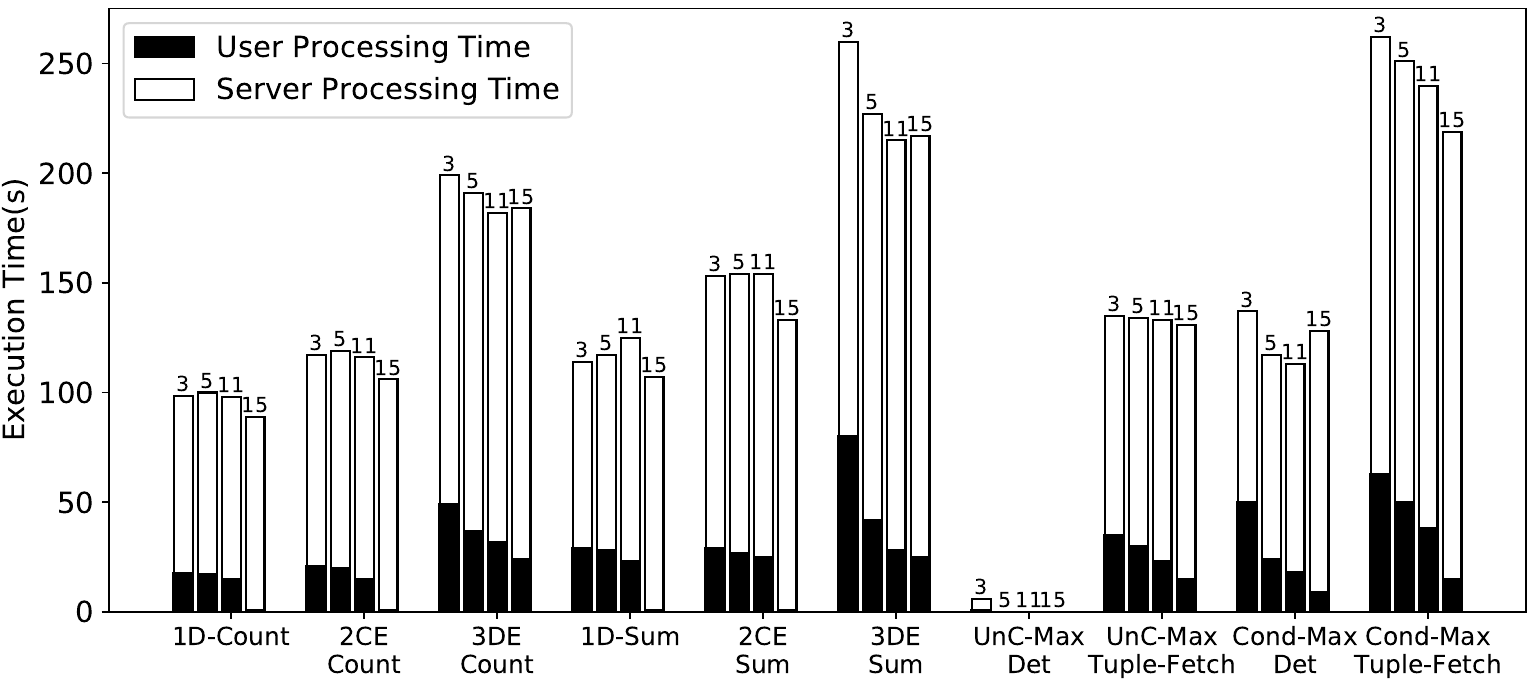}
	   \caption{Impact of the number of shares, using a single threaded implementation on 32GB RAM, 2.5GHz Intel Xeon CPU.}
	   \label{fig:Impact of the number of shares}
\end{figure}

\smallskip
\noindent\textbf{\textit{1D-sum.}} Sum queries behave differently than count queries, with an increasing number of shares. 1D-sum queries include query predicate on OK attribute, wherein each value has six digits, which we represent as $\langle d_1,d_2,d_3,d_4,d_5,d_6 \rangle$. Generally, in a sum query, if the server does not have enough shares, they need to communicate with the user, who reduces the degree of the polynomial of searching predicate attribute, and then, the user again sends the shares of string-matching resultant to the server. Afterwards, the server performs sum operations by multiplying $i^{\mathit{th}}$ secret-shared result with the $i^{\mathit{th}}$ value of desired attribute on which the user is executing a sum query, and then, the server adds all the values.

As we increase the number of servers from three to fifteen, the server computation time increases, due to more computations, like count 1D-count queries, and the user time decreases. Also, here, the total time increases when going to three to five shares, like 1D-count query. However, note that the time when using eleven shares, in the case of 1D-sum query is higher than five shares, unlike 1D-count query. 
In the case of eleven shares, the server checks $\langle d_1,d_2,d_3,d_4,d_5\rangle$ and sends their output to reduce the degree of the polynomial. However, the server does not send the output of string-matching operation over $\langle d_6\rangle$ digit. Now, the user creates five shares of each value after reducing the degree. On receiving new shares, the server  multiplies the $i^{\mathit{th}}$ new share to the output of $\langle d_6 \rangle$ digit string-matching, whose resultant is multiplied by $i^{\mathit{th}}$ share of the attribute on which the sum operation is carried out. Note that after this multiplication, the degree of the polynomial is four; thus, the user sent five shares to recover the secret value. Note that while user time is almost same when using five or eleven shares, the server time is higher in the case of eleven shares, since the server is matching almost the entire query predicate, except the last digit. Thus, the server time in the case of eleven shares is higher than five shares. In the case of fifteen shares, the user time is minimum, since servers sends the final answer to the sum query, while server time is maximum.

\smallskip
\smallskip
\noindent\textbf{\textit{2CE-sum query.}} Now, consider a 2CE-sum query on OK and LN attributes, where the number of digits in OK and LN were six and one, respectively. In order to execute this 2CE-sum query, we need at most fifteen shares. A 2CE-sum query execution time follows a similar trend like of a 1D-sum query (see Figure~\ref{fig:Impact of the number of shares}). Hence, we omit details here.

\smallskip
\noindent\textbf{\textit{3DE-sum query.}} 3DE-sum query is also executed similar to a 1D-sum query; hence, we omit details here.

\smallskip
\noindent\textbf{\textit{UnC-Max-Tuple-Fetch query.}} While retrieving a tuple having the maximum value for an attribute, say $A_c$, the server joins two relations $S(R^1)$ and $S(R^2)$, based on \texttt{TID}, as mentioned in \S\ref{subsec: Unconditional Maximum Query}. For both 1M and 6M tuples dataset, each \texttt{TID} consists of seven digits, which we represent as $\langle d_1,d_2,d_3,d_4,d_5,d_5,d_6,d_7\rangle$. When increasing number of shares from three to fifteen, the server time increases and the user time decreases, similar to all the above-mentioned queries. Moreover, in this case, the total time of computation is decreasing. For this the reason is as follows:

During the join operation, the degree of the polynomial used to create shares of \texttt{TID} values increases. In order to execute string-matching over \texttt{TID} values and tuple retrieval in one round, we need at least fifteen and sixteen shares, respectively. Thus, in our setting, the user needs to reduce the degree of string-matching resultant and re-generate shares of this to fetch the tuple, regardless of three, five, eleven, or fifteen shares.

In the case of three shares, the server compares only each digit and sends partial results to the user. After interpolating the partial results, which consists of zeros for every tuple except for one, for each tuple, the user creates three secret-shares of this vector and sends to the servers to retrieve the desired tuple. This operation requires interpolating seven shares, and then, generating three new shares.

In the case of five shares, the server compares and sends partial results of string-matching over $\langle d_1,d_2\rangle$, $\langle d_3,d_4\rangle$, $\langle d_5,d_6\rangle$ to the user. Note that the user interpolates three shares and generates five new shares. Hence, the user time decreases as compared to the case of three shares.

In the case of eleven shares, each server sends partial results of string-matching over $\langle d_1,d_2,d_3,d_4,d_5\rangle$ its share, and the user generates five new shares. Thus, the user time again decreases in this case. In the case of fifteen servers, each server checks the entire \texttt{TID} value and sends partial results for degree reduction. After interpolating the values, the user generates three new secret-shared files. Thus, the user time again decreases in this case as compared to eleven servers. In addition, as the servers check more number of digits in the \texttt{TID} value, their time increases.

\smallskip
\noindent\textbf{\textit{Cond-Max-Det.}} As mentioned in \S\ref{subsec:conditional Maximum Query}, finding the maximum value for conditional query requires at least two rounds of communication, when having enough shares. For this query, we set query predicate on OK attribute, which has six digits in every value. Hence, checking the query predicate on OK values in only one communication round, requires at least thirteen shares.

In the case of less number of shares (\textit{e}.\textit{g}., three, five, or eleven), the server first checks partial query predicates and sends results to the user for degree reduction, like 1D-count query. The user decreases the degree of string-matching resultant and sends new shares, where the new $i^{\mathit{th}}$ share gets multiplied by $i^{\mathit{th}}$ value of \texttt{Index} attribute to know the tuple-ids. Finally, the server sends the resultant to the user. After interpolation, the user knows the tuple-ids that satisfy the query predicate. Hence, in the case of three, five, or eleven shares, the user executes the interpolation operation two times, while in the case of fifteen shares, the user executes interpolation operation only one time. Hence, the user computation time reduced when increasing the number of shares. After knowing tuple-ids, the user asks the server to find the maximum value in the given tuple-ids using the relation $S(R^2)$, and this operation takes same time regardless of the number of shares. While increasing the number of shares, the server computation time increases, as it happened in all above-mentioned queries. 

\smallskip
\noindent\textbf{\textit{Cond-Max-Tuple-Fetch.}} Fetching a tuple having maximum value according to a conditional query requires two rounds, as stated in \S\ref{subsec:conditional Maximum Query}.
The first round at the server is identical to Cond-Max-Det queries. However, in the second round, the server joins $S(R^1)$ with one of the tuples of $S(R^2)$ based on \texttt{TID} attribute of $S(R^1)$. In this query, we set a condition on OK attribute, which requires six digits to represent a value. Hence, we need at least $2\times 6+1=13$ shares to know the tuple-ids in one communication round. Further, to get the desired tuple, based on join over \texttt{TID}, that has seven digits, we need at least sixteen shares (fifteen shares for string-matching operations ($2\times7+1=15$) and one more share for reconstructing the tuple values).

In the first round, the user interpolates  at least twice in the case of three, five, and eleven shares, and at least once in the case of fifteen shares to know the tuple-ids. Further, we use at most fifteen shares in our experiments; hence, the user needs to reduce the degree at least once, of string-matching resultant in the second round to get the desired tuple. Thus, user computation time decreases as the number of shares increases. Further, the server time decreases, as we increase the number of shares, similar to other queries.

\section{Security Proof Outline}
\label{app_sec:Security Proof}
Now, we provide the security proof outline for \textsc{Obscure}.  In our context, we, first, need to show that an adversary cannot distinguish any two queries of the same type based on the output size, \textit{i}.\textit{e}., the query/user privacy will be maintained. Once we can prove the query privacy, we will show how the server privacy (\textit{i}.\textit{e}., not revealing more information to the user) is achieved.

\begin{theorem}
If the adversarial cloud can distinguish two input queries, then either the random polynomials  used for creating shares of a query is not correct or \textsc{Obscure} does not provide query privacy.
\end{theorem}
In order to show that the adversary can never know the exact query value, we consider two instances of the datasets, as follows: $D_1$ and $D_2$, where $D_1$ differs from $D_2$ only at one value each, say $v_1$ and $v_2$, \textit{i}.\textit{e}., $v_1$ is in $D_1$ but $D_2$ and $v_2$ is in $D_2$ but $D_1$. Here, we show that if the adversary can distinguish the single different value in $D_1$ and $D_2$, she can break \textsc{Obscure}. In this setting, the server executes the input queries on $D_1$ and $D_2$.

By our assumption of ciphertext indistinguishability (mentioned in \S\ref{subsec:Security Properties}), the adversary cannot distinguish that $D_1$ and $D_2$ are identical or different. Note that if the DB owner uses only one polynomial (\textit{i}.\textit{e}., a weak cryptographic plan), then the adversary can find which value is the only single values of $D_1$  that is different from values of $D_2$. Moreover, it reveals frequency-count of values.

Now assume the queries for the value $v_1$ and $v_2$ that will be mapped to secret-shared queries, $q_{v1}(D_1)$ and  $q_{v1}(D_2)$, respectively. Further, assume that $q_{v1}(D_1)$ and $q_{v2}(D_2)$ are identical. Hence, the adversary will consider both of them as an identical query, while they are for different queries. Hence, the adversary cannot distinguish two queries. Now, assume that $q_{v1}(D_1)$ and $q_{v2}(D_2)$ are different, and here the adversary objective is to deduce which tuple of relations satisfy the query or not. If the adversary cannot know which tuple is satisfying the query or not, the adversary can distinguish two queries, as well as, the two datasets. This violates our assumption of ciphertext indistinguishability of the dataset. Thus, the adversary cannot distinguish two datasets or two queries.

\smallskip Now, we provide an intuition that how does the server privacy is maintained. Recall that we assumed a trusted user. In response to a query, the user obtains some numbers. Since the servers cannot distinguish between two queries and they follow the algorithm on the entire dataset, the server sends only the desired answer to the query.

\end{document}